\documentclass[10pt]{IEEEtran}  

\newcommand{\onlyphd}[1]{}
\newcommand{\onlypaper}[1]{#1}
\newcommand{\excluded}[1]{}
\newcommand{\selector}[2]{\onlypaper{#1}\onlyphd{#2}} 

\newcommand{\furtherdetails}[1]{}
\usepackage[pdftex]{hyperref} 

\usepackage{amsmath,amsthm, amssymb, epsfig, verbatim, color, amsfonts} 
\usepackage{graphicx}

\theoremstyle{plain}
\newtheorem{theorem}{Theorem}
\newtheorem{lemma}{Lemma}

\newtheorem{corollary}{Corollary} 
\theoremstyle{definition}
\newtheorem{definition}{Definition}
\newtheorem{example}{Example}
\newtheorem{proposition}{Proposition}

\def\vr{\mathbf}

\def\Pr{\mathrm{Pr}}
\def\E{\mathbb{E}}
\def\Ber{\mathrm{Ber}}

\def\conv{\mathrm{conv}}

\def\Ind{\mathrm{Ind}} 
\def\Ber{\mathrm{Ber}}
\def\vol{\mathrm{vol}}
\def\dim{\mathrm{dim}}
\def\half{\tfrac{1}{2}}

\def\msg{\mathrm{\mathbf{m}}} 
\def\msg{\mathbf{m}} 

\def\endofproof{\hspace{\stretch{1}}$\Box$}
\def\defeq{\triangleq} 

\def\etal{\textit{et al}}
\def\hstr{\hspace{\stretch{1}}}

\def\Holder{H\"older}
\newcommand{\tsubs}[1]{{\scriptscriptstyle \mathrm{#1}}}

\newcommand{\fpr}[1]{\breve{#1}} 
\def\ntoinfty{\arrowexpl{n \to \infty}}

\newcommand{\argmax}[1] {\underset{#1}{\textstyle \mathrm{argmax}}\hspace{0.5ex}}
\newcommand{\argmin}[1] {\underset{#1}{\textstyle \mathrm{argmin}}\hspace{0.5ex}}
\newcommand{\arrowexpl}[1] {\underset{#1}{\textstyle \longrightarrow}}

\newcommand{\arrowabovebelow}[2] {\underset{#2}{\overset{#1}{\textstyle \longrightarrow}}}
\newcommand{\toinprobl}[1] {\arrowabovebelow{Prob.}{#1}}

\newcommand{\todo}[1]{\textcolor{red}{TODO:} \begin{color}{blue}#1\end{color}\\}

\allowdisplaybreaks[1]

\def\Cavc{C_{\scriptscriptstyle \mathrm{AVC}}}
\def\Dpred{\Delta_\tsubs{pred}}
\newcommand{\Dpredsup}[1]{\Delta_\tsubs{pred}^{#1}}

\def\Rtarget{{R_T}}

\title{Universal Communication over Arbitrarily Varying Channels}

\author{Yuval Lomnitz, Meir Feder \\
Tel Aviv University, Dept. of EE-Systems  \\
Email: \{yuvall,meir\}@eng.tau.ac.il}

\begin{document}
\maketitle

\begin{abstract}
We consider the problem of universally communicating over an unknown and arbitrarily varying channel, using feedback. The focus of this paper is on determining the input behavior, and specifically, a prior distribution which is used to randomly generate the codebook. We pose the problem of setting the prior as a sequential universal prediction problem, that attempts to approach a given target rate, which depends on the unknown channel sequence. The main result is that, for a channel comprised of an unknown, arbitrary sequence of memoryless channels, there is a system using feedback and common randomness that asymptotically attains, with high probability, the capacity of the time-averaged channel, universally for every sequence of channels. While no prior knowledge of the channel sequence is assumed, the rate achieved meets or exceeds the traditional arbitrarily varying channel (AVC) capacity for every memoryless AVC defined over the same alphabets, and therefore the system universally attains the random code AVC capacity, without knowledge of the AVC parameters. The system we present combines rateless coding with a universal prediction scheme for the prior. We present rough upper bounds on the rates that can be achieved in this setting and lower bounds for the redundancies.
\end{abstract}

\section{Introduction}\label{sec:intro}
We consider the problem of communicating over an unknown and arbitrarily varying channel, with the help of feedback. We would like to minimize the assumptions on the communication channel as much as possible, while using the feedback link to learn the channel. The main questions with respect to such channels are how to define the expected communication rates, and how to attain them universally, without channel knowledge.

The traditional models for unknown channels \cite{Lapidoth_AVC} are compound channels, in which the channel law is selected arbitrarily out of a family of known channels, and arbitrarily varying channels ({AVC's}), in which a sequence of channel states is selected arbitrarily. The well known results for these models \cite{Lapidoth_AVC} do not assume adaptation. Therefore, the AVC capacity, which is the supremum of the communication rates that can be obtained with vanishing error probability over any possible occurrence of the channel state sequence, is in essence a worst-case result. For example, if one assumes that $y_i$, the channel output at time $i$, is determined by the probability law $W_i(y_i | x_i)$ where $x_i$ is the channel input, and $W_i$ is an arbitrary sequence of conditional distributions, clearly no positive rate can be guaranteed a-priori, as it may happen that all $W_i$ have zero capacity, and therefore the AVC capacity is zero. This capacity may be non-zero only if a constraint on $W_i$ is defined. In this paper we use the term ``arbitrarily varying channel'' in a loose manner, to describe any kind of unknown and arbitrary change of the channel over time, and the acronym ``AVC'' to refer to the traditional model \cite{Lapidoth_AVC}.

Other communication models, which allow positive communication rates over such AVC's were proposed by the authors and others \cite{Ofer_ModuloAdditive, Eswaran, YL_individual_full, YL_UnivModuloAdditive}. Although the channel models considered in these papers are different, the common feature distinguishing them from the traditional AVC setting is that the communication rate is adaptively modified using feedback. The target rate is known only a-posteriori, and is gradually learned throughout the communication process. By adapting the rate, one avoids worst case assumptions on the channel, and can achieve positive communication rates when the channel is good. However, in the aforementioned communication models, the distribution of the transmitted signal is fixed and independent of the feedback, and only the rate is adapted. Specifically in the ``individual channel'' model \cite{YL_individual_full} for reasons explained therein, the distribution of the channel input is fixed to a predefined prior. Likewise, Eswaran~\etal~\cite{Eswaran} show that for a fixed prior, the mutual information of the averaged channel can be attained. Clearly, with this limitation these systems are incapable of universally attaining the channel capacity in many cases of interest. For example, consider even the simple case where the channel is a compound memoryless channel, i.e. the conditional distributions $W_i = W$ are all constant but unknown.

In the last paper \cite{YL_UnivModuloAdditive}, the problem of universal communication was formulated as that of a competition against a reference system, comprised of an encoder and a decoder with limited capabilities. For the case where the channel is modulo-additive with an individual, arbitrary noise sequence, it was shown possible to asymptotically perform at least as well as any finite-block system (which may be designed knowing the noise sequence), without prior knowledge of the noise sequence. However, this result crucially relies on the property of the modulo-additive channel, that the capacity achieving prior is the uniform i.i.d. prior for any noise distribution. To extend the result to more general models, we would like to be able to adapt the input behavior. The key parameter to be adapted is the ``prior'', i.e. the distribution of the codebook (or equivalently the channel input), since it plays a vital role in the converse as well as the attainability proof of channel capacity and is the main factor in adapting the message to the channel \cite{Shannon48}.

In a crude way we may say that previous works achieve various kinds of ``mutual information'' for a fixed prior and any channel from a wide class, by mainly solving problems of universal decoding and rate adaptation. However to obtain more than the ``mutual information'', i.e. the ``capacity'', one would need to select the prior in a universal way.

Prior adaptation using feedback is a well known practice for static or semi-static channels. Two familiar examples are bit and power loading performed in Digital Subscriber Lines ({DSL-s}) \cite{DSL_ChowCioffi}, and precoding for in multi-antenna systems \cite{MIMO_Precoding_Love08} which is performed in practice in wireless standards such as WiFi, WiMAX and LTE. If the channel can be assumed to be static for a period of time sufficient to close a loop of channel measurement, feedback and coding, then an input prior close to the optimal one can be chosen. In the theoretical setting of the compound memoryless channel where $\Pr(Y_i | X_i) = W(Y_i | X_i)$, where $W$ is unknown but fixed, a system with feedback can asymptotically attain the channel capacity of $W$, without prior knowledge of it, by using an asymptotically small portion of the transmission time to estimate the channel, and using an estimate of the optimal prior and the suitable rate during the rest of the time \cite{Mahajan09}. All models for prior adaptation that we are aware of, use the assumption that the knowledge of the channel at a given time yields non trivial statistical information about future channel states, but do not deal with arbitrary variation.

The question that we deal with in this paper is: assuming a channel which is \emph{arbitrarily} changing over time, is there any merit in using feedback to adapt the input distribution, and what rates can be guaranteed? As a target, we would have liked to consider the most general variation of the channel (as in the unknown vector channel model \cite{YL_UnivModuloAdditive}), however to start our exploration, we focus on channel models which are memoryless in the input, i.e. whose behavior at a certain time does not depend on any previous channel \emph{inputs}. The most general model that does not include memory of the input is that of an unknown sequence of memoryless channels (which is in essence an AVC without constraints) and this is the main model considered in this paper. The motivation for avoiding memory of the input can be appreciated by considering the negative examples in \cite{YL_UnivModuloAdditive}.

We now give a brief overview of the structure and the results of this paper. In Section~\ref{sec:problem_statement_and_notation} we state the problem, and define several communication rates (as a function of the channel sequence) that would be of interest. In order to focus thoughts on questions related to the problem of determining the \emph{prior}, we initially adopt an abstract model of the communication system, stripping off the details of communication, such as decoding, channel estimation, overheads, error probability, etc. We begin by presenting an easier synthetic problem, in which all previous channels are known (Section~\ref{sec:toy_problem}). This problem may represent a channel which changes its behavior in a block-wise manner and remains i.i.d. memoryless during each block (a subset of the original problem). This problem is related to standard prediction problems (Section~\ref{sec:toy_categorization}), and used as a tool to gain insight into the prediction problem involved, present bounds on what can be achieved universally, and develop the techniques that will be used later on. Furthermore, we show that even for this easier problem there is no hope to attain the channel capacity universally and we would have to settle for lower rates (Section~\ref{sec:regret_LB}). The attained rate is the maximum over the prior, of the averaged mutual information (Theorem~\ref{theorem:prior_predictor_exp}). In Section~\ref{sec:arbitrary_channel_var}, we return to the main problem, and show that the rate that can be attained when the past channel is not known, but is estimated from the output, is lower. We focus on the capacity of the time-averaged channel. We show this rate is the best achievable rate that does not depend on the order of the channel sequence (Theorem~\ref{theorem:C_overlineW_optimality}), and present the main result showing that this rate is indeed achievable (Theorem~\ref{theorem:C_overlineW_achievability}). Furthermore, this rate meets or exceeds the AVC capacity, and essentially equals the ``empirical capacity'' defined by Eswaran~\etal~\cite{Eswaran}. We present a scheme based on rateless coding and combines a prior predictor that attains this rate. In Section~\ref{sec:arbitrary_var_exp_prior_predictor}, the prior predictor is developed under abstract assumptions regarding the channel estimation and decoding rate. In Section~\ref{sec:mainproof}, we present and analyze the full communication system and prove the main result. Finally, Section~\ref{sec:discussion} is devoted to discussion and comments.

\section{Notation and problem statement}\label{sec:problem_statement_and_notation}
\subsection{Notation}\label{sec:notation}
We denote random variables by capital letters and vectors by boldface. However for probabilities which are sometimes treated as vectors we use regular capital letters. We apply superscript and subscript indices to vectors to define subsequences in the standard way, i.e. $\vr x_i^j \defeq (x_i, x_{i+1}, ... , x_j)$, $\vr x^i \defeq \vr x_1^i$

$I(Q,W)$ denotes the mutual information obtained when using a prior $Q$ over a channel $W$, i.e. it is the mutual information $I(Q,W)=I(X;Y)$ between two random variables with the joint probability $\Pr(X,Y)=Q(X) \cdot W(Y|X)$. $C(W)$ denotes the channel capacity $C(W) = \max_Q I(Q,W)$. For discrete channels, the channel $W(y|x)$ is sometimes presented as a matrix where $W(y|x)$ is in the $x$-th column and the $y$-th row. Logarithms and all information quantities are base $2$ unless specified otherwise.

We denote by $\Delta_{\mathcal{X}}$ the unit simplex $\Delta_{\mathcal{X}} \defeq \{Q: \sum_{x \in \mathcal{X}} Q(x) = 1\}$, i.e. the set of all probability measures on $\mathcal{X}$.

$\Ber(p)$ denotes a Bernoulli random variable with probability $p$ to be $1$. $\Ind(\cdot)$ denotes an indicator function of an event or a condition, and equals $1$ if the event occurs and $0$ otherwise. We use ``$\ldots$'' to denote simple mathematical inductions, where the same rule is repeatedly applied, for example $a_n \leq n \cdot a_{n-1} \leq \ldots \leq n! \cdot a_0$.

A hat $\hat \Box$ denotes an estimated value, and a line $\overline \Box$ denotes an average value. The empirical distribution of a vector $\vr x$ of length $n$ is a function representing the relative frequency of each letter,
\begin{equation}\label{eq:325}
\hat P_{\vr x}(x) = \frac{\sum_{i=1}^n \Ind(x_i = x)}{n},
\end{equation}
where the subscript identifies the vector. The conditional empirical distribution of two equal length vectors $\vr x, \vr y$ is defined as
\begin{equation}\label{eq:331}
\hat P_{\vr y|\vr x}(y|x) = \frac{\hat P_{\vr x, \vr y}(x,y)}{\hat P_{\vr x}(x)}
.
\end{equation}

\subsection{Problem setting}\label{sec:problem_setting}
Let $\mathcal{X},\mathcal{Y}$ be sets defining the input and output alphabets, respectively. Both $\mathcal{X},\mathcal{Y}$ are assumed to be finite, unless stated otherwise.\footnote{Note that the results in Section~\ref{sec:toy_problem},\ref{sec:arbitrary_channel_var} do not require $\mathcal{Y}$ to be finite} Let $\{W_i\}_{i=1}^n$ be a sequence of memoryless channels over $n$ channel uses. Each $W_i$ is a conditional distribution $W_i(y|x)$ where $x \in \mathcal{X}$ and $y \in \mathcal{Y}$ represent an input and output symbol respectively. The conditional distribution of the output vector $\vr Y$ given the input vector $\vr X$ is given by:
\begin{equation}\label{eq:687}
\Pr(\vr Y | \vr X) = \prod_{i=1}^n W_i(Y_i | X_i)
.
\end{equation}
The sequence of channels $W_i$ is arbitrary and unknown to the transmitter and the receiver. We assume the existence of common randomness (i.e. that the transmitter and the receiver both have access to some random variable of choice). There exists a feedback link between the receiver and the transmitter. To simplify, we assume the feedback is completely reliable, has unlimited bandwidth and is instantaneous, i.e. arrives to the encoder before the next symbol.\footnote{The asymptotical results hold also when feedback is band limited and delayed.} We assume the system is rate adaptive, which means that the message is represented by an infinite bit sequence $\msg_0^\infty$, and the system may choose how many bits to send. The error probability is measured only over the bits which were actually sent (i.e. over the first $\lceil n R \rceil$ bits, where $R$ is the rate reported by the receiver). The system setup is \selector{presented in Figure~\ref{fig:system_adaptive}}{the same as presented in Figure~\ref{fig:system_adaptive}}.

\onlypaper{
\begin{figure*}[t]
\setlength{\unitlength}{1mm}
\hspace{\stretch{1}}
\begin{picture}(140, 30)
\put(23,16){Transmitter}\put(20,10){\line(1,0){20}}\put(40,10){\line(0,1){15}}\put(40,25){\line(-1,0){20}}\put(20,25){\line(0,-1){15}}
\put(63,20){Channel}\put(60,19){\line(1,0){20}}\put(80,19){\line(0,1){5}}\put(80,24){\line(-1,0){20}}\put(60,24){\line(0,-1){5}}
\put(103,16){Receiver}\put(100,10){\line(1,0){20}}\put(120,10){\line(0,1){15}}\put(120,25){\line(-1,0){20}}\put(100,25){\line(0,-1){15}}
\put(0,17.5){\vector(1,0){20}}\put(6,18.5){$\vr w$}\put(0,14){(message)}
\put(40,21.5){\vector(1,0){20}}\put(42,22.5){$x_i \in \mathcal{X}$}
\put(80,21.5){\vector(1,0){20}}\put(82,22.5){$y_i \in \mathcal{Y}$}
\put(100,15){\vector(-1,0){60}}\put(55,10){$f_i \in \mathcal{F}$ (feedback)}
\put(120,21.5){\vector(1,0){20}}\put(125,22.5){$R$ (rate)}
\put(120,15){\vector(1,0){20}}\put(125,16){$\hat{\vr w}$ (message)}
\put(30,5){\vector(0,1){5}}\put(30,0){$S$ (common randomness)}
\put(110,5){\vector(0,1){5}}\put(110,0){$S$}
\end{picture}
\hspace{\stretch{1}}
\caption{A rate adaptive system with feedback}\label{fig:system_adaptive}
\end{figure*}
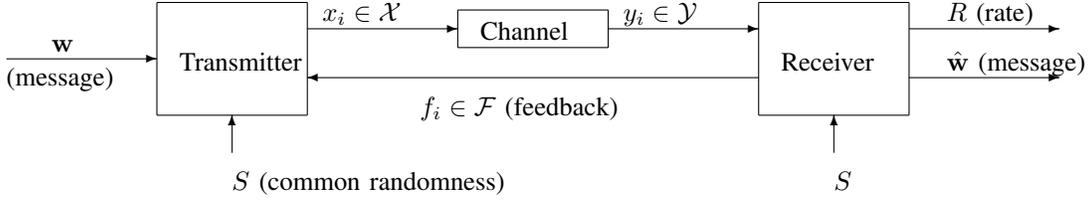
}

To simplify, we assume that there are no constraints on the channel input (such as power constraints). If such constraints exist they can be accommodated by changing the set of potential priors.

Since the channel sequence is arbitrary there is no positive rate which can be guaranteed a-priori. Instead, we define a target rate $R(W_1^n)$ as a function of the channel sequence $W_1^n$.
\begin{definition}\label{def:attainability_of_RW}
We say that a sequence of rate functions $R(W_1^n)$ is asymptotically attinable, if for every $\epsilon, \delta, \Delta > 0$ there is $n$ large enough such that there is a system with feedback and common randomness over $n$ channel uses, in which, for \emph{every} sequence $\{W_i\}_{i=1}^n$, the rate is $R(W_1^n) - \Delta$ or more, with probability of at least $1-\delta$, while the probability of error is at most $\epsilon$.
\end{definition}
In the next section we propose several potential target rates and then we would ask which of these are attainable.

\subsection{Potential target rates}\label{sec:target_rates_Ci}
With respect to the sequence $\{W_i\}$ we can define various meaningful information theoretic measures. The maximum possible rate of reliable communication is the capacity when the sequence is known a-priori (in other words, the capacity with full, non causal, channel state information at the transmitter and the receiver) and is given by:
\begin{equation}\begin{split}\label{eq:185}
C_1(W_1^n)
&=
\max_{\{Q_i\}} \frac{1}{n} \sum_{i=1}^n I(Q_i, W_i)
\\&=
\frac{1}{n} \sum_{i=1}^n \max_{Q} I(Q, W_i)
=
\frac{1}{n} \sum_{i=1}^n C(W_i)
.
\end{split}\end{equation}
Note that if constraints on the sequence $\{Q_i\}$ existed, then we would have an equality \cite{YL_PriorPrediction_ISIT2011}. The maximum rate that can be obtained with a single \emph{fixed} prior when the sequence is known is:
\begin{equation}\label{eq:189}
C_2(W_1^n) = \max_{Q}  \frac{1}{n} \sum_{i=1}^n I\left(Q, W_i\right)
.
\end{equation}
Lastly, the capacity of the time-averaged channel is:
\begin{equation}\label{eq:193}
C_3(W_1^n) = \max_{Q} I\left(Q, \frac{1}{n} \sum_{i=1}^n W_i\right) = C(\overline W)
,
\end{equation}
where we define the time-averaged channel as
\begin{equation}\label{eq:204}
\overline W(y|x) = \frac{1}{n} \sum_{i=1}^n W_i(y|x)
.
\end{equation}

Clearly, $C_1 \geq C_2 \geq C_3$ where the first inequality results from the order of maximization and the other results from the convexity of the mutual information with respect to the channel. For each of the above target rates we would like to find out whether it is achievable under the definitions above. As we shall see, $C_1$ is not achievable, $C_3$ is achievable, and $C_2$ is achievable only under further constraints imposed on the problem.

A rigorous proof that $C_1$ is the capacity of the channel sequence is left out of the scope of this paper. For our purpose, it is sufficient to observe that $C_1$ is an upper bound on the achievable rate, because the mutual information between channel input and output is maximized by a memoryless (not i.i.d.) input distribution $\prod_{i=1}^n Q_i(x_i)$. To see intuitively how $C_1$ can be achieved, consider that since $n$ can be arbitrarily large while the input and output alphabets, and thus the set of channels, remain constant, we may sort the channels into groups of similar channels, and apply block coding to each group. A close result pertaining to stationary ergodic channels appears in \cite[(3.3.5)]{Biglieri98fadingchannels}.

\section{A synthetic ``toy'' problem}\label{sec:toy_problem}
In this section we present a synthetic problem, which will help us examine the achievability of the target rates defined above in a simplified scenario, draw the links to universal prediction, and introduce the techniques that will be used in the sequel.

\subsection{Problem description}
We focus on the problem of setting a prior $\hat Q_i$ at time $i$. We assume that at each time instance $i$, the system has full knowledge of the sequence of past channels $W_1^{i-1}$. The prior prediction mechanism sets $\hat Q_i$ based on the knowledge of $W_1^{i-1}$. Then, we assume that $I(\hat Q_i, W_i)$ bits are conveyed during time instance $i$. A predictor $\hat Q_i(W_1^{i-1})$ attains a given target rate $R(W_1^n)$ if for all sequences $W_1^n$ we have $\frac{1}{n} \sum_{i=1}^n I(\hat Q_i, W_i)  \geq R(W_1^n) - \delta_n$ , and $\delta_n \ntoinfty 0$.

This abstract problem can apply to a situation where the channel sequence is constant during long blocks, and changes its value only from block to block, or from one transmission to another. In this case $i$ denotes the block index, and denoting by $m$ the constant block length, at most $m \cdot I(\hat Q_i, W_i)$ bits can be sent in block $i$. If the channel is constant over long blocks it is reasonable to assume that past channels can be estimated. Note that in addition we made the assumption that $I(\hat Q_i, W_i)$ is achievable, although this communication rate is unknown to the transmitter in advance, i.e. we ignored the problem of rate adaptation. Therefore the synthetic problem is a subset of the original problem and upper bounds that we show here apply also to the original problem.

\subsection{Classification as a universal prediction problem}\label{sec:toy_categorization}
We begin by discussing the achievability of $C_2$ for the synthetic problem. The target rate $C_2$ is special in being an additive function for each value of $Q$. Universally attaining $C_2$ under the conditions specified above, falls into a widely studied category of universal prediction problems \cite{FederMerhav98,Nicolo,Haussler,Vovk97}. Below, we present this class of problems and review some results that will be important for our discussion.

These prediction problems have the following form: let $b \in \mathcal{B}$ be a strategy in a set of possible strategies $\mathcal{B}$, and $x \in \mathcal{X}$ be a state of nature. A loss function $l(b,x)$ associates a loss with each combination of a strategy and a state of nature. The total loss over $n$ occurrences is defined as $L = \sum_{i=1}^n l(b_i, x_i)$. The universal predictor $\hat b_i(\vr x_1^{i-1})$ assigns the next strategy given the past values of the sequence, and before seeing the current value. There is a set of reference strategies $\{b_{i}^{(k)}\}_{k=1}^N$ (sometimes called experts), which are visible to the universal predictor. The target of universal prediction is to provide a predictor $\hat b_i$ which is asymptotically and universally better than any of the reference strategies, in the sense defined below.

For a given sequence $\vr x_1^n$, denote the losses of the universal predictor and the reference strategies as $\hat L \defeq \sum_{i=1}^n l(\hat b_i, x_i)$ and $L_k \defeq \sum_{i=1}^n l(b_{i}^{(k)}, x_i)$, respectively. Denote the regret of the universal predictor with respect a specific reference strategy as the excessive loss:
\begin{equation}\label{eq:239}
\mathcal{R}(k) \defeq \hat L - L_k .
\end{equation}
$\mathcal{R}_k$ is a function of the sequence $\vr x_1^n$ and the predictor. The target of the universal predictor is to minimize the worst case regret, i.e. attain
\begin{equation}\label{eq:240}
\mathcal{R}_{\mathrm{minimax}} \defeq \min_{\{\hat b_i(\cdot)\}} \max_{k} \max_{\vr x_1^n}  \mathcal{R}(k) .
\end{equation}

The reference strategies may be defined in several different ways. In the simplest form of the problem the competition is against the set of fixed strategies $b_i^{(k)} = b(k)$. The exact minimax solution is known only for very specific loss functions \cite[\S 8]{Nicolo}, and a solution guaranteeing $\max_{\vr x_1^n, k} \mathcal{R}(k) \arrowexpl{n \to \infty} 0$ is not known for general loss functions. However there are many prediction schemes which perform well for a wide range of loss functions (see references above).

In the information theoretic framework, the log-loss $l(b,x) = \log \left( \frac{1}{b(x)} \right)$, where $b(x)$ is a probability distribution over $\mathcal{X}$ is the most familiar loss function, and used in analyzing universal source encoding schemes \cite{FederMerhav98}, since $l(b,x)$ represents the optimal encoding length of the symbol $x$ when assigned a probability $b(x)$. It exhibits an asymptotical minimax regret of $\frac{1}{n} \mathcal{R}_{\mathrm{minimax}} = O \left( \frac{\log n}{n} \right)$. However in the more general setting the asymptotical minimax regret decreases in a slower rate of $\frac{1}{n} \mathcal{R}_{\mathrm{minimax}} = O \left( \frac{1}{\sqrt{n}} \right)$.  There are several loss functions which are characterized by a ``smoother'' behavior for which better minimax regret is obtained \cite[Theorem 3.1, Proposition 3.1]{Nicolo}. For some of these loss functions, a simple forecasting algorithm termed ``Follow the leader'' (FL) can be used \cite[\S 3.2]{Nicolo} \cite[Theorem 1]{FederMerhav93}. In FL, the universal forecaster picks at every iteration $i$ the strategy that performed best in the past, i.e. minimizes the cumulative loss over the instances from $1$ to $i-1$.

The archetype of loss functions for which it is not possible to obtain a better convergence rate than  $O \left( \frac{1}{\sqrt{n}} \right)$ is the absolute loss $l(b,x) = |b-x|$, where $x \in \mathcal{X} = \{0,1\}$ and $b \in \mathcal{B} = [0,1]$. The proof for the lower bound on the minimax regret \cite[Theorem 3.7]{Nicolo} is based on generating the sequence $\vr x_1^n$ randomly, and calculating the minimum \emph{expected} regret (over $\vr x$). This value is a lower bound for the minimum-maximum regret \eqref{eq:240}. To show that the regret is $\omega(\sqrt{n})$ it is enough to consider only two competitors -- one forecasting a constant zero, and one a constant one, and observe that since the cumulative losses of the two competitors always sum up to $n$, the minimum loss of the two competitors is a random variable with a standard deviation of $O(\sqrt{n})$ which is upper bounded by $\frac{n}{2}$, and therefore its expected value is $\frac{n}{2} - O(\sqrt{n})$, whereas the expected loss of the best single strategy over the random sequence cannot be better than $\frac{n}{2}$. We will use a similar idea to prove lower bounds on the regret in the current problems. For general loss functions, and specifically for the absolute loss, the simple FL strategy does not converge.

The problem of asymptotically attaining $C_2(W_1^n)$ is analogous to the standard prediction problem, where the prior $Q_i$ represents a strategy, and the channel $W_i$ represents a state of nature. Our problem is given in terms of gains rather than losses, so we may consider the loss to be $l(Q,W) = - I(Q,W)$. The regret is therefore:
\begin{equation}\label{eq:304}
\mathcal{R}_n(Q) = \sum_{i=1}^n I(Q,W_i) - \sum_{i=1}^n I(\hat Q_i,W_i).
\end{equation}
Note that the regret is defined in terms of bits rather than rates (i.e. it is not normalized), from technical reasons.

\subsection{A lower bound on the regret}\label{sec:regret_LB}
A natural question to ask is, then: what is the asymptotical form of the minimax regret expected in our case? As we will show, the prior prediction problem we posed, includes as a special case the prediction problem with the absolute loss function. Therefore, the asymptotical behavior cannot be better than $O(\sqrt{n})$, and it is not possible to apply the simple FL strategy.

The following example shows why the problem of attaining $C_2$ includes as a particular case the absolute loss function:
\begin{example}\label{example:prediction_channel1}
Consider the quaternary to binary channel ($|\mathcal{X}|=4, |\mathcal{Y}|=2$), which may be in one of two states $s \in \{0,1\}$, which define two conditional probability functions (shown as $|\mathcal{Y}| \times |\mathcal{X}|$ matrices below):
\begin{equation}\begin{split}\label{eq:305}
& W_{0}(Y|X) = \left[ \begin{array}{cccc} 1 & 0 & \half & \half \\  0 & 1 & \half & \half \end{array} \right]
\\& W_{1}(Y|X) = \left[ \begin{array}{cccc} \half & \half & 1 & 0 \\  \half & \half & 0 & 1 \end{array} \right]
.
\end{split}\end{equation}

\begin{figure}
\centering
\ifpdf
  \setlength{\unitlength}{1bp}%
  \begin{picture}(206.79, 112.54)(0,0)
  \put(0,0){\includegraphics{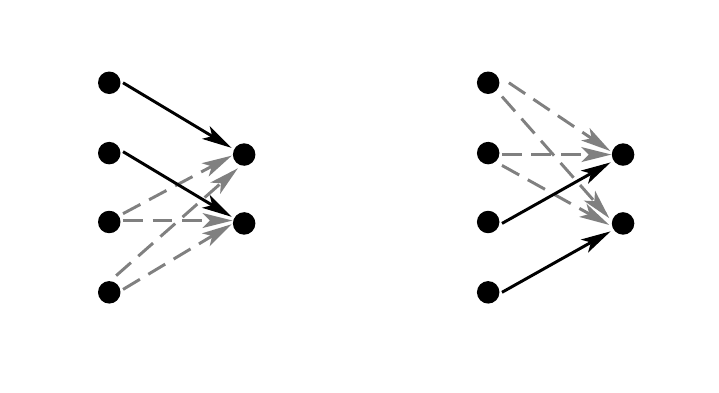}}
  \put(5.67,87.18){\fontsize{9.96}{11.95}\selectfont $0$}
  \put(5.67,65.36){\fontsize{9.96}{11.95}\selectfont $1$}
  \put(5.67,45.51){\fontsize{9.96}{11.95}\selectfont $2$}
  \put(5.67,25.67){\fontsize{9.96}{11.95}\selectfont $3$}
  \put(23.53,99.09){\fontsize{9.96}{11.95}\selectfont $X$}
  \put(63.21,81.23){\fontsize{9.96}{11.95}\selectfont $Y$}
  \put(77.10,67.34){\fontsize{9.96}{11.95}\selectfont $0$}
  \put(77.10,45.51){\fontsize{9.96}{11.95}\selectfont $1$}
  \put(25.51,7.81){\fontsize{9.96}{11.95}\selectfont $W_0(Y|X)$}
  \put(114.80,87.18){\fontsize{9.96}{11.95}\selectfont $0$}
  \put(114.80,65.36){\fontsize{9.96}{11.95}\selectfont $1$}
  \put(114.80,45.51){\fontsize{9.96}{11.95}\selectfont $2$}
  \put(114.80,25.67){\fontsize{9.96}{11.95}\selectfont $3$}
  \put(132.66,99.09){\fontsize{9.96}{11.95}\selectfont $X$}
  \put(172.35,81.23){\fontsize{9.96}{11.95}\selectfont $Y$}
  \put(186.24,67.34){\fontsize{9.96}{11.95}\selectfont $0$}
  \put(186.24,45.51){\fontsize{9.96}{11.95}\selectfont $1$}
  \put(134.65,7.81){\fontsize{9.96}{11.95}\selectfont $W_1(Y|X)$}
  \end{picture}%
\else
  \setlength{\unitlength}{1bp}%
  \begin{picture}(206.79, 112.54)(0,0)
  \put(0,0){\includegraphics{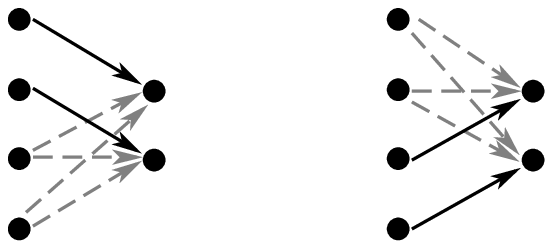}}
  \put(5.67,87.18){\fontsize{9.96}{11.95}\selectfont $0$}
  \put(5.67,65.36){\fontsize{9.96}{11.95}\selectfont $1$}
  \put(5.67,45.51){\fontsize{9.96}{11.95}\selectfont $2$}
  \put(5.67,25.67){\fontsize{9.96}{11.95}\selectfont $3$}
  \put(23.53,99.09){\fontsize{9.96}{11.95}\selectfont $X$}
  \put(63.21,81.23){\fontsize{9.96}{11.95}\selectfont $Y$}
  \put(77.10,67.34){\fontsize{9.96}{11.95}\selectfont $0$}
  \put(77.10,45.51){\fontsize{9.96}{11.95}\selectfont $1$}
  \put(25.51,7.81){\fontsize{9.96}{11.95}\selectfont $W_0(Y|X)$}
  \put(114.80,87.18){\fontsize{9.96}{11.95}\selectfont $0$}
  \put(114.80,65.36){\fontsize{9.96}{11.95}\selectfont $1$}
  \put(114.80,45.51){\fontsize{9.96}{11.95}\selectfont $2$}
  \put(114.80,25.67){\fontsize{9.96}{11.95}\selectfont $3$}
  \put(132.66,99.09){\fontsize{9.96}{11.95}\selectfont $X$}
  \put(172.35,81.23){\fontsize{9.96}{11.95}\selectfont $Y$}
  \put(186.24,67.34){\fontsize{9.96}{11.95}\selectfont $0$}
  \put(186.24,45.51){\fontsize{9.96}{11.95}\selectfont $1$}
  \put(134.65,7.81){\fontsize{9.96}{11.95}\selectfont $W_1(Y|X)$}
  \end{picture}%
\fi
\caption{\label{fig:W0W1_example_channels}%
 Example channels $W_0, W_1$}
\end{figure}

By writing the input as two binary digits $X = [X_1, X_2]$, the channel can be defined as follows: if $X_2 = s$ then $Y=X_1$, otherwise, $Y = \Ber \left(\half \right)$. These channels are depicted in Figure~\ref{fig:W0W1_example_channels}, where transitions are denoted by solid lines for probability $1$, and dashed lines for probability $\half$. We consider the same prediction problem, under the simplifying assumption that the channel $W_i = W_{s_i}$ is chosen only between the two channels above, and the forecaster knows this limitation, i.e. only the sequence of states $s_i \in \{0,1\}$ is unknown.

It is clear from convexity of the mutual information, and the symmetry with respect to $X_1$ (interchanging the values of $X_1$ leads to the same mutual information), that any solution can only be improved by taking a uniform distribution over $X_1$. Therefore, without loss of generality, the input distribution $Q$ can be defined by a single value $q = \Pr(X_2 = 1) \in [0,1]$, and be written $Q = [\half (1-q), \half (1-q), \half q, \half q]$. For this choice the output will always be uniformly distributed $\Ber \left( \half \right)$. We have:
\begin{equation}\begin{split}\label{eq:330}
I(Q,W_0)
& =
H(Y) - H(Y | X)
\\ & =
1 - \sum_{x} Q(x) H(Y|X=x)
=
1 - q ,
\end{split}\end{equation}
and similarly $I(Q,W_1) = q$, therefore we can write:
\begin{equation}\label{eq:341}
I(Q,W_s) = 1 - |s-q|
.
\end{equation}
Hence, even under this limited scenario, the loss function $1 - I(Q,W)$ behaves like the absolute loss function, and therefore the normalized minimax regret (and the redundancy in attaining $C_2$) is at least $O \left(\sqrt{\frac{1}{n}} \right)$.
\end{example}

Note that the relation to the absolute loss implies that the simple FL predictor $\hat Q_i = \argmax{Q} \sum_{t=1}^{i-1} I(Q,W_t)$, cannot be applied to our problem. An example to illustrate this and some further details are given in Appendix~\ref{sec:FL_failure_example}.

Since in the rest of the paper we will focus on the rate function $C_3$, it is interesting to note that, although this rate is smaller, in general, than $C_2$, the minimum redundancy in obtaining it cannot be better than $O \left(\sqrt{\frac{1}{n}} \right)$. To show this, we only need to show that in the context of the counter-example shown above, $C_2=C_3$. For a specific sequence of channels, denote by $p$ the relative frequency with which channel $W_1$ appears. The averaged channel is $(1-p) W_0 + p W_1$. It is easy to see that the capacity of this channel is obtained by placing the entire input probability on the two useful inputs of the channel that appears most of the time. That is, if $p \geq \half$ we place the input probability on the useful inputs of $W_1$ and obtain the rate $p \cdot C(W_1) = p$, and otherwise obtain $(1-p) \cdot C(W_0) = 1-p$. Hence the capacity of the averaged channel is $C_3 = \max(p,1-p)$. On the other hand,
\begin{equation}\begin{split}\label{eq:342}
C_2
&=
\max_Q \left( (1-p) \cdot I(Q,W_0) + p \cdot I(Q,W_1) \right)
\\&=
\max_{q \in [0,1]} \left( (1-p) \cdot (1-q) + p q \right)
=
\max(p,1-p)
.
\end{split}\end{equation}
Using the example above, we can also see why $C_1$ is not universally achievable with an asymptotically vanishing normalized regret by a sequential predictor. In the example, the capacities of the two channels are $C(W_s) = 1$. Suppose the sequence of channel states $\vr s_1^n \in \{0,1\}^n$ is generated randomly i.i.d. $\Ber \left( \half \right)$. Then for any sequential predictor of $q$, the expected loss in each time instance is $\E [I(Q,W_s)] = \half (1-q) + \half q = \half$, while the target rate is $C_1 = 1$. Therefore the expected normalized regret with respect to $C_1$ is $\half$, and the maximum regret (maximum over the sequence $\{W_i\}$) is lower bounded by the expected regret.

To summarize, we have seen why $C_1$ is not universally achievable, and therefore $C_2$ constitutes a reasonable target. Furthermore, the minimax regret with respect to $C_2$ is at least $O \left(\sqrt{\frac{1}{n}} \right)$, and the simple FL predictor following the best a-posteriori strategy does yield a vanishing regret.

\subsection{A prediction algorithm}\label{sec:toy_exp_prior_predictor}
The prediction algorithm proposed below is based on a well known technique of a weighted average predictor, using exponential weighting \cite[\S 2.1]{Nicolo}. A minor difference with respect to known results is the extension to a continuous set of reference strategies.

A weight function $w(Q)$ is any non-negative function $w: \Delta_{\mathcal{X}} \to \mathbb{R}^+$ with $\int_{\Delta_{\mathcal{X}}} w(Q) dQ = 1$. All integrals in the sequel are by default over $\Delta_{\mathcal{X}}$.

\begin{figure}
\centering
\ifpdf
  \setlength{\unitlength}{1bp}%
  \begin{picture}(152.28, 85.26)(0,0)
  \put(0,0){\includegraphics{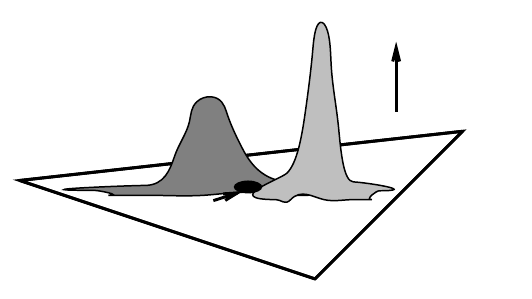}}
  \put(70.31,14.00){\fontsize{7.11}{8.54}\selectfont $Q \in \Delta_{\mathcal{X}}$}
  \put(119.06,62.36){\fontsize{7.11}{8.54}\selectfont $w_i(Q)$}
  \put(43.77,22.71){\fontsize{7.11}{8.54}\selectfont $\hat Q_i$}
  \end{picture}%
\else
  \setlength{\unitlength}{1bp}%
  \begin{picture}(152.28, 85.26)(0,0)
  \put(0,0){\includegraphics{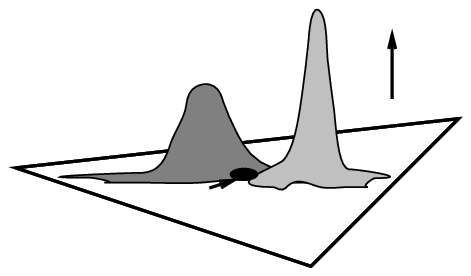}}
  \put(70.31,14.00){\fontsize{7.11}{8.54}\selectfont $Q \in \Delta_{\mathcal{X}}$}
  \put(119.06,62.36){\fontsize{7.11}{8.54}\selectfont $w_i(Q)$}
  \put(43.77,22.71){\fontsize{7.11}{8.54}\selectfont $\hat Q_i$}
  \end{picture}%
\fi
\caption{\label{fig:exponential_weighting_illustration}%
 An illustration of exponential weighting. The triangle represents the unit simplex. The two peaks represent two priors $Q$ which have a relatively large gain $\sum_{t=1}^{i-1} I(Q,W_i)$. The weight function $w_i(Q)$ combines them exponentialy, and the predictor $\hat Q_i$ (represented as a black spot) is the weighted average. }
\end{figure}

Define the following weight function:
\begin{equation}\label{eq:w398}
w_i(Q) =
\frac{e^{\eta \sum_{t=1}^{i-1} I(Q, W_t)}}{\int_{\Delta_{\mathcal{X}}} e^{\eta \sum_{t=1}^{i-1} I(\tilde Q, W_t)} d \tilde Q}
,
\end{equation}
and the predictor:
\begin{equation}\label{eq:p404}
\hat Q_i = \int_{\Delta_{\mathcal{X}}} Q \cdot w_i(Q) \cdot d Q
.
\end{equation}
The weighting function gives a higher weight to priors that succeeded in the past and the predictor averages the potential priors with respect to the weight. This is illustrated in Fig.~\ref{fig:exponential_weighting_illustration}.
The following theorem gives a bound on the regret of this predictor, which is proven in the next section.

\begin{theorem}\label{theorem:prior_predictor_exp}
Let $I(Q,W), Q \in \Delta_{\mathcal{X}}$ be bounded function $0 \leq I(Q,W) \leq I_{\max}$ which is concave in its first argument. Then for $n$ large enough so that $\frac{\ln (n)}{n} \leq e^{-2}$, the predictor defined by \eqref{eq:w398} and \eqref{eq:p404} with $\eta = \sqrt{\frac{|\mathcal{X}| \ln n}{n}} \cdot I_{\max}^{-1}$ yields
\begin{equation}\label{eq:388}
R = \frac{1}{n} \sum_{i=1}^n I(\hat Q_i,W_i) \geq C_2(W_1^n) - \delta
,
\end{equation}
with
\begin{equation}\label{eq:RBt1}
\delta
=
2  I_{\max} \cdot \sqrt{\frac{(|X| - 1) \ln n}{n}}
.
\end{equation}
\end{theorem}
Note that the theorem applies to gain functions more general than the mutual information, since it uses only the properties of concavity and boundness. In the case of mutual information we have
\begin{equation}\label{eq:Imax_def}
I_{\max} = \log \min (|\mathcal{X}|, |\mathcal{Y}|)
.
\end{equation}

We obtained a convergence rate of $O \left( \sqrt{\frac{\ln n}{n}} \right)$ which is slightly worse than the asymptotic bound of $O \left( \sqrt{\frac{1}{n}} \right)$ from Section~\ref{sec:regret_LB}. The additional $\sqrt{\ln n}$ may be attributed to the fact the space of reference predictors is continuous (it results from Lemma~\ref{lemma:F_exp_weight_UB} stated below), but we do not know if this is the best convergence rate.

\subsection{Proof of Theorem~\ref{theorem:prior_predictor_exp}}\label{sec:toy_performance}
In this section we analyze the performance of the predictor \eqref{eq:p404} and prove Theorem~\ref{theorem:prior_predictor_exp}. Define the instantaneous regret $r_i(Q)$ and the cumulative regret $\mathcal{R}_i(Q)$ as functions of $Q$:
\begin{equation}\label{eq:r426}
r_i(Q) = I(Q, W_i) - I(\hat Q_i, W_i)
,
\end{equation}
\begin{equation}\label{eq:R429}
\mathcal{R}_i(Q) = \sum_{t=1}^i r_t(Q) = \sum_{t=1}^i I(Q, W_t) - \sum_{t=1}^i I(\hat Q_i, W_t)
.
\end{equation}

These functions express the regret with respect to a fixed competing prior $Q$. The claim of the theorem is equivalent to the claim that for all $Q$, $\mathcal{R}_n(Q) \leq n \delta$. We sometimes omit the dependence on $Q$ for brevity.

For $\eta > 0$ of our choice, we define the following potential function:
\begin{equation}\label{eq:defPhi427}
\Phi(u) = \int_{\Delta_{\mathcal{X}}} e^{\eta u(Q)} d Q
,
\end{equation}
where $u: \Delta_{\mathcal{X}} \to \mathbb{R}$ is an arbitrary function defined over the unit simplex. Note that for large values of $\eta \cdot u$, $\Phi(u)$ approximates $\max_Q(u)$. As customary in this prediction technique, the proof consists of two parts:
\begin{enumerate}
\item Bounding the growth rate of $\Phi(\mathcal{R}_i(Q))$ over $i=1,2,\ldots,n$ for any $Q$.
\item Relating $\max_{Q} \{ \mathcal{R}_n(Q) \}$ to $\Phi(\mathcal{R}_n(Q))$.
\end{enumerate}
The techniques we use are based on Cesa-Bianchi and Lugosi's \cite{Nicolo} (see Theorem 2.1, Corollary 2.2, Theorem 3.3).

From the concavity of $I(Q,W)$ with respect to $Q$ we have that for any weight function $w(Q)$ and any $W_i$:
\begin{equation}\begin{split}\label{eq:w467}
\int w(Q) r_i(Q) dQ
& =
\int w(Q) I(Q, W_i) dQ - I(\hat Q_i, W_i)
\\ & \leq
I \Bigg(\underbrace{\int w(Q) Q dQ}_{\hat Q_i}, W_i \Bigg) - I(\hat Q_i, W_i)
\\ & =
0.
\end{split}\end{equation}
Following \cite{Nicolo} we term this inequality the ``Blackwell condition''. The meaning of this condition is that by choice of $w(Q)$ we can prevent an increase of  $\mathcal{R}_i(Q)$ in a chosen direction ($w(Q)$ can be thought of as a unit vector in the Hilbert space of functions over $\Delta_{\mathcal{X}}$). For the specific choice of the weight function \eqref{eq:w398}, this direction is proportional to the gradient of $\Phi(R)$ with respect to $R$, thus preventing any growth in this direction and leaving only second order terms that contribute to the increase of $\Phi(\mathcal{R}_n(Q))$. Since the factor $\sum_{t=1}^i I(\hat Q_i, W_t)$ in \eqref{eq:R429} does not depend on $Q$, the weight function \eqref{eq:w398} can be alternatively written as:
\begin{equation}\label{eq:091}
w_i(Q) = \frac{e^{\eta \mathcal{R}_{i-1}(Q)}}{\int e^{\eta \mathcal{R}_{i-1}(Q)} d Q}.
\end{equation}
$w_i(Q)$ is indifferent to any constant addition to $\mathcal{R}_{i-1}(Q)$ due to the normalization. The growth of the potential can be bounded as follows:
\begin{equation}\begin{split}\label{eq:Phi492}
\Phi(\mathcal{R}_{i})
&=
\Phi(\mathcal{R}_{i-1} + r_i)
=
\int e^{\eta \mathcal{R}_{i-1} + \eta r_i} d Q
\\ & =
\int e^{\eta \mathcal{R}_{i-1}} \cdot e^{\eta r_i} d Q
\\& \stackrel{\eqref{eq:091}}{=}
\int e^{\eta \mathcal{R}_{i-1} } d Q \cdot \int w_i(Q) e^{\eta r_i} d Q
\\ & =
\Phi(\mathcal{R}_{i-1}) \cdot \int w_i(Q) e^{\eta r_i} d Q,
\end{split}\end{equation}
Notice that $r_i \leq I_{\max}$. We take $\eta$ small enough that $\eta r_i \leq \eta I_{\max} \leq 1$ and use the following inequality (proven in Appendix~\ref{sec:proofs_of_small_lemmas}):
\begin{lemma}\label{lemma:exp_second_order_bound}
For $x \in [-1,1]$:
\begin{equation}\label{eq:ex483}
1 + x \leq e^x \leq 1 + x + x^2.
\end{equation}
\end{lemma}

Returning to \eqref{eq:Phi492} we have:
\begin{multline}\label{eq:w524}
\int w_i(Q) e^{\eta r_i} d Q
\stackrel{\eqref{eq:ex483}}{\leq}
\int w_i(Q) \left( 1 + \eta r_i + (\eta r_i)^2 \right) d Q
\hstr \\ =
\int w(Q) d Q + \eta  \underbrace{\int w(Q) r_i d Q}_{\leq 0, \eqref{eq:w467}} + \eta^2 \int w(Q) r_i^2 d Q
\hstr \\ \stackrel{\eqref{eq:w467}}{\leq}
1 + \eta^2 I_{\max}^2
\stackrel{\eqref{eq:ex483}}{\leq}
e^{\eta^2 I_{\max}^2}.
\hstr
\end{multline}

Therefore recursively applying \eqref{eq:Phi492}:
\begin{equation}\label{eq:Phi523}
\Phi(\mathcal{R}_{n}) \stackrel{\eqref{eq:Phi492}, \eqref{eq:w524}}{\leq} e^{\eta^2 I_{\max}^2} \Phi(\mathcal{R}_{n-1}) \leq \ldots \leq e^{n \eta^2 I_{\max}^2} \cdot \Phi(0).
\end{equation}
Notice that $\Phi(0) = \int 1 d Q = \vol(\Delta_{\mathcal{X}})$. This completes the first part of showing that the increase in $\Phi(\mathcal{R}_{n})$ is bounded. For the second part we shall use the following lemma which relates the exponential weighting of a function to its maximum, and is proven in Appendix~\ref{sec:proof_of_lemma1}:
\begin{lemma}\label{lemma:F_exp_weight_UB}
Let $F(\vr x)$ be a real non-negative bounded function $F: S \to [a,b]$ concave in $S$, where $S$ is a closed convex vector region of dimension $d$, and let $\eta$ satisfy $\eta (b-a) \geq d$, then
\begin{equation}\begin{split}\label{eq:exp_weight_UB_2603}
\max_{\vr x \in S} F(\vr x)
& \leq
\frac{1}{\eta} \ln \left[ \frac{\displaystyle \int_S e^{\eta F(\vr x)} d \vr x}{\vol(S)} \right] + \frac{d}{\eta} \ln \left( \frac{\eta e (b-a)}{d} \right)
\\ & =
\frac{1}{\eta} \ln \left[ \frac{\Phi(F)}{\Phi(0)} \right] + \frac{d}{\eta} \ln \left( \frac{\eta e (b-a)}{d} \right).
\end{split}\end{equation}
\end{lemma}

Let $F(Q) = \mathcal{R}_n(Q)$. In this case the convex region is $\Delta_{\mathcal{X}}$ and therefore $d=\dim(\Delta_{\mathcal{X}})=|X|-1$. By \eqref{eq:R429} we can bound $F$ by:
\begin{equation}\label{eq:580}
\underbrace{- \sum_{i=1}^n I(\hat Q_i, W_i)}_{\defeq a} \leq F(Q) \leq \underbrace{n I_{\max} - \sum_{i=1}^n I(\hat Q_i, W_i)}_{\defeq b},
\end{equation}
where the factor $\sum_{i=1}^n I(\hat Q_i, W_i)$ is constant in $Q$. We have $b-a= n I_{\max}$. Assuming $\eta n I_{\max} \geq d$ to satisfy the conditions of Lemma~\ref{lemma:F_exp_weight_UB}, we obtain from \eqref{eq:exp_weight_UB_2603}:
\begin{equation}\begin{split}\label{eq:568}
\mathcal{R}_n(Q)
& \leq
\frac{1}{\eta} \ln \frac{\Phi(\mathcal{R}_n(Q))}{\Phi(0)} + \frac{d}{\eta}  \ln \left( \frac{\eta e n I_{\max}}{d} \right)
\\ & \stackrel{\eqref{eq:Phi523}}{\leq}
n \eta I_{\max}^2 + \frac{d}{\eta} \ln \left( \frac{\eta e n I_{\max}}{d} \right).
\\ & \leq
n \eta I_{\max}^2 + \frac{d}{\eta} \ln \left( n \right) \defeq \Delta
,
\end{split}\end{equation}
where in the last inequality we assumed $\frac{\eta e I_{\max}}{d} \leq 1$ (this would hold for $\eta$ small enough). We use the following lemma to optimize the RHS of \eqref{eq:568} with respect to $\eta$:
\begin{lemma}\label{lemma:ab_alphabeta}
The unique minimum over $t \in \mathbb{R}^+$ of $f(t) = a \cdot t^{\alpha} + b \cdot t^{-\beta}$ ($a,b,\alpha,\beta > 0$) is obtained at $t^* = \left( \frac{b \beta}{a \alpha} \right)^{\frac{1}{\alpha+\beta}}$ and equals
\begin{equation}\label{eq:941}
f(t^*) = \left( \frac{\beta}{\alpha} \right)^{\frac{\alpha}{\alpha+\beta}} \left[ 1
+ \frac{\alpha}{\beta} \right] \cdot a^{\frac{\beta}{\alpha+\beta}} \cdot b^{\frac{\alpha}{\alpha+\beta}}
.
\end{equation}
Particularly, for $\alpha=\beta=1$, i.e. $f(t) = a \cdot t + \frac{b}{t}$ we have $t^* = \sqrt{\frac{b}{a}}$ and $f(t^*) = 2 \sqrt{ab}$.
\end{lemma}
The proof of the lemma is simple by a direct derivation (see Appendix~\ref{sec:proofs_of_small_lemmas}). Applying the lemma to the optimization of $\eta$ in \eqref{eq:568} we obtain:
\begin{equation}\label{eq:505}
\eta^* = \sqrt{\frac{d \ln (n)}{n  I_{\max}^2}}
,
\end{equation}
and
\begin{equation}\label{eq:509}
\Delta^* = \Delta \Big|_{\eta=\eta^*} = 2 I_{\max} \sqrt{d n \ln (n)}
.
\end{equation}

We now verify the assumptions we made along the way. In \eqref{eq:w524} we assumed that $\eta I_{\max} \leq 1$. If the contrary holds $\eta I_{\max} > 1$ then considering the first term in the RHS of \eqref{eq:568}, we have $\Delta > n I_{\max}$, and therefore the theorem holds in a void way. To apply Lemma~\ref{lemma:F_exp_weight_UB} we required $\eta n I_{\max} \geq d$. If the opposite is true, i.e. $\eta n I_{\max} < d$ then the second term the RHS of \eqref{eq:568} becomes $\frac{d}{\eta} \ln (n) > n I_{\max} \ln (n)$, and so for $n \geq e$ we would have again $\Delta > n I_{\max}$ and the theorem will hold in a void way. Thus for the two last conditions, it is enough that $n \geq 3$, since in this case if either of the conditions does not hold, the theorem becomes true automatically (in a void way). Lastly, in \eqref{eq:568} we assumed $\frac{\eta e I_{\max}}{d} \leq 1$. Substituting $\eta^*$ we have $\frac{\eta e I_{\max}}{d} = e \cdot \sqrt{\frac{e^2 \ln (n)}{d n}} \leq e \cdot \sqrt{\frac{\ln (n)}{n}}$, which becomes smaller than $1$ for $n$ large enough. The last condition supersedes $n \geq e$, and is specified as a requirement in the theorem.
\endofproof

\section{Arbitrary channel variation}\label{sec:arbitrary_channel_var}
In this section we return to the problem defined in Section~\ref{sec:problem_setting} and present the main results of the paper: the achievability of the capacity of the averaged channel, and a converse showing that this is the best rate, under some conditions. We give the outline of the communication system attaining this rate, while leaving out some of the technical details, such as decoding and channel estimation (these will be completed in the next section). We show that under abstract assumptions, the system achieves the desired rate. The same communication scheme and predictor will be used, with slight modifications, to prove the main result in Section~\ref{sec:mainproof}.

\subsection{Target rate}\label{sec:arbitrary_var_target_rate}
The synthetic problem differs from the problem defined in Section~\ref{sec:problem_setting}, in two main aspects:
\begin{enumerate}
\item It assumes that the sequence of past channels is fully known. Since the receiver observes only one output sample from each channel, this assumption is not realistic. On the other hand, the time-averaged channel over ``large'' chunks of symbols can be measured.
\item It assumes that a rate corresponding to a sum of the per-symbol mutual information can be attained, whereas with an arbitrarily varying channel, the amount of mutual information between the input and output vectors is potentially lower.
\end{enumerate}
Therefore, as we shall see, $C_2$ is no longer achievable in the context of the arbitrarily varying channel defined in Section~\ref{sec:problem_setting}. In Appendix~\ref{sec:example_prediction_channel2} we show that even imposing on the synthetic problem only the limitation that the past channels are not given, but need to be estimated, leads to the conclusion that $C_2$ is not attainable. Therefore we compromise on an alternative target: obtaining $C_3 = C(\overline W)$, i.e. the capacity of the averaged channel. As we shall show in this section and the next, this rate is indeed asymptotically achievable.

The rate $C(\overline W)$ is certainly not the maximum achievable target rate. As an example, if $C(\overline W)$ is achievable for large $n$ then by operating the same scheme on two halves of the transmission time one could attain $R = \half C \left( \overline {W_1^{n/2}} \right) + \half C \left( \overline {W_{n/2+1}^{n \phantom{/}}} \right)$, where $\overline {W_1^{n/2}}, \overline {W_{n/2+1}^{n \phantom{/}}}$ denote the averaged channels on the two halves. This rate is in general higher, because due to the convexity of the mutual information with respect to the channel $C(\overline W) = \max_Q I(Q, \overline W) \leq \max_Q \left[ \half I \left( Q, \overline {W_1^{n/2}} \right) + \half I \left( Q, \overline {W_{n/2+1}^{n \phantom{/}}} \right) \right] \leq R$.

On the other hand, $C(\overline W)$ is the maximum achievable rate which is independent of the order of the sequence $\{W_i\}$, or, in other words, which is fixed under permutation of the sequence. This observation is formalized in the following theorem:
\begin{theorem}\label{theorem:C_overlineW_optimality}
Let $R(W_1^n)$ (for $n=1,2,..$) be a sequence of rate functions, which are oblivious to the order of $W_1^n$. If the sequence is asymptotically attainable according to Definition~\ref{def:attainability_of_RW}, then there exists a sequence $\delta_n \ntoinfty 0$ such that $R(W_1^n) \leq C(\overline W) + \delta_n$.
\end{theorem}
Note that $C(\overline W)$ depends on $n$ through the average over $n$ channels $\{W_i\}_1^n$. Since both $C_1$ and $C_2$ are oblivious to the order of $W_1^n$, Theorem~\ref{theorem:C_overlineW_optimality} implies they are not achievable.

Following is a rough outline of the proof. Consider the channel generated by uniformly drawing a random permutation $\pi$ of the indices $i=1,\ldots,n$, using the channels $W_i$ in a permuted order. If a system guarantees a rate $R(W_1^n)$, which is fixed under permutation, then this rate would be fixed for all drawing of $\pi$, and therefore for the channel we described, the system can guarantee the rate $R(W_1^n)$ a-priori. Hence, the capacity of this channel must be at least $R(W_1^n)$. The next stage is to show that the feedback capacity of this channel is at most $C(\overline W)$. Due to the fact we select the channels from the set $\{W_i\}_{i=1}^n$ without replacement, the proof is a little technical and will be deferred to Appendix~\ref{sec:proof_CW_max_rate}. However to give an intuitive argument, if we replace the channel described above, by a similar channel, obtained by randomly drawing at each time instance one of $\{W_i\}_{i=1}^n$, this time \emph{with} replacement, then this new channel is simply the DMC with channel law $\overline W$. Therefore feedback does not increase the capacity and its feedback capacity is simply $C(\overline W)$. The main point in the proof is to show there is no difference in feedback-capacity between the two channels, and the main tool is Hoeffding's bounds on sampling without replacement \cite{Hoeffding}.

Another interesting property of the rate $C(\overline W)$ is that it meets or exceeds the random-code capacity of any memoryless AVC defined over the same alphabet, and thus attaining $C(\overline W)$ yields universality over all AVC's (see Section~\ref{sec:discussion_AVC}). Through the relation to AVC capacity we can see that common randomness is essential to obtain $C(\overline W)$, as it is essential for obtaining the random-code capacity \cite{Lapidoth_AVC}. 

After settling for $C(\overline W)$, the next question that naturally arises is: what is the best convergence rate of the regret, with respect to this target? In Section~\ref{sec:regret_LB} we have shown that even in the context of the synthetic problem of Section~\ref{sec:toy_problem} (with full knowledge of past channels), the regret with respect to $C_3$ is at least $O(n^{-\half})$, and this lower bound naturally holds in the current problem, where only partial knowledge of past channels is available.

The following theorem formalizes claim that $C(\overline W)$ is achievable according to Definition~\ref{def:attainability_of_RW}:
\begin{theorem}\label{theorem:C_overlineW_achievability}
For every $\epsilon, \delta > 0$ there exists $N$ and a constant $c_{\Delta}$, such that for any $n \geq N$ there is an adaptive rate system with feedback and common randomness, where for the problem of Section~\ref{sec:problem_setting}, over any sequence of channels $\{ W_i(y|x) \}_{i=1}^n$:
\begin{enumerate}
  \item The probability of error is at most $\epsilon$
  \item The rate satisfies $R \geq C(\overline W) - \Delta_C$ with probability at least $1-\delta$
  \item $\displaystyle \Delta_C = c_{\Delta} \cdot \left( \frac{\ln^2 (n)}{n} \right)^{\tfrac{1}{4}}$
\end{enumerate}
\end{theorem}

\begin{corollary}\label{corollary:symbolwise_random_numerical}
Specific values for $\epsilon, \delta, \Delta_C$ can be obtained as follows. Let $d_\epsilon, \delta_0, c_\lambda > 0$ be parameters of choice. Then the constants $n_{\min}$ and $c_{\Delta}$ are given in the proof, by \eqref{eq:1858}, \eqref{eq:1970}, where constants used in these equations are defined in \eqref{eq:Imax_def}, \eqref{eq:795}, \eqref{eq:k1227}, \eqref{eq:1919-1}-\eqref{eq:1919-3}, \eqref{eq:1804a}. For any $n \geq n_{\min}$, $\epsilon = n^{-d_{\epsilon}}$ and $\delta = \epsilon + \delta_0$.
\end{corollary}

\begin{corollary}\label{corollary:symbolwise_random_prior_predictor1}
The same holds if $W_i$ is determined (e.g. by an adversary) as a function of the message and all previous channel inputs and outputs $\vr X^{i-1}, \vr Y^{i-1}$.
\end{corollary}

A numerical example is given after the proof (Example~\ref{example:symbolwise_random_prior_predictor}). The proof of the theorem is given in Section~\ref{sec:mainproof}.

\subsection{The communication scheme}\label{sec:arbitrary_var_rateless_scheme}
In this section give the communication scheme, up to some details which will be completed later on (Section~\ref{sec:mainproof_decoding_cond}). One of the issues that we ignored in the synthetic problem is the determination of the rate $R$ before knowing the channel. To solve this problem we use rateless codes \cite{Shulman}. We divide the available time into multiple such blocks as done by Eswaran~\etal~\cite{Eswaran} and \selector{in \cite{YL_individual_full}}{in Section~\ref{sec:rate_adaptive_scheme}}.

We fix a number $K$ of bits per block. In each block, $K$ bits from the message string are sent. At each block $i=1,2,\ldots$, a codebook of $\exp(K)$ codewords is generated randomly and i.i.d. (in time and message index) according to the prior $\hat Q_i(x)$. $\hat Q_i(x)$ is determined by a prediction scheme which is specified below. The random drawing of the codewords is carried out by using the common randomness, and the codebook is known to both sides. The relevant codeword matching the message sub-string is sent to the receiver symbol by symbol. At each symbol of the block and for each codeword $\vr x_l, l=1,\ldots,\exp(K)$ in the codebook, the receiver evaluates a decoding condition \eqref{eq:ter1315} that will be specified later on. Roughly speaking, the condition measures whether there is enough information from the channel output to reliably decode the message.

The receiver decides to terminate the block if the condition \eqref{eq:ter1315} holds, and informs the transmitter. When this happens, the receiver determines the decoded codeword as one of the codewords that satisfied \eqref{eq:ter1315}. Then, using the known channel output $\vr y$, and the decoded input $\vr x$ over the block which was decoded, the receiver computes an estimate of the averaged channel over the block. The specific estimation scheme will be specified in Section~\ref{sec:mainproof_decoding_cond}.

The receiver calculates a new prior for the next block according to the prediction scheme that will be specified below. The receiver sends the new prior to the transmitter. Alternatively, the receiver may send the estimated channel, and the new prior can be calculated at each side separately. The new block $i+1$ starts at the next symbol, and the process continues, until symbol $n$ is reached. The last block may terminate before decoding.

\begin{figure}
\centering
\ifpdf
  \setlength{\unitlength}{1bp}%
  \begin{picture}(229.77, 131.75)(0,0)
  \put(0,0){\includegraphics{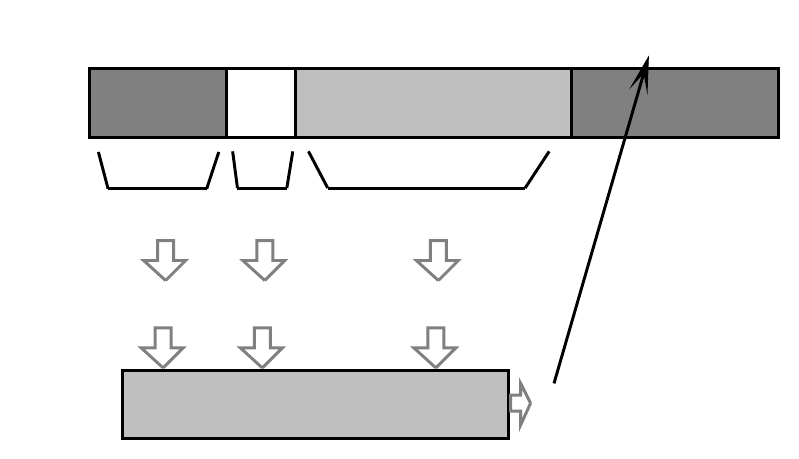}}
  \put(75.28,118.30){\fontsize{9.96}{11.95}\selectfont \makebox[0pt]{$\hat Q_2$}}
  \put(45.52,118.30){\fontsize{9.96}{11.95}\selectfont \makebox[0pt]{$\hat Q_1=U$}}
  \put(122.90,118.30){\fontsize{9.96}{11.95}\selectfont \makebox[0pt]{$\hat Q_3$}}
  \put(190.37,118.30){\fontsize{9.96}{11.95}\selectfont \makebox[0pt]{$\hat Q_4$}}
  \put(123.56,66.71){\fontsize{9.96}{11.95}\selectfont \makebox[0pt]{$\hat {\vr x}, \vr y$}}
  \put(125.33,41.63){\fontsize{9.96}{11.95}\selectfont \makebox[0pt]{$\overline W_3$}}
  \put(71.53,12.69){\fontsize{9.96}{11.95}\selectfont Predictor}
  \put(160.84,12.18){\fontsize{9.96}{11.95}\selectfont \makebox[0pt]{$\hat Q_4$}}
  \put(75.34,41.63){\fontsize{9.96}{11.95}\selectfont \makebox[0pt]{$\overline W_2$}}
  \put(73.58,66.71){\fontsize{9.96}{11.95}\selectfont \makebox[0pt]{$\hat {\vr x}, \vr y$}}
  \put(46.78,41.63){\fontsize{9.96}{11.95}\selectfont \makebox[0pt]{$\overline W_1$}}
  \put(45.02,66.71){\fontsize{9.96}{11.95}\selectfont \makebox[0pt]{$\hat {\vr x}, \vr y$}}
  \end{picture}%
\else
  \setlength{\unitlength}{1bp}%
  \begin{picture}(229.77, 131.75)(0,0)
  \put(0,0){\includegraphics{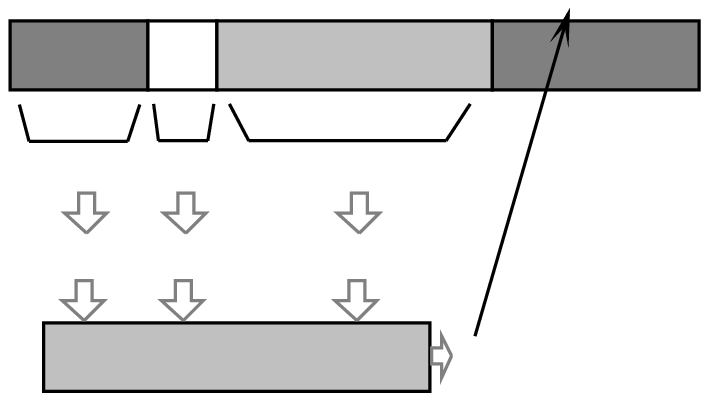}}
  \put(75.28,118.30){\fontsize{9.96}{11.95}\selectfont \makebox[0pt]{$\hat Q_2$}}
  \put(45.52,118.30){\fontsize{9.96}{11.95}\selectfont \makebox[0pt]{$\hat Q_1=U$}}
  \put(122.90,118.30){\fontsize{9.96}{11.95}\selectfont \makebox[0pt]{$\hat Q_3$}}
  \put(190.37,118.30){\fontsize{9.96}{11.95}\selectfont \makebox[0pt]{$\hat Q_4$}}
  \put(123.56,66.71){\fontsize{9.96}{11.95}\selectfont \makebox[0pt]{$\hat {\vr x}, \vr y$}}
  \put(125.33,41.63){\fontsize{9.96}{11.95}\selectfont \makebox[0pt]{$\overline W_3$}}
  \put(71.53,12.69){\fontsize{9.96}{11.95}\selectfont Predictor}
  \put(160.84,12.18){\fontsize{9.96}{11.95}\selectfont \makebox[0pt]{$\hat Q_4$}}
  \put(75.34,41.63){\fontsize{9.96}{11.95}\selectfont \makebox[0pt]{$\overline W_2$}}
  \put(73.58,66.71){\fontsize{9.96}{11.95}\selectfont \makebox[0pt]{$\hat {\vr x}, \vr y$}}
  \put(46.78,41.63){\fontsize{9.96}{11.95}\selectfont \makebox[0pt]{$\overline W_1$}}
  \put(45.02,66.71){\fontsize{9.96}{11.95}\selectfont \makebox[0pt]{$\hat {\vr x}, \vr y$}}
  \end{picture}%
\fi
\caption{\label{fig:prior_prediction_comm_scheme_illustration}%
 An illustration of the combination of a rateless scheme with prior prediction. Each box represents a rateless block in which $K$ bits are transmitted.}
\end{figure}

\subsection{The prediction algorithm}\label{sec:arbitrary_var_exp_prior_predictor}
In this section we present the prediction algorithm. We denote by $i$ the index of the block, and by $\overline W_i$ the averaged channel over the block, i.e. if the block $i$ starts at symbol $k_i$ and ends at $k_{i+1}-1$, then $\overline W_i(y|x) \defeq \frac{1}{k_{i+1} - k_i} \sum_{t=k_i}^{k_{i+1}-1} W_t(y|x)$. The length of the $i$-th block is denoted $m_i = k_{i+1} - k_i$. We use an exponentially weighted predictor mixed with a uniform prior. The motivation for using the uniform prior is explained in the next section. Let $U = \frac{1}{|\mathcal{X}|} \vr 1$ be the uniform prior over $\mathcal{X}$. We define the predictor as:
\begin{equation}\label{eq:Qi3382}
\hat Q_i = (1-\lambda) \int_{\Delta_{\mathcal{X}}} w_i(Q) Q d Q + \lambda U.
\end{equation}
where
\begin{equation}\label{eq:8385}
w_i(Q) = \frac{1}{\Phi \left(\sum_{j=1}^{i-1} m_j \cdot F_j(\tilde Q) \right)} \cdot e^{\eta \sum_{j=1}^{i-1} m_j \cdot F_j(Q)},
\end{equation}
where $F_i(Q)$ is an estimate of the mutual information of the averaged channel over block $i$, $I(Q, \overline W_j)$, and is interpreted as an estimate of the number of bits that would have been sent with the alternative prior $Q$. This estimate is defined later on in Section~\ref{sec:mainproof_attained_rate}. The parameters $\lambda, \eta$ and $K$ will be chosen later on. $\Phi$ is the potential function defined in \eqref{eq:defPhi427}. The term $\frac{1}{\Phi(\ldots)}$ normalizes $w_i(Q)$ to $\int_{\Delta_{\mathcal{X}}} w_i(Q) d Q = 1$.

The following Lemma formalizes the claim that the predictor resulting of  \eqref{eq:Qi3382}-\eqref{eq:8385}, asymptotically achieves a rate $R \geq \sum_{i=1}^{B+1} \frac{m_i}{n} F_i(Q)$:
\begin{lemma}\label{lemma:rateless_prior_predictor_lemma}
Let $F_i(Q)$, $i=1,\ldots,B+1$ be a set of $B+1$ non-negative concave functions of the prior $Q \in \Delta_{\mathcal{X}}$, let $\{m_i\}_{i=1}^{B+1}$ denote a set of non-negative numbers, and $K,n, I_{\max}$ be arbitrary positive constants satisfying $n > e$ and $K \geq 2 I_{\max}$.

Define the target rate
\begin{equation}\label{eq:683t}
\Rtarget = \max_{Q \in \Delta_{\mathcal{X}}} \sum_{i=1}^{B+1} \frac{m_i}{n} F_i(Q)
.
\end{equation}
Define the actual rate $R$ over $n$ channel uses as:
\begin{equation}\label{eq:683}
R = \frac{KB}{n}
.
\end{equation}
Define the sequential predictor $\hat Q_i$ as the result of \eqref{eq:Qi3382} and \eqref{eq:8385}.
Let $\{m_i\}_{i=1}^{B+1}$ satisfy:
\begin{equation}\label{eq:mi728}
m_i F_i(\hat Q_i) \leq K
.
\end{equation}
Then for the value of $\eta$ specified below \eqref{eq:1060} it is guaranteed that:
\begin{equation}\label{eq:688}
R \geq \min(\Rtarget, I_{\max}) - \Dpred
,
\end{equation}
where
\begin{equation}\label{eq:743}
\Dpred = \frac{K}{n} + I_{\max} \cdot \lambda + c_1 \sqrt{\frac{\ln (n)}{n}} \lambda^{-\half}
,
\end{equation}
and
\begin{equation}\label{eq:795}
c_1 = 2 \sqrt{K \cdot |\mathcal{X}| (|\mathcal{X}|-1) \cdot I_{\max}}
.
\end{equation}
The value of $\eta$ attaining the result above is:
\begin{equation}\label{eq:1060}
\eta = \sqrt{\frac{|\mathcal{X}|-1}{K \cdot |\mathcal{X}| \cdot I_{\max}} \cdot \frac{\ln (n) \cdot \lambda}{n}}
.
\end{equation}
\end{lemma}

The lemma is proven in Appendix~\ref{sec:proof_rateless_prior_predictor_lemma}. The proof uses similar techniques to those introduced in Section~\ref{sec:toy_performance}, however, different from the previous analysis, due to mixing with the uniform prior, the ``Blackwell'' condition (\eqref{eq:w467} in the previous case) only approximately holds. On the other hand, the use of the uniform prior enables relating $F_i(\hat Q_i)$ to $F_i(Q)$ for any other $Q$, and thus obtain from \eqref{eq:mi728} an upper bound on the gain $m_i F_i(Q)$ related to an alternative prior $Q$. The trade-off between the two is expressed in the two last factors in \eqref{eq:743}, one of which is increasing with $\lambda$ and the other decreasing.

Since by \eqref{eq:mi728}, $R \geq \sum_{i=1}^{B+1} \frac{m_i}{n} F_i(\hat Q_i)  - \frac{K}{n}$, the claim of the lemma appears similar to Theorem~\ref{theorem:prior_predictor_exp}, with $m_i F_i(Q)$ taking the place of the function $I(Q,W_i)$. However two important properties of the lemma, distinguishing it from the rather standard claim of Theorem~\ref{theorem:prior_predictor_exp} are that the bound does not depend on the number of blocks (i.e. the number of prediction steps), and that no upper bound on $F_i(Q)$ is assumed.

The rate $I_{\max}$ represents a bound on mutual information, but in the context of the lemma it enough to consider it as an arbitrary rate that caps $\Rtarget$. It affects the setting of $\eta$ and the resulting loss. Also, $n$ does not have to correspond to the actual number of symbols and serves here merely as a scaling parameter for the communication rate. The lemma sets a value of $\eta$ but not for $\lambda$, since $\lambda$ will have additional roles in the next section.

\subsection{Motivation for the prediction algorithm}\label{sec:arbitrary_var_predictor_motivation}
In this section a motivation for the prediction algorithm, and especially for the use of the uniform prior is given. Under abstract assumptions it is shown to achieve the capacity of the averaged channel. This section is intended merely to give motivation and is not formally necessary for the proof of Theorem~\ref{theorem:C_overlineW_achievability}.

To simplify the discussion, let us make abstract assumptions regarding the decoding condition and the channel estimation:
\begin{enumerate}
\item The decoding condition yields block lengths satisfying:
\begin{equation}\label{eq:3352}
m_i \leq \frac{K}{I(\hat Q_i, \overline W_i)}
,
\end{equation}
with an equality for all blocks except the last one which is not decoded. This implies the rate $\frac{K}{m_i}$ equals the mutual information of the averaged channel.

\item The averaged channels over all previous blocks are known and available for the predictor
\end{enumerate}
With these assumptions, the prediction problem can be considered separately from decoding and channel estimation issues. Supposing that $B$ blocks were transmitted, the achieved rate is $R = \frac{KB}{n}$. Since $n \approx \sum_i m_i$, using \eqref{eq:3352} this can be written as $R \approx \left( \frac{1}{B} \sum_{i=1}^B \frac{1}{I(\hat Q_i, \overline W_i)} \right)^{-1}$. The target is to find a prediction scheme for $\hat Q_i$, such that for any sequence $W_i$, one will have $R \geq C(\overline W) - \delta_n$ with $\delta_n \to 0$. There are two main difficulties compared to the prediction problem discussed in Section~\ref{sec:toy_problem}:
\begin{enumerate}
\item The problem is not directly posed as a prediction problem with an additive loss.
\item The loss is not bounded: if for some $i$, $I(\hat Q_i, \overline W_i)=0$ then the rate becomes zero regardless of other blocks.
\end{enumerate}

The first issue is resolved by posing an alternative problem which has an additive loss, and using the convexity of the mutual information with respect to the channel (as will be exemplified below in the abstract case). Regarding the second issue, notice that if the channel has zero capacity (always, or from some point in time onward), it is possible that one of the blocks will extend forever and will never be decoded. However we must avoid a situation where the channel has non-zero capacity (which our competition enjoys), while a badly chosen prior yields $I(\hat Q_i,\overline W_i)=0$. This may happen for example in the channels of Example~\ref{example:prediction_channel1}, if the predictor selects to use the pair of inputs that yield zero capacity. If this happens then the scheme will get stuck since the block will never be decoded, and hence there will be no chance to update the prior. In addition, notice that selecting some inputs with zero probability makes the predictor blind to the channel values over these inputs. To resolve these difficulties we construct the predictor as a mixture between an exponentially weighted predictor and a uniform prior. We use a result by Shulman and Feder \cite{Shulman_Prior}, which bounds the loss from capacity by using the uniform prior $U$:
\begin{equation}\label{eq:3397}
I(U;W) \stackrel{\cite[(3)]{Shulman_Prior}} \geq C  \cdot  \beta(C) \stackrel{\cite[(17)]{Shulman_Prior}}{\geq} \frac{C}{|\mathcal{X}| \cdot (1-e^{-1})},
\end{equation}
where $C$ is the channel capacity and $\beta(C)$ is defined therein. This guarantees that if the capacity is non-zero, then the uniform prior will yield a non-zero rate, and hence the block will not last indefinitely.

Under the abstract assumptions made here, the following $F_i$ is known and can be substituted in Lemma~\ref{lemma:rateless_prior_predictor_lemma}:
\begin{equation}\label{eq:1910}
F_i(Q) = I(Q,\overline W_i)
.
\end{equation}
This yields the following result:
\begin{lemma}\label{lemma:C_overlineW_achievability_w_side_info}
For the scheme of Section~\ref{sec:arbitrary_var_rateless_scheme} under the abstraction specified above, with $n \geq 3$ and $K \geq 2 I_{\max}$ and properly chosen $\eta, \lambda$, the following holds: for any sequence of channels, the rate satisfies:
\begin{equation}\label{eq:917R}
R = \frac{K \cdot B}{n} \geq C(\overline W) - \Dpred,
\end{equation}
where $C(\overline W)$ is the capacity of the averaged channel and
\begin{equation}\label{eq:686}
\Dpred = 4 \cdot I_{\max}^{\tfrac{2}{3}} \cdot  |\mathcal{X}|^{\tfrac{2}{3}} \cdot  K^{\tfrac{1}{3}} \cdot \left( \frac{\ln (n)}{n} \right)^{\tfrac{1}{3}} \ntoinfty 0
,
\end{equation}
where $I_{\max} = \log \min (|\mathcal{X}|, |\mathcal{Y}|)$.
The parameters of the scheme $\eta, \lambda$ required to attain the result are specified in \eqref{eq:1060} and \eqref{eq:988} respectively.
\end{lemma}

Note that the bound \eqref{eq:686} is increasing with $K$, so it appears that that it can be improved by taking the minimal value of $K$. However in the actual system, there are be fixed overheads related to the communication scheme, and a large block size would be needed to overcome them. Taking any fixed and large enough $K$, the normalized regret is bounded by $O \left(\frac{\ln n}{n} \right)^{\frac{1}{3}}$, which converges to zero, but at a worse rate than we had in Section~\ref{sec:toy_exp_prior_predictor}.

Note that the claims of Lemma~\ref{lemma:C_overlineW_achievability_w_side_info} are stronger than the claims that appeared in the conference paper on the subject \cite{YL_PriorPrediction_ISIT2011}, for the same problem, mainly in terms of the improved convergence rate with $n$. Also, the scheme used here is slightly different than the one in the conference paper (in Equation~\eqref{eq:8385}). The proof corresponding to the scheme presented in the conference paper can be found in an early version uploaded to arXiv \cite{YL_PriorPredictionV2}.

To prove Lemma~\ref{lemma:C_overlineW_achievability_w_side_info}, Lemma~\ref{lemma:rateless_prior_predictor_lemma} is used with $F_i$ defined in \eqref{eq:1910}. The rate guaranteed by Lemma~\ref{lemma:rateless_prior_predictor_lemma} is approximately $\Rtarget \geq \sum_{i=1}^{B+1} \frac{m_i}{n} I(Q,\overline W_i)$. Using convexity of the mutual information with respect to the channel this is at least $I \left(Q, \sum_{i=1}^{B+1} \frac{m_i}{n} \overline W_i \right) = I \left(Q, \overline W \right)$, and since this is true for any $Q$, the rate is at least $C \left(\overline W \right)$. The detailed proof appears in Appendix~\ref{sec:proof_C_overlineW_achievability_w_side_info}.

\section{Proof of the main result}\label{sec:mainproof}
In this section we prove Theorem~\ref{theorem:C_overlineW_achievability}, regarding the attainability of $C(\overline W)$. The principles of the prediction scheme have been laid in the previous section, and here we plug-in a suitable decoding condition and a channel estimator.

\subsection{Preliminaries}\label{sec:mainproof_preliminaries}
Suppose that during a certain block of length $m$ we have used the i.i.d. prior $Q(x)$. In order to estimate the channel after the block has ended and $\vr x$ was decoded, we use the following estimate:
\begin{equation}\label{eq:W1159}
\fpr{W}(y|x) = \frac{\hat P_{\vr x, \vr y}(x,y)}{Q(x)}  ,
\end{equation}
where here and throughout the current section, $\vr x, \vr y$ denote the $m$-length input and output vectors over the block, and $\hat P_{\vr x, \vr y}(x,y)$ is the empirical distribution of the pair $(x_i, y_i)$ (for $i=1,\ldots,m$). The estimator is the joint empirical distribution divided by the (known) marginal distribution of the input $X$. Since we mix a uniform prior into $Q(x)$ \eqref{eq:Qi3382}, all $Q(x)$ are bounded away from zero, which makes the estimator \eqref{eq:W1159} statistically stable, in comparison with the more natural estimator given by the empirical conditional distribution:
\begin{equation}\label{eq:1171}
\hat W(y|x) = \hat P_{\vr y | \vr x}(x,y) = \frac{\hat P_{\vr x, \vr y}(x,y)}{\hat P_{\vr x}(x)}
,
\end{equation}
in which the denominator may turn out to be zero. A drawback of the proposed estimator \eqref{eq:W1159} is, that it does not generally yield a legitimate probability distribution, i.e. $\sum_{y} \fpr{W}(y|x) \neq 1$. The result of using this estimator is that in the calculations, we will see values that formally appear like probabilities but are not. To distinguish them from legitimate probabilities we term these values ``false'' probabilities, and mark them with a $\fpr{\square}$. These functions usually approximate or estimate a legitimate probability. Formally, a false probability $\fpr{p}(y)$ or $\fpr{p}(y|x)$ can be any non-negative function of $y$ or $x,y$ (respectively). Note that until this point we did not need the assumption that the output alphabet $\mathcal{Y}$ is finite, since the channel was given to the predictor rather than being estimated, and it is the first time this assumption is used.

The function that we use as an optimization target for selecting the prior for the next block is, as before, the mutual information. The reason is that since our aim is to achieve the capacity of the averaged channels, the ``competing'' schemes, for each prior $Q$, achieve the mutual information of the averaged channel. However, since the estimates of past channels are false-probabilities, we need to define how to apply the mutual information to them. We do this by simply plugging-in the false channel into the standard formula of $I(Q,W)$. This substitution results in what we define as the \textit{false mutual information} $\fpr{I}(Q,\fpr{W})$:
\begin{equation}\label{eq:fMI_def3386}
\fpr{I}(Q,\fpr{W}) \defeq \sum_{x,y} Q(x) \fpr{W}(y|x) \log \left( \frac{\fpr{W}(y|x)}{\sum_{x'} Q(x') \fpr{W}(y|x')} \right)
,
\end{equation}
where cases of $Q(x)=0$ or $\fpr{W}(y|x)$ are resolved using the convention $0 \cdot \log 0 = 0$.
The following lemma shows that most of the properties of the mutual information function $I(P,W)$ needed for our previous analysis in Section~\ref{sec:arbitrary_var_exp_prior_predictor} are maintained.

\begin{lemma}[Properties of false mutual information]\label{lemma:fMI_properties}
The function $\fpr{I}(Q,\fpr{W})$ defined in \eqref{eq:fMI_def3386} is
\begin{enumerate}
\item Non negative
\item Concave with respect to $Q$
\item Convex with respect to $\fpr{W}$
\item Upper bounded by $\sigma \cdot \log |\mathcal{X}|$, where $\sigma = \max_x \left[ \sum_y \fpr{W}(y|x) \right]$.
\end{enumerate}
\end{lemma}
The proof is technical and appears in Appendix~\ref{sec:proof_of_lemma_fMI_properties}. In addition to the properties above, our proof relies on the next property which is more surprising. When the prior $Q$ used for estimating the channel in \eqref{eq:W1159} is the same prior $Q$ used as input in \eqref{eq:fMI_def3386}, the false mutual information attains a form which is familiar from \selector{\cite{YL_PhdThesis}}{Part~I (see Section~\ref{sec:examples_eMI}, and also Section~\ref{sec:eMI_optimality}, Equation~\ref{eq:A1739})} as a prototype of the zero order rate function. As in \selector{\cite{YL_PhdThesis}}{Part~I}, we use this form to obtain a bound on the probability of $\fpr{I}(Q,\fpr{W})$ to exceed a threshold for a random drawing of $\vr x$. This bound, in turn, allows us to construct the rate-adaptive system attaining a block length $m_i$ that depends on $\fpr{I}(Q,\fpr{W})$.

Following \selector{\cite{YL_PhdThesis}}{Section~\ref{sec:empirical_distributions_and_information_measures}}, we define conditional empirical \emph{probability} of the discrete sequence $\vr x$ given the sequence $\vr y$ as $\hat p(\vr x | \vr y) \defeq \prod_{i=1}^m \hat P_{\vr x|\vr y}(x_i|y_i)$, i.e. the probability of the sequence $\vr x$ under the conditionally i.i.d. distribution $P(y|x) = \hat P_{\vr x| \vr y}(y|x)$. Also, when vectors are substituted into $Q$ we explicitly extend $Q$ in an i.i.d. fashion, i.e. $Q(\vr x) \defeq \prod_{i=1}^m Q(x_i)$. We will use the following result:
\begin{lemma}[False mutual information as a decoding metric]\label{lemma:fMI_as_Remp}
The false MI with prior $Q(x)$ and $\fpr{W}(y|x) = \frac{\hat P_{\vr x, \vr y}(x,y)}{Q(x)}$ where $\vr x, \vr y$ are $m$-length vectors can be written as:
\begin{equation}\label{eq:1219}
\fpr{I}(Q,\fpr{W}) = \fpr{I} \left(Q(x), \frac{\hat P_{\vr x \vr y}(x,y)}{Q(x)} \right) = \frac{1}{m} \log \frac{\hat p(\vr x | \vr y)}{Q(\vr x)}
.
\end{equation}
Furthermore, for any $Q$ and any $\vr y$, when $\vr X$ is distributed i.i.d. $\vr X \sim Q^n$,
\begin{equation}\begin{split}\label{eq:1227}
\Pr \left( \fpr{I}(Q,\fpr{W}) \geq T | \vr y \right)
& =
\Pr \left( \frac{\hat p(\vr X | \vr y)}{Q(\vr X)} \geq  \exp(m T) \Big| \vr y \right)
\\& \leq
\exp(-(m T - k_0 \log m - k_1))
,
\end{split}\end{equation}
where
\begin{equation}\label{eq:k1227}
k_0=k_1 = (|\mathcal{X}|-1) \cdot |\mathcal{Y}|
.
\end{equation}
\end{lemma}

\onlypaper{Note that from the results in \cite[Theorem 9?]{YL_PhdThesis}\footnote{Reference is to be updated in the final revision.} (by using the result of the Theorem and the definition of intrinsic redundancy therein) we can obtain a tighter upper bound with $k_0 = \frac{|\mathcal{Y}| \cdot (|\mathcal{X}|-1)}{2} \log m$ ($k_0 \log m + k_1 = r_m$ where $r_m$ is explicitly stated in \cite[Theorem 9?]{YL_PhdThesis}). For the sake of simplicity we prove here a looser result above, as this does not change the asymptotical results significantly.}
\onlyphd{\todo{write the respective paragraph for thesis}}

\textit{Proof of Lemma~\ref{lemma:fMI_as_Remp}: }
The first part is shown by direct substitution. When $\fpr{W} = \frac{\hat P_{\vr x \vr y}(x,y)}{Q(x)}$ we have
\begin{equation}\begin{split}\label{eq:1240}
\sum_{x'} Q(x') \fpr{W}(y|x')
&=
\sum_{x'} Q(x') \frac{\hat P_{\vr x \vr y}(x',y)}{Q(x')}
\\& =
\sum_{x'} \hat P_{\vr x \vr y}(x',y)
=
\hat P_{\vr y}(y)
.
\end{split}\end{equation}
Therefore
\begin{equation}\begin{split}\label{eq:3455}
\fpr{I}(Q,\fpr{W})
&=
\fpr{I} \left(Q(x), \frac{\hat P_{\vr x \vr y}(x,y)}{Q(x)} \right)
\\ & \stackrel{\eqref{eq:fMI_def3386}, \eqref{eq:1240}}{=}
\sum_{x,y} Q(x) \frac{\hat P_{\vr x \vr y}(x,y)}{Q(x)} \log \left( \frac{\hat P_{\vr x \vr y}(x,y)}{Q(x) \hat P_{\vr y}(y)} \right)
\\&=
\sum_{x,y} \hat P_{\vr x \vr y}(x,y) \log \left( \frac{\hat P_{\vr x|\vr y}(x|y)}{Q(x)} \right)
\\&=
\frac{1}{m} \sum_{i=1}^m \log \left( \frac{\hat P_{\vr x|\vr y}(x_i|y_i)}{Q(x_i)} \right)
\\&=
\frac{1}{m} \log \frac{\hat p(\vr x | \vr y)}{Q(\vr x)}
.
\end{split}\end{equation}

As for the second claim, by Markov bound \onlyphd{(Section~\ref{sec:chernoff_markov_exposition})} we have:
\begin{equation}\begin{split}\label{eq:1270}
&
\Pr \left( \frac{\hat p(\vr X | \vr y)}{Q(\vr X)} \geq  \exp(m T) \Big| \vr y  \right)
\\& \leq
\frac{1}{\exp(m T)} \E \left[ \frac{\hat p(\vr X | \vr y)}{Q(\vr X)} \Big| \vr y \right]
\\ & \stackrel{(a)}{=}
\exp(-m T) \sum_{\vr x \in \mathcal{X}^m} \frac{\hat p(\vr x | \vr y)}{Q(\vr x)} Q(\vr x)
\\& =
\exp(-m T) \sum_{\vr x \in \mathcal{X}^m} \hat p(\vr x | \vr y)
,
\end{split}\end{equation}
where in (a) we have used the fact $\vr X$ is distributed $Q$ independently of $\vr y$. To bound the sum above we split the set of sequences $\vr x$ to sub-sets having the same conditional empirical probability $\hat P_{\vr x|\vr y}(x|y)$ (i.e. same conditional type \cite{MethodOfTypes}\cite[\S 11]{CoverThomas_InfoTheoryBook}). In a subset having $\hat P_{\vr x|\vr y}(x|y) = p(x|y)$, the empirical probability $\hat p(\vr x | \vr y) = \prod_i p(x_i | y_i)$ equals the (legitimate) probability of the sequence under the i.i.d. distribution $p$, and as a result we have $\sum_{\vr x: \hat P_{\vr x|\vr y}(x|y) = p(x|y)} \hat p(\vr x | \vr y) \leq 1$. The number of subsets is upper bounded (similarly to bounds on the number types \cite[Theorem 11.1.1]{CoverThomas_InfoTheoryBook} \onlyphd{see also Section~\ref{sec:1588}}) by which is upper bounded by $(m+1)^{(|\mathcal{X}|-1) \cdot |\mathcal{Y}|}$, since $p(x|y) \in \left\{0,\tfrac{1}{m},\tfrac{2}{m},\ldots, 1 \right\}$ is completely defined by $(|\mathcal{X}|-1) \cdot |\mathcal{Y}|$ integers in $\{0,\ldots,m\}$.

\begin{equation}\begin{split}\label{eq:1284}
\sum_{\vr x \in \mathcal{X}^m} \hat p(\vr x | \vr y)
&=
\sum_{p} \sum_{\vr x: \hat P_{\vr x|\vr y}(x|y) = p(x|y)} \hat p(\vr x | \vr y)
\\& \leq
(m+1)^{(|\mathcal{X}|-1) \cdot |\mathcal{Y}|}
.
\end{split}\end{equation}
Substituting in \eqref{eq:1270} and using $m+1 \leq 2m$ yields the desired result.

\subsection{Decoding condition and estimated channel}\label{sec:mainproof_decoding_cond}
When the communication scheme was described in Section~\ref{sec:arbitrary_var_rateless_scheme}, the details of the decoding condition and channel estimation were omitted. These are specified below. At each symbol of the block and for each codeword $\vr x_l, l=1,\ldots,\exp(K)$ in the codebook, the receiver evaluates the following decoding condition:
\begin{equation}\label{eq:ter1315}
\log \frac{\hat p(\vr x_l | \vr y)}{\hat Q_i(\vr x_l)} > \beta K
,
\end{equation}
where $\beta$ is a parameter to be specified later on, and the vectors $\vr x_l$ and $\vr y$ are taken over the symbols of the block.

Equivalently, by Lemma~\ref{lemma:fMI_as_Remp}, the decoding condition can be written as:
\begin{equation}\label{eq:ter1315alt}
m \cdot \fpr{I}(\hat Q_i,\fpr{W}) > \beta K
,
\end{equation}
where $m$ is the number of the symbol in the block and $\fpr{W}(y|x)$ is a channel estimate according to \eqref{eq:W1159}, where $\vr x$ is substituted with the hypothesized input $\vr x_l$ and $\vr y$ is known output vector over the block.

After decoding, the receiver sets the estimated channel $\fpr{W}_i$ as the false channel $\fpr{W}(y|x)$ measured according to \eqref{eq:W1159}, where $\vr x, \vr y$ are the $m$ length vectors denoting the (hypothesized) input and output vectors over the duration of the block.

To produce the next prior, this false channel is fed into the prediction scheme of Lemma~\ref{lemma:rateless_prior_predictor_lemma}, with $F_i(Q) = \fpr{I}(Q,\fpr{W}_i)$ and where $m_i$ denotes the length of block $i$. The parameters $\beta, \eta, \lambda$ (the latter are required for the prediction scheme of Lemma~\ref{lemma:rateless_prior_predictor_lemma}) will be determined in the course of the proof.

\subsection{Proof outline}\label{sec:mainproof_outline}
The following proof outline conveys the main ideas in the proof, while some details were intentionally dropped, for simplicity.
\begin{enumerate}
\item Using the results of Lemma~\ref{lemma:fMI_as_Remp} we show that the block lengths can satisfy the inequality \eqref{eq:mi728} required by Lemma~\ref{lemma:rateless_prior_predictor_lemma}, up to a small overhead term in $K$, while still attaining a small probability of error.
\item Operating the prior prediction scheme of Lemma~\ref{lemma:rateless_prior_predictor_lemma}, with $F_i(Q) =\fpr{I}(Q, \fpr{W}_i)$ as the metric with $\fpr{W}_i$ the \emph{measured} channels, guarantees that if no errors were made, the rate achieved by the system exceeds $\max_Q \sum_{i=1}^{B+1} \frac{m_i}{n} \fpr{I}(Q, \fpr{W}_{i})$ up to vanishing factors, where $B$ is the number of blocks that were sent.
\item Due to the convexity of the false mutual information with respect to the channel, the rate above exceeds $\max_Q \fpr{I}(Q, \fpr{W}_A)$ where $\fpr{W}_A = \sum_{i=1}^{B+1} \frac{m_i}{n} \fpr{W}_{i}$.
\item Since the rate above exceeds $\fpr{I}(Q, \fpr{W}_A)$ for any $Q$, it exceeds  $\fpr{C}(\fpr{W}_A) = \max_Q \fpr{I}(Q, \fpr{W}_A)$.
\item All is left is to show the convergence in probability of $\fpr{W}_A$ to the true average channel $\overline W$, and by using the continuity of the capacity this proves the convergence in probability of $\fpr{C}(\fpr{W}_A)$ to the capacity of the averaged channel $C(\overline W)$.
\item In order to attain explicit bounds on the convergence rate we develop bounds relating the difference in capacity to the difference in the channels, and optimize the system parameters.
\end{enumerate}

Note that there are several delicate issues caused by the relations between $\fpr{W}_i$, $m_i$ and $\hat Q_i$. For example, the correct operation of the prior predictor relies on the assumption of correct decoding which is required to obtain the correct channel estimators (i.e. that $\vr x$ used in \eqref{eq:W1159} is the true channel input). However, conditioning on the event of correct decoding changes the distribution of the average estimated channel $\fpr{W}_A$. Another example is that, although the convergence of $\sum_{i=0}^{B+1} \frac{m_i}{n} \fpr{W}_{i}$ to $\overline W$ appears to be trivial at first sight, the proof is complicated by the fact that the block lengths $m_i$ are random variables, which themselves depend on the estimated channels $\fpr{W}_i$. One embodiment of this dependence is that the block would never end with an estimated channel which has zero capacity. Another dependence is between $m_i, \fpr{W}_i$ of different blocks, created through the prior prediction $\hat Q_i$.

We start with a set of definitions and propositions formalizing the claims made in the proof outline above. We use $k$ to denote the symbol index and $i$ to denote the block index. We denote by $i=b_k$ the block index of a certain symbol (i.e. $i=b_k$ if symbol $k$ belongs to block $i$). We define $m_i$ ($i=1,\ldots,B+1$) as the length of each block including the last one. The last block is not accounted for in the rate, even if it is decoded.

\subsection{Error probability}\label{sec:mainproof_error_prob}
\begin{proposition}[Error probability and decoding thresholds]\label{prop:symbolwise_scheme_err_probability}
For the value of $\beta$ given below \eqref{eq:1356}, the probability of any decoding error occurring in any of the blocks is at most $\epsilon$.
\end{proposition}

\onlyphd{This part is similar to Section~\ref{sec:rate_adaptive_performance}.}
\textit{Proof:} Consider a specific block and denote by $m$ the number of the symbol inside the block. Since codewords other than the one which is actually transmitted are independent of $\vr x, \vr y$, the probability to decide in favor of a specific erroneous codeword $\vr X_l$, at any specific symbol $k$ (i.e. that \eqref{eq:ter1315} will hold with respect to it), is upper bounded using \eqref{eq:1227} by:
\begin{equation}\begin{split}\label{eq:1342}
P_{err}(l,k)
&=
\Pr \left( \log \frac{\hat p(\vr X_l | \vr y)}{Q(\vr X_l)} > \beta K \big| \vr y \right)
\\&=
\Pr \left( \frac{\hat p(\vr X_l | \vr y)}{Q(\vr X_l)} > \exp(mT) \big| \vr y \right) \Big|_{T=\beta K/m}
\\&\leq
\exp(-(\beta K - k_0 \log m - k_1))
,
\end{split}\end{equation}
where $k_0,k_1$ are defined in Lemma~\ref{lemma:fMI_as_Remp}. And by taking expected value over $\vr Y$ we have that the same bound holds when not conditioning on $\vr y$. Since there are $\exp(K)-1$ competing codewords, and $n$ symbols, the probability to decide in favor of any erroneous codeword at any symbol (i.e. to make any decoding error), is upper bounded using the union bound, by:
\begin{equation}\begin{split}\label{eq:1347}
P_{err}
&\leq
\exp(K) \cdot n \cdot \exp(-(\beta K - k_0 \log n - k_1))
\\&=
\exp(-((\beta-1) K - (k_0+1) \log n - k_1))
,
\end{split}\end{equation}
where we replaced $\log m$ by $\log n \geq \log m$. We now determine $\beta$ so as to make the RHS equal $\epsilon$, and thereby guarantee the error probability is at most $\epsilon$:
\begin{equation}\label{eq:1356}
\beta = 1 + \frac{\log (\epsilon^{-1}) + (k_0+1) \log n + k_1}{K}
.
\end{equation}
Note that with a suitable choice of $K$ we would have $\beta \ntoinfty 1^+$.
\endofproof

\subsection{Attained rate}\label{sec:mainproof_attained_rate}
The following lemma relates the rate to the averaged estimated channel $\fpr{W}_A$:
\begin{proposition}[Rate as a function of average estimated channel]\label{prop:symbolwise_scheme_predictor_rate}
If there are no decoding errors, the rate of the scheme satisfies:
\begin{equation}\label{eq:1411}
R = \frac{K B}{n} \geq \left( 1 - \delta_1 \right) \cdot \min \left( \fpr{C}\left(\fpr{W}_A \right), I_{\max} \right) - \Dpred
,
\end{equation}
where $\fpr{C}\left(\fpr{W} \right) \defeq \max_{Q \in \Delta_\mathcal{X}} \fpr{I}(Q, \fpr{W})$ is the false capacity, $\fpr{W}_A$ is the averaged estimated channel
\begin{equation}\label{eq:1417}
\fpr{W}_A(y|x) = \frac{1}{n} \sum_{i=1}^{B+1} m_i \fpr{W}_i(y|x)
,
\end{equation}
$\Dpred$ is defined in Lemma~\ref{lemma:rateless_prior_predictor_lemma} (for the relevant parameters $n,K,\lambda$), and
\begin{equation}\label{eq:1418}
\delta_1 = \frac{1}{K} \left[\log (\epsilon^{-1}) + (k_0+1) \log n + k_1 + \log \left( \frac{|\mathcal{X}|}{\lambda} \right) \right]
.
\end{equation}
\end{proposition}

\textit{Proof:}
Denote by $\fpr{W}_i^{(l)}(y|x)$ the channel estimate according to \eqref{eq:W1159}, taken over the symbols of the $i$-th block, with respect to the hypothesized input sequence $\vr x_l$. By our definition of $\fpr{W}_i$ (Section~\ref{sec:mainproof_decoding_cond}), $\fpr{W}_i = \fpr{W}^{(l)}(y|x)$ when $l$ is the index of the correct codeword. Denote by $\fpr{W}^*_i$ the value of $\fpr{W}_i^{(l)}(y|x)$ when $l$ is the index of the hypothesized codeword. When there are no errors, $\fpr{W}^*_i = \fpr{W}_i$.

We use the prediction scheme of Lemma~\ref{lemma:rateless_prior_predictor_lemma} with $F_i(Q) = \fpr{I}(Q,\fpr{W}^*_i)$. By Lemma~\ref{lemma:fMI_properties}, this choice satisfies the conditions of the lemma with respect to $F_i(Q)$. Assuming there are no errors, we can equivalently write $F_i(Q) = \fpr{I}(Q,\fpr{W}_i)$.

We now use the decoding condition to show the requirements of Lemma~\ref{lemma:rateless_prior_predictor_lemma} with respect to the block length \eqref{eq:mi728} hold.

Denote by $\fpr{W}_i^{(B)}$ and $\fpr{W}_i^{(E)}$, the channel estimates taken with respect to the true $\vr x$ over the first $m_i-1$ symbols of the block $i$, and over the last symbol of the block, respectively. In other words, if block $i$ spans symbols $[k_i, l_i]$ where $l_i - k_i + 1 = m_i$ then
\begin{eqnarray}\label{eq:1483}
\fpr{W}_i(y|x) &=& \frac{\sum_{k=k_i}^{l_i} \Ind(X_k = x, Y_k = y)}{m_i \cdot \hat Q_i(x)} \\
\fpr{W}_i^{(B)}(y|x) &=& \frac{\sum_{k=k_i}^{l_i - 1} \Ind(X_k = x, Y_k = y)}{(m_i - 1) \hat Q_i(x)} \\
\fpr{W}_i^{(E)}(y|x) &=& \frac{\Ind(X_{l_i} = x, Y_{l_i} = y)}{\hat Q_i(x)}
,
\end{eqnarray}
where in the equations above we wrote the empirical distribution in \eqref{eq:W1159} explicitly as a normalized sum of indicator functions. We currently assume $m_i > 1$ and we'll return to the case of $m_i=1$ at the end. From the above we have that:
\begin{equation}\label{eq:1491}
\fpr{W}_i(y|x) = \frac{m_i-1}{m_i} \fpr{W}_i^{(B)}(y|x) + \frac{1}{m_i} \fpr{W}_i^{(E)}(y|x)
.
\end{equation}

Since at symbol $m_i-1$ in the block , which is one symbol before decoding, none of the codewords satisfies the decoding condition \eqref{eq:ter1315alt}, including the correct codeword (which corresponds to the true channel input $\vr X$), we have
\begin{equation}\label{eq:mi1413rand0}
(m_i-1) \cdot \fpr{I} \left(\hat Q_i,\fpr{W}_i^{(B)} \right) \leq \beta K
.
\end{equation}
The same holds for the last block $i=B+1$. As for $\fpr{W}_i^{(E)}$, from \eqref{eq:Qi3382} we have that
\begin{equation}\label{eq:QiMin1503}
\hat Q_i(x) \geq \frac{\lambda}{|\mathcal{X}|}
,
\end{equation}
and because $\fpr{W}_i^{(E)}$ is measured on a single symbol, we can bound:
\begin{equation}\label{eq:1507}
\fpr{I} \left(\hat Q_i,\fpr{W}_i^{(E)} \right) = \log \left( \frac{1}{\hat Q_i(X_{l_i})} \right) \leq \log \left( \frac{|\mathcal{X}|}{\lambda} \right)
.
\end{equation}
The equality above can be obtained using Lemma~\ref{lemma:fMI_as_Remp}, or by definition \eqref{eq:fMI_def3386}, using the fact that only for a single pair $(x,y)$, $\fpr{W}_i^{(E)}(y|x) > 0$.
Combining \eqref{eq:mi1413rand0} and \eqref{eq:1507} using \eqref{eq:1491} we have:
\begin{equation}\begin{split}\label{eq:mi1413rand}
m_i \cdot \fpr{I} \left(\hat Q_i,\fpr{W}_i \right)
& =
m_i \cdot \fpr{I} \left(\hat Q_i, \frac{m_i-1}{m_i} \fpr{W}_i^{(B)} + \frac{1}{m_i} \fpr{W}_i^{(E)} \right)
\\ & \leq
(m_i-1) \cdot \fpr{I} \left(\hat Q_i,\fpr{W}_i^{(B)} \right) + 1 \cdot \fpr{I} \left(\hat Q_i,\fpr{W}_i^{(E)} \right)
\\ & \leq
\beta K + \log \left( \frac{|\mathcal{X}|}{\lambda} \right)
\defeq
\tilde K
.
\end{split}\end{equation}
In the case of $m_i=1$, $\fpr{W}_i = \fpr{W}_i^{(E)}$ and \eqref{eq:mi1413rand} holds due to \eqref{eq:1507}.
The last inequality means the conditions of Lemma~\ref{lemma:rateless_prior_predictor_lemma} with respect to $m_i$ are satisfied, with $K$ replaced by $\tilde K$.
Under the conditions of the lemma, it guarantees that:
\begin{equation}\label{eq:688p}
\tilde R \defeq \frac{\tilde K B}{n} \geq \min \left( \max_Q \sum_{i=1}^{B+1} \frac{m_i}{n} \cdot \fpr{I}(Q, \fpr{W}_i), I_{\max} \right) - \tilde \Dpred
,
\end{equation}
where $\tilde \Dpred = \Dpred(\tilde K)$ is the offset defined in the lemma, with $K$ replaced by $\tilde K$. We use the convexity of $\fpr{I}$ with respect to the channel (Lemma~\ref{lemma:fMI_properties}) in order to relate the sum above to the capacity of the estimated averaged channel $\fpr{W}_A$:
\begin{equation}\label{eq:1450}
\sum_{i=1}^{B+1} \frac{m_i}{n} \cdot \fpr{I}(Q, \fpr{W}_i)
\geq
\fpr{I}\left(Q, \sum_{i=1}^{B+1} \frac{m_i}{n} \cdot  \fpr{W}_i \right)
=
\fpr{I}\left(Q, \fpr{W}_A \right)
.
\end{equation}
Substituting in \eqref{eq:688p} we obtain:
\begin{equation}\begin{split}\label{eq:1460}
\tilde R
& \geq
\min \left( \max_Q \fpr{I}\left(Q, \fpr{W}_A \right), I_{\max} \right) - \tilde \Dpred
\\ & =
\min \left( \fpr{C}\left(\fpr{W}_A \right), I_{\max} \right) - \tilde \Dpred
.
\end{split}\end{equation}
Because the actual rate that the scheme achieves is not $\tilde R$ but $R = \frac{K \cdot B}{n}$, we have:
\begin{equation}\label{eq:1465}
R = \tilde R \cdot \frac{K}{\tilde K}
\geq
\frac{K}{\tilde K} \cdot \min \left( \fpr{C}\left(\fpr{W}_A \right), I_{\max} \right) - \frac{K}{\tilde K} \tilde \Dpred
.
\end{equation}
Considering the second term, notice that the expression for $\Dpred(K)$ in Lemma~\ref{lemma:rateless_prior_predictor_lemma}, is sublinear in $K$, i.e. $\frac{1}{K} \Dpred(K)$ is decreasing with $K$, and therefore $\frac{K}{\tilde K} \tilde \Dpred(\tilde K) \leq \frac{K}{K} \tilde \Dpred(K)$, and we can replace the offset term in \eqref{eq:1465} by $\Dpred(K)$.

As for the factor $\frac{K}{\tilde K}$ we have
\begin{equation}\begin{split}\label{eq:1569}
\frac{\tilde K}{K}
&=
\beta + \frac{1}{K} \log \left( \frac{|\mathcal{X}|}{\lambda} \right)
\\& =
1 + \underbrace{\frac{1}{K} \left[\log (\epsilon^{-1}) + (k_0+1) \log n + k_1 + \log \left( \frac{|\mathcal{X}|}{\lambda} \right) \right]}_{\delta_1}
.
\end{split}\end{equation}
and by using $\frac{K}{\tilde K} = \frac{1}{1 + \delta_1} \geq 1 - \delta_1$ we have the desired result.
\endofproof

\subsection{Channel convergence}
We would now like to show the convergence of $\fpr{W}_A$ to $\overline W$. As mentioned above, $m_i$ and $\fpr{W}_i$ are statistically dependent. To avoid conditioning on $m_i$, we first write $\fpr{W}_A$ in an alternative form. Plugging the explicit form of $\fpr{W}_i$ from \eqref{eq:1483} into the definition of $\fpr{W}_A$ \eqref{eq:1417}, we have:

\begin{equation}\begin{split}\label{eq:1583}
\fpr{W}_A
&=
\frac{1}{n} \sum_{i=1}^{B+1} m_i \fpr{W}_i(y|x)
\\&=
\frac{1}{n} \sum_{i=1}^{B+1} m_i \frac{\sum_{k=k_i}^{l_i} \Ind(X_k = x, Y_k = y)}{m_i \cdot \hat Q_i(x)}
\\&=
\frac{1}{n} \sum_{k=1}^{n} \frac{\Ind(X_k = x, Y_k = y)}{\hat Q_{b_k}(x)}
.
\end{split}\end{equation}
Recall that the averaged channel is
\begin{equation}\label{eq:1596}
\overline W = \frac{1}{n} \sum_{k=1}^{n} W_k(y|x)
.
\end{equation}
We would like to show that $\fpr{W}_A - \overline{W} \toinprobl{n \to \infty} 0$. Define
\begin{equation}\label{eq:1555}
\gamma_k(x,y) \defeq \frac{1}{n} \left[ \frac{\Ind(X_k = x, Y_k = y)}{\hat Q_{b_k}(x)} - W_k(y|x) \right]
,
\end{equation}
then
\begin{equation}\label{eq:1544}
\fpr{W}_A - \overline{W}
=
 \sum_{k=1}^n \gamma_k(x,y)
.
\end{equation}
Although $\gamma_k(x,y)$ are not i.i.d., they constitute a bounded martingale difference sequence, where the martingale is $\sum_{j=1}^k \gamma_j$, as we will show below. First, by \eqref{eq:QiMin1503}, each component $\gamma_k(x,y)$ is bounded $-\frac{1}{n} \leq \gamma_k(x,y) \leq \frac{1}{n} |\mathcal{X}| \lambda^{-1} \defeq \gamma_{\max}$, so they be bounded in absolute value by $\gamma_{\max}$. On average over the common randomness, each symbol $X_k$ is generated $X_k \sim \hat Q_{b_k}(x)$ independent of the past (given $\hat Q_{b_k}(x)$). In other words, for someone not knowing the specific codebook, the knowledge of past values of $\vr X_1^{k-1}, \vr Y_1^{k-1}$ does not yield any information about $X_k$ when $\hat Q_{b_k}(x)$ is given. Define the state variable $S_{k-1} = \left(\vr X_1^{k-1}, \vr Y_1^{k-1}, \{ \hat Q_{b_j} \}_{j=1}^k \right)$. Note that $\hat Q_{b_k}$ is only generated as a function of past symbols and therefore can be considered as part of the state at time $k$. We have:
\begin{equation}\begin{split}\label{eq:1564}
\E \left[ \gamma_k(x,y) \Big| S_{k-1} \right]
& =
\frac{\Pr(X_k = x, Y_k = y | S_{k-1})}{n \cdot \hat Q_{b_k}(x)} - \frac{W_k(y|x)}{n}
\\& =
\frac{\hat Q_{b_k}(x) \cdot W_k(y|x)}{n \cdot \hat Q_{b_k}(x)} - \frac{W_k(y|x)}{n}
=
0
.
\end{split}\end{equation}
Now, since the previous value of the sum $\sum_{j=1}^{k-1} \gamma_j$ is only a function of $S_{k-1}$, by applying the iterated expectations law we have
\begin{equation}\begin{split}\label{eq:1575}
&
\E \left[ \gamma_k(x,y) \Bigg| \sum_{j=1}^{k-1} \gamma_j \right]
\\ & =
\E \left\{ \E \left[ \gamma_k(x,y) \Bigg| S_{k-1}, \sum_{j=1}^{k-1} \gamma_j \right] \Bigg| \sum_{j=1}^{k-1} \gamma_j \right\} = 0
,
\end{split}\end{equation}
which shows $\sum_{j=1}^{k} \gamma_j$ is a martingale. We can now apply Hoeffding-Azuma Inequality \cite[A.1.3]{Nicolo}\cite{Wikipedia}\cite{Hoeffding} and obtain:
\begin{equation}\begin{split}\label{eq:1580}
&
\Pr \left\{ \left| \fpr{W}_A(y|x) - \overline{W}(y|x) \right| > t \right\}
=
\Pr \left\{ \left| \sum_{k=1}^n \gamma_k(x,y) \right| > t \right\}
\\& \leq
2 e^{-\frac{2 t^2}{n \gamma_{\max}^2}}
=
2 e^{-\frac{2 n \lambda^2 t^2}{|\mathcal{X}|^2}}
.
\end{split}\end{equation}
The above holds for each value of $(x,y)$ separately. To bound the $L_{\infty}$ norm we use the union bound:
\begin{equation}\begin{split}\label{eq:1598}
&
\Pr \left\{ \| \fpr{W}_A - \overline{W} \|_{\infty} > t \right\}
\\ & =
\Pr \left\{ \bigcup_{x,y} \left[ \left| \fpr{W}_A(y|x) - \overline{W}(y|x) \right| > t \right] \right\}
\\& \leq
\sum_{x,y} \Pr \left\{ \left| \fpr{W}_A(y|x) - \overline{W}(y|x) \right| > t \right\}
\\& \stackrel{\eqref{eq:1580}}{\leq}
2 |\mathcal{X}| \cdot |\mathcal{Y}| \cdot e^{-\frac{2 n \lambda^2 t^2}{|\mathcal{X}|^2}}
.
\end{split}\end{equation}
To guarantee the above holds with probability at most $\delta_0$ we choose $t$ to make the RHS equal $\delta_0$:
\begin{equation}\label{eq:1617}
t = \delta_W = \frac{|\mathcal{X}|}{\lambda} \sqrt{\frac{1}{2n} \ln \left( \frac{2 |\mathcal{X}| \cdot |\mathcal{Y}|}{\delta_0} \right)}
.
\end{equation}
This is summarized in the following proposition:
\begin{proposition}[Average estimated channel convergence]\label{prop:symbolwise_scheme_channel_convergence2}
For any $\delta_0 > 0$, and for $\delta_W$ defined above,
\begin{equation}\label{eq:1615p}
\Pr \left\{ \| \fpr{W}_A - \overline{W} \|_{\infty} > \delta_W  \right\} \leq \delta_0
.
\end{equation}
\end{proposition}
Observe that a large $\lambda$ improves the channel estimate convergence (reduces $\delta_W$), since it increases the minimum rate at which each input symbol is sampled. This is the additional role of $\lambda$ that we did not have in Lemma~\ref{lemma:C_overlineW_achievability_w_side_info}.

\subsection{Convergence of capacity}
The final step is to link the difference in the channels $\| \fpr{W}_A  - \overline W \|$ to the difference in capacities. For this purpose we use the following lemma:
\begin{lemma}[$L_p$ bound on difference of false mutual information and capacity]\label{lemma:fMI_Lp_bound}
Let $Q(x)$ be an input distribution on the discrete alphabet $\mathcal{X}$, $W(y|x), y \in \mathcal{Y}$ a conditional distribution, and $\fpr{W}(y|x)$ a false conditional distribution. Define
\begin{equation}\label{eq:1630}
\Delta_p
=
\| \fpr{W}(y|x) - W(y|x) \|_{p}
,
\end{equation}
where
\begin{equation}\label{eq:1630b}
\| f(x,y) \|_{p}
\defeq
\begin{cases}
    \displaystyle \left( \sum_{x,y} |f(x,y)|^p \right)^{1/p}   & p < \infty \\
    \displaystyle \max_{x,y} |f(x,y)|                          & p = \infty
\end{cases}
.
\end{equation}
Assuming $\Delta_p \leq \tfrac{1}{4}$ we have:
\begin{equation}\label{eq:1541}
\forall Q: \left| \fpr{I}(Q,\fpr{W}) - I(Q,W) \right| \leq 2 f_p (\Delta_p)
,
\end{equation}
and
\begin{equation}\label{eq:1545}
\left| \fpr{C}(\fpr{W}) - C(W) \right| \leq 2 f_p (\Delta_p)
,
\end{equation}
where
\begin{equation}\label{eq:1469t}
f_p(t) = -t \cdot |\mathcal{Y}|^{1 - 1/p} \log \left( \frac{t}{|\mathcal{Y}|^{1/p}} \right)
.
\end{equation}
For $p=\infty$, by convention $1/p=0$. Furthermore $f_p(t)$ is concave and monotonically non-decreasing for $t \leq \tfrac{1}{4}$.
\end{lemma}
Note that the lemma is also true with respect to legitimate distributions. The proof of the lemma is based on Cover and Thomas' $L_1$ bound on entropy \cite{CoverThomas_InfoTheoryBook}, and \Holder's~inequality, and appears in Appendix~\ref{sec:Lp_bound_on_capacity}.

\subsection{Main argument of the proof}
We now combine the results above as follows:
Choose a value of $\delta_0$. We denote by $E$ the event of any decoding error occurring in any of the blocks, and by $D$ the event $\|  \fpr{W}_A - \overline W \|_{\infty} > \delta_W$. We use and overline $\overline \Box$ to denote complementary events.

Consider the event $\overline D \cap \overline E$. In this case, we have $\|  \fpr{W}_A - \overline W \|_{\infty} \leq \delta_W$ and from Lemma~\ref{lemma:fMI_Lp_bound} this implies
$| \fpr{C}(\fpr{W}_A)  - C(\overline W) | \leq \delta_C$ where $\delta_C = 2 f_\infty (\delta_W) = - 2 \delta_W \cdot |\mathcal{Y}| \log (\delta_W)$. From Proposition~\ref{prop:symbolwise_scheme_predictor_rate} we have that:
\begin{equation}\begin{split}\label{eq:1689}
R
&\geq
\left( 1 - \delta_1 \right) \cdot \min \left( \fpr{C}\left(\fpr{W}_A \right), I_{\max} \right) - \Dpred
\\& \geq
\left( 1 - \delta_1 \right) \cdot \min \left( C(\overline W) - \delta_C, I_{\max} \right) - \Dpred
\\& \geq
\left( 1 - \delta_1 \right) \cdot \left( \min \left( C(\overline W), I_{\max} \right) - \delta_C \right) - \Dpred
\\& =
\left( 1 - \delta_1 \right) \cdot \left( C(\overline W)  - \delta_C \right) - \Dpred
\\& =
C(\overline W) - \delta_1 \cdot C(\overline W) - \delta_C \cdot \left( 1 - \delta_1 \right) - \Dpred
\\& \geq
C(\overline W) - \underbrace{\left( \delta_1 \cdot I_{\max} + \delta_C  + \Dpred \right)}_{\defeq \Delta_C}
.
\end{split}\end{equation}
To summarize, if $\overline D \cap \overline E$ then $R \geq C(\overline W) - \Delta_C$. By the union bound and Propositions~\ref{prop:symbolwise_scheme_channel_convergence2},\ref{prop:symbolwise_scheme_err_probability}, we have:
\begin{equation}\begin{split}\label{eq:1702}
\Pr \{ R < C(\overline W) - \Delta_C \}
& \leq
\Pr \{  D \cup E \}
\leq
\Pr \{  D \} + \Pr \{ E \}
\\& \leq
\delta_0 + \epsilon
.
\end{split}\end{equation}

Note that although Lemma~\ref{lemma:fMI_Lp_bound} is stated for general $L_p$ norms, we have used it here only with respect to the $L_{\infty}$ norm, since it is relatively simple to obtain bounds on the convergence of $\fpr{W}_A - \overline W$ by using the well known Hoeffding-Azuma inequality per channel element ($x,y$) and the union bound. However as the distribution of $\fpr{W}_A$ tends to a mutlivariate Gaussian distribution, using $L_2$ norm seems to be more suited. Indeed, applying Lemma~\ref{lemma:fMI_Lp_bound} with $L_2$ norm, together with the (yet unpublished) bound on the $L_2$ convergence of vector martingales due to Hayes \cite{HayesVectorMartingale} yields tighter bounds on the probability of having a small difference $\fpr{C}(\fpr{W}_A) - C(\overline W)$ for large alphabet sizes.

\subsection{Choice of the parameters}
We now substitute the numerical expressions for the various overheads, and set the parameters of the scheme to optimize the convergence rate. $\delta_0, \epsilon$ are parameters of choice, and together with $\lambda, K$ they determine $\Delta_C$. Our purpose is to choose $\lambda, K$ that will approximately minimize $\Delta_C$. This part is rather tedious. We write $\Delta_C$ and collect all the relations below:

\begin{eqnarray}
\Delta_C &=& \delta_1 \cdot I_{\max} + \delta_C  + \Dpred
\\ \delta_1 &=& \frac{1}{K} \Big[ \log (\epsilon^{-1}) + (k_0+1) \log n + k_1
\nonumber \\          & & \qquad + \log \left( \frac{|\mathcal{X}|}{\lambda} \right) \Big]
\\ \delta_C  &=& - 2 \delta_W \cdot |\mathcal{Y}| \log (\delta_W)
\\ \delta_W &=& \frac{|\mathcal{X}|}{\lambda} \sqrt{\frac{1}{2n} \ln \left( \frac{2 |\mathcal{X}| \cdot |\mathcal{Y}|}{\delta_0} \right)}
\\ \Dpred &=& \frac{K}{n} + I_{\max} \cdot \lambda + c_1 \sqrt{\frac{\ln (n)}{n}} \lambda^{-\half}  .
\\ c_1 &=& 2 \sqrt{K \cdot |\mathcal{X}| (|\mathcal{X}|-1) \cdot I_{\max}}
\end{eqnarray}

Since $\delta_W \geq \sqrt{\frac{1}{n}}$, $-2 \log (\delta_W) \leq \log n$, therefore $\delta_C \leq \delta_W \cdot |\mathcal{Y}| \log (n)$. To make $\delta_W \ntoinfty 0$ we need $\frac{|\mathcal{X}|}{\lambda} \leq \sqrt{n}$, and making this assumption, we have that the last element in $\delta_1$ is bounded by $\log \left( \frac{|\mathcal{X}|}{\lambda} \right) \leq \half \log n$. Further assuming that $k_1 \leq \tfrac{1}{4} k_0 \log n$ (this holds trivially for the values of $k_0,k_1$ of Lemma~\ref{lemma:fMI_as_Remp} when $n > 2^4$), and $\epsilon \geq \frac{1}{n^{d_{\epsilon}}}$ (for some arbitrary polynomial decay rate $d_{\epsilon}$) we have
\begin{equation}\begin{split}\label{eq:1781}
\delta_1
&\leq
\frac{1}{K} \left[d_{\epsilon} \log (n) + (k_0+1) \log n + \tfrac{1}{4} k_0 \log n + \half \log n \right]
\\&=
\frac{\log n}{K} (d_{\epsilon} + \tfrac{5}{4} k_0+ \tfrac{3}{2})
.
\end{split}\end{equation}
Using these bounds and extracting the constants we can upper bound $\Delta_C$ by:
\begin{equation}\label{eq:1785}
\Delta_C \leq
\underbrace{c_2 \frac{\ln n}{K}}_{(1)}  +
\underbrace{\frac{c_3}{\lambda} \frac{\ln (n)}{\sqrt{n}}}_{(2)} +
\underbrace{I_{\max} \cdot \lambda}_{(3)} +
\underbrace{c_4 \sqrt{\frac{\ln (n)}{n} \cdot \frac{K}{\lambda}}}_{(4)}  +
\underbrace{\frac{K}{n}}_{(5)}
,
\end{equation}
where element $(1)$ stems from $\delta_1$, $(2)$ from $\delta_C$ and $(3)-(5)$ from $\Dpred$, and the constants are:
\begin{eqnarray}
c_2 &=& \left(d_{\epsilon} + \tfrac{5}{4} k_0+ \tfrac{3}{2} \right) \cdot I_{\max} \cdot \log e \label{eq:1919-1} \\
c_3 &=& |\mathcal{X}| \cdot |\mathcal{Y}| \cdot \log(e) \cdot \sqrt{\frac{1}{2} \ln \left( \frac{2 |\mathcal{X}| \cdot |\mathcal{Y}|}{\delta_0} \right)} \label{eq:1919-2} \\
c_4 &=& \frac{c_1}{\sqrt{K}} = 2 \sqrt{|\mathcal{X}| (|\mathcal{X}|-1) \cdot I_{\max}} \label{eq:1919-3}
.
\end{eqnarray}
As we shall see, element $(5)$ is negligible. Therefore we first optimize the sum of $(1)$ and $(4)$ with respect to $K$, using Lemma~\ref{lemma:ab_alphabeta}. We write the sum as $a K^{\alpha} + b K^{-\beta}$ with $\alpha=\half, \beta=1, a = c_4 \sqrt{\frac{\ln (n)}{n} \cdot \frac{1}{\lambda}}, b = c_2 \ln n$. Since $K$ is required to be integer, we write it as a function of a real valued parameter $t$: $K=\lfloor t \rfloor$, and assume $t \geq 5$. Then $\frac{1}{K} \leq \frac{1}{t-1} = \frac{1}{t} \frac{t}{t-1} \leq \tfrac{5}{4} \frac{1}{t}$, and therefore $a K^{\alpha} + b K^{-\beta} \leq a t^{\alpha} + \underbrace{b \left(\tfrac{5}{4} \right)^{\beta}}_{b'} \cdot t^{-\beta}$. By optimizing the bound with respect to $t$ using Lemma~\ref{lemma:ab_alphabeta}, we obtain
\begin{equation}\label{eq:1804}
t^*
=
\left( \frac{b' \beta}{a \alpha} \right)^{\frac{1}{\alpha+\beta}}
=
\underbrace{\left( \tfrac{5}{2} c_2 c_4^{-1} \right)^{\frac{2}{3}}}_{c_5} \cdot \left( \lambda \cdot n \ln n \right)^{\frac{1}{3}}
,
\end{equation}
where we defined
\begin{equation}\label{eq:1804a}
c_5 \defeq \left( \tfrac{5}{2} c_2 c_4^{-1} \right)^{\frac{2}{3}}
.
\end{equation}

\begin{equation}\begin{split}\label{eq:1810}
a K^{\alpha} + b K^{-\beta}
& \stackrel{\eqref{eq:941}}{\leq}
2^{\tfrac{1}{3}}  \tfrac{3}{2} \cdot a^{\tfrac{2}{3}} \cdot (b')^{\tfrac{1}{3}}
\\& =
\underbrace{\tfrac{3}{2} \cdot \left( \tfrac{5}{2} c_2 c_4^2  \right)^{\tfrac{1}{3}}}_{c_6} \cdot  \left(  \frac{\ln^2 (n)}{n} \cdot \frac{1}{\lambda} \right)^{\tfrac{1}{3}}
.
\end{split}\end{equation}
Substituting in \eqref{eq:1785} (and upper bounding element $(5)$ by $t^*/n$), we obtain:
\begin{equation}\begin{split}\label{eq:1785b}
\Delta_C
&\leq
\underbrace{c_6 \cdot  \left(  \frac{\ln^2 (n)}{n} \cdot \frac{1}{\lambda} \right)^{\tfrac{1}{3}}}_{(1)+(4)}  +
\underbrace{\frac{c_3}{\lambda} \frac{\ln (n)}{\sqrt{n}}}_{(2)}
\\ & \qquad +
\underbrace{I_{\max} \cdot \lambda}_{(3)} +
\underbrace{c_5 \cdot \left( \lambda \cdot \frac{\ln n}{n^2} \right)^{\frac{1}{3}}}_{(5)}
,
\end{split}\end{equation}
To determine $\lambda$ we notice that it is a trade-off between element $(3)$ which is increasing in $\lambda$ and either $(1)+(4)$ or $(2)$ which are decreasing. Minimizing any combination separately (i.e. $((1)+(4))+(3)$ or $(2)+(3)$) using Lemma~\ref{lemma:ab_alphabeta}, yields the same decay rate $O \left( \left( \frac{\ln^2 (n)}{n} \right)^{\tfrac{1}{4}} \right)$, and $\lambda$ of the form \begin{equation}\label{eq:1861}
\lambda = c_{\lambda} \cdot \left( \frac{\ln^2 (n)}{n} \right)^{\tfrac{1}{4}}
.
\end{equation}
Therefore this determines the best decay rate possible for \eqref{eq:1785b}.
Note that we do not have to worry about the case $\lambda > 1$, since in this case the term $\lambda I_{\max}$ in \eqref{eq:1785} will exceed $I_{\max}$ and Theorem~\ref{theorem:C_overlineW_achievability} will be true in a void way. Substituting $\lambda$ we have:
\begin{equation}\begin{split}\label{eq:1785c}
\Delta_C
&\leq
\underbrace{\frac{c_6}{c_{\lambda}^{\tfrac{1}{3}}} \cdot  \left(  \left( \frac{\ln^2 (n)}{n} \right)^{1 - \tfrac{1}{4}}  \right)^{\tfrac{1}{3}}}_{(1)+(4)}  +
\underbrace{\frac{c_3}{c_{\lambda}} \left( \frac{\ln^2 (n)}{n} \right)^{\half - \tfrac{1}{4}}}_{(2)}
\\ & \qquad +
\underbrace{I_{\max} \cdot c_{\lambda} \cdot \left( \frac{\ln^2 (n)}{n} \right)^{\tfrac{1}{4}} }_{(3)}+
\underbrace{c_5 \cdot \left( \lambda \frac{\ln n}{n^2} \right)^{\frac{1}{3}}}_{(5)}
\\&\leq
\left[ \frac{c_6}{c_{\lambda}^{\tfrac{1}{3}}} + \frac{c_3}{c_{\lambda}} + I_{\max} \cdot c_{\lambda} \right] \cdot
\left( \frac{\ln^2 (n)}{n} \right)^{\tfrac{1}{4}}
+ c_5 \cdot \left( \frac{\ln n}{n^2} \right)^{\frac{1}{3}}
\\&\leq
\left[ \tfrac{3}{2} \cdot \left( \tfrac{5}{2} \frac{c_2 c_4^2}{c_{\lambda}}  \right)^{\tfrac{1}{3}} + \frac{c_3}{c_{\lambda}} + I_{\max} \cdot c_{\lambda}  + 1\right] \cdot
\left( \frac{\ln^2 (n)}{n} \right)^{\tfrac{1}{4}}
\\&=
c_{\Delta} \cdot
\left( \frac{\ln^2 (n)}{n} \right)^{\tfrac{1}{4}}
,
\end{split}\end{equation}
where in the last inequality we substituted the expression for $c_6$ and assumed $c_5 \cdot \left( \frac{\ln n}{n^2} \right)^{\frac{1}{3}} \leq \left( \frac{\ln^2 (n)}{n} \right)^{\tfrac{1}{4}}$. In the last step we defined
\begin{equation}\label{eq:1858}
c_{\Delta} \defeq  \tfrac{3}{2} \cdot \left( \tfrac{5}{2} \frac{c_2 c_4^2}{c_{\lambda}}  \right)^{\tfrac{1}{3}} + \frac{c_3}{c_{\lambda}} + I_{\max} \cdot c_{\lambda}  + 1
.
\end{equation}

We now revisit the assumptions we have made along the way.
\begin{itemize}
\item In \eqref{eq:1785c}, we assumed $c_5 \cdot \left( \frac{\ln n}{n^2} \right)^{\frac{1}{3}} \leq \left( \frac{\ln^2 (n)}{n} \right)^{\tfrac{1}{4}}$. This requires that
$(\ln n)^{\tfrac{1}{6}} n^{\tfrac{5}{12}} \geq c_5 = \left( \tfrac{5}{2} \frac{c_2}{c_4} \right)^{\frac{2}{3}}$, and a sufficient condition is $n \geq \left( \tfrac{5}{2} \frac{c_2}{c_4} \right)^{\frac{8}{5}}$.
\item For \eqref{eq:1781} we assumed $\frac{|\mathcal{X}|}{\lambda} \leq \sqrt{n}$. Substituting $\lambda$ leads to $n \ln^2 (n) \geq \frac{|\mathcal{X}|^4}{c_{\lambda}^4 }$, and a sufficient condition is
\begin{equation}\label{eq:1901}
n \geq \frac{|\mathcal{X}|^4}{c_{\lambda}^4 }
.
\end{equation}
\item For \eqref{eq:1781} we assumed $\epsilon \geq \frac{1}{n^{d_{\epsilon}}}$. We may simply determine $d_{\epsilon}$ and set $\epsilon = \frac{1}{n^{d_{\epsilon}}}$.
\item For \eqref{eq:1781} we assumed $k_1 \leq \tfrac{1}{4} k_0 \log n$, i.e. $n \geq \exp(4 k_1 / k_0)$
\item The application of Lemma~\ref{lemma:rateless_prior_predictor_lemma} to obtain Proposition~\ref{prop:symbolwise_scheme_predictor_rate} requires that $n \geq e$ and $\tilde K \geq 2 \cdot I_{\max}$. Since $\tilde K > K$ it is sufficient that $K \geq 2 I_{\max}$, or  $t^* \geq 2 I_{\max} + 1$. Furthermore for \eqref{eq:1810} we assumed $t^* \geq 5$, so we require $t^* \geq \max(2 I_{\max} + 1,5)$. Substituting $t^* = c_5 \cdot \left( \lambda \cdot n \ln n \right)^{\frac{1}{3}} = c_5 \cdot  c_{\lambda}^{\frac{1}{3}} (n \ln^2 n)^{1/4} \geq  c_5 \cdot c_{\lambda}^{\frac{1}{3}} n^{1/4}$ leads to the sufficient condition:
\begin{equation}\label{eq:1902}
    n \geq \left( \max(2 I_{\max} + 1,5) \cdot c_5^{-1} \cdot c_{\lambda}^{-\frac{1}{3}} \right)^4
.
\end{equation}

To summarize, the results holds for $n \geq n_{\min}$ where $n_{\min}$ is the maximum of the conditions of \eqref{eq:1901},\eqref{eq:1902}, \eqref{eq:1781} and of $n \geq e$:
\begin{equation}\begin{split}\label{eq:1970}
n_{\min}
&=
\max \bigg[e,
\frac{|\mathcal{X}|^4}{c_{\lambda}^4 }  ,
\left( \max(2 I_{\max} + 1,5) \cdot c_5^{-1} \cdot c_{\lambda}^{-\frac{1}{3}} \right)^4    ,
\\ & \qquad
\exp(4 k_1 / k_0)
\bigg]
.
\end{split}\end{equation}
\end{itemize}
This proves Corollary~\eqref{corollary:symbolwise_random_numerical}. \endofproof

The claims of the Theorem are milder and are easily deduced from this Corollary. Given $\epsilon, \delta$, let $\delta_0 = \half \delta$, and choose any $d_{\epsilon}>0$ and $c_{\lambda} > 0$. Choose $N$ large enough so that the error probability given by the Corollary satisfies $\epsilon(N) = N^{-d_{\epsilon}} < \min(\epsilon, \half \delta)$, and $N \geq n_{\min}$. This guarantees that for $n \geq N$, the requirements of the Corollary are met the error probability is $\epsilon(n) \leq \epsilon$, and the probability to fall short of the rate is at most $\epsilon(n) + \delta_0 \leq \delta$.
This concludes the proof of Theorem~\ref{theorem:C_overlineW_achievability}.
\endofproof

Following is a numerical example for the calculation of $c_{\Delta}$ and $n_{\min}$ in Theorem~\ref{theorem:C_overlineW_achievability}.
\begin{example}\label{example:symbolwise_random_prior_predictor}
For $|\mathcal{X}|=4, |\mathcal{Y}|=6$, $d_\epsilon=1$ and $\delta_0=10^{-10}$
we obtain $I_{\max}=2$ and $c_2=72.1, c_3=127, c_4=9.8, c_5=6.97$.
Choosing $c_{\gamma}=10$ we obtain $c_{\Delta}=51.7$ and $n_{\min}=\min(e, 0.0256,0.0123,16)=16$.
The convergence rate is rather slow and we have $\Delta_C \leq 0.2$ only for $n > 3.98 \cdot 10^{12}$.
\end{example}

\subsection{Proof of Corollary~\ref{corollary:symbolwise_random_prior_predictor1}}
During the proof of Theorem~\ref{theorem:C_overlineW_achievability} we assumed the channel sequence is unknown but fixed. It is easy to see that the same proof holds even if the channel sequence is determined by an online adversary.

The error probability (Proposition~\ref{prop:symbolwise_scheme_err_probability}) is maintained regardless of channel behavior, because the probabilistic assumptions made \eqref{eq:1342} refer to the distribution of codewords that were \emph{not} transmitted. Proposition~\ref{prop:symbolwise_scheme_predictor_rate} does not make any assumptions on the channel as it connects the communication rate with the \emph{measured} channel. The main difference is with respect to channel convergence. For
the proof of Proposition~\ref{prop:symbolwise_scheme_channel_convergence2} to hold we need to show that $\gamma_k$ remains a bounded martingale difference sequence, which boils down to verifying \eqref{eq:1564} still holds, i.e. that $\gamma_k$ has zero mean conditioned on the past. Adding the message to the state variable $S_{k-1}$ defined before \eqref{eq:1564}, i.e. redefining $S_{k-1} = \left(\vr X_1^{k-1}, \vr Y_1^{k-1}, \{ \hat Q_{b_j} \}_{j=1}^k, \vr b_1^\infty \right)$, where $\vr b_1^\infty$ is the message bit sequence, we have that \eqref{eq:1564} holds even when the channel $W_k(y|x)$ is a function of $S_{k-1}$. \endofproof

\subsection{A result for channels with memory of the input}
Although channels with memory of the input are not considered in this paper, the scheme presented above can be used over such channels as well. In this case, the performance of the scheme can be characterized as follows:
\begin{lemma}\label{lemma:prior_prediction_channel_with_memory}
When the scheme of Theorem~\ref{theorem:C_overlineW_achievability} is operated over a general channel $\Pr(\vr Y_1^n | \vr X_1^n)$, the results of the theorem hold if the averaged channel is redefined as follows:
\begin{equation}\label{eq:1631}
\overline W = \frac{1}{n} \sum_{k=1}^n \Pr(Y_k = y | X_k = x, \vr X^{k-1}, \vr Y^{k-1})
\end{equation}
\end{lemma}

Note that for each pair $x,y$, $\Pr(Y_k = y | X_k = x, \vr X^{k-1}, \vr Y^{k-1})$ is a random variable depending on the history  $\vr X^{k-1}, \vr Y^{k-1}$, and therefore, different from the main setting considered in this paper, $\overline W$ is also a random variable. The definition above \eqref{eq:1631} coincides with the previous definition of $\overline W$ \eqref{eq:204} when the channel is memoryless in the input. This lemma is used in \cite{YL_Univ_w_Memory} to show competitive universality for channels with memory of the input.

\textit{Proof:}
As in the proof of Corollary~\ref{corollary:symbolwise_random_prior_predictor1} it is easy to see that assumptions on the channel apply only to Proposition~\ref{prop:symbolwise_scheme_channel_convergence2} showing the convergence of the average estimated channel $\fpr{W}_A$ to $\overline W$. To show
Proposition~\ref{prop:symbolwise_scheme_channel_convergence2} holds, we need to show that $\gamma_k$ remains a bounded martingale difference sequence, where now $\gamma_k$ is defined as:
\begin{equation}\begin{split}\label{eq:1645g}
\gamma_k(x,y)
& \defeq
\frac{1}{n} \Big[ \frac{\Ind(X_k = x, Y_k = y)}{\hat Q_{b_k}(x)}
\\ & \qquad - \Pr(Y_k = y | X_k = x, \vr X^{k-1}, \vr Y^{k-1}) \Big]
.
\end{split}\end{equation}
As in \eqref{eq:1544}, we have $\fpr{W}_A - \overline{W} =  \sum_{k=1}^n \gamma_k(x,y)$. Equation \eqref{eq:1564} now becomes
\begin{equation}\begin{split}\label{eq:1564}
\E \left[ \gamma_k \Big| S_{k-1} \right]
& =
\frac{\Pr(X_k = x, Y_k = y | S_{k-1})}{n \cdot \hat Q_{b_k}(x)}
\\ & \qquad - \frac{1}{n} \Pr(Y_k = y | X_k = x, \vr X^{k-1}, \vr Y^{k-1})
\\& =
\frac{\hat Q_{b_k}(x) \cdot \Pr(Y_k = y | X_k = x, \vr X^{k-1}, \vr Y^{k-1})}{n \cdot \hat Q_{b_k}(x)}
\\ & \qquad - \frac{1}{n} \Pr(Y_k = y | X_k = x, \vr X^{k-1}, \vr Y^{k-1})
\\ & =
0
.
\end{split}\end{equation}
The rest of the proof of Proposition~\ref{prop:symbolwise_scheme_channel_convergence2} remains the same. \endofproof

\section{Discussion and comments}\label{sec:discussion}
In this section we discuss the relation of the current results to existing results pertaining to unknown channels and make some comments on schemes presented here.

\subsection{A comparison with AVC capacity}\label{sec:discussion_AVC}
It is interesting to compare the target rate $C(\overline W)$ with the AVC capacity. We will give a short background on the AVC and the relation to the current problem.

In the traditional AVC setting \cite{Lapidoth_AVC}, the channel model is similar to the setting assumed here, but slightly more constrained. The channel in each time instance is assumed to be chosen arbitrarily out of a set of channels, each of which is determined by a state. Frequently, constrains on the state sequence (such as maximum power, number of errors) are defined. The AVC capacity is the maximum rate that can be transmitted reliably, for every sequence of states that obeys the constraints.

The AVC capacity may be different depending on whether the maximum or the average error probability over messages is required to tend to zero with block length, on the existence of feedback, and on whether common randomness is allowed, i.e. whether the transmitter and the receiver have access to a shared random variable. The last factor has a crucial effect on the achievable rate as well as on the complexity of the underlying mathematical problem: the characterization of AVC capacity with randomized codes is relatively simple and independent on whether maximum or average error probability is considered, while the characterization of AVC capacity for deterministic codes is, in general, still an open problem. Randomization has a crucial role, since we consider the worst-case sequence of channels. This sequence of channels is chosen after the deterministic code was selected (and therefore sometimes viewed as an adversary), enabling the worst-case sequence of channels to exploit vulnerabilities that exist in the specific code. As an example, for every symmetrizable AVC \cite[Definition 2]{Csiszar_AVC}, the AVC capacity for deterministic codes is zero \cite[Theorem 1]{Csiszar_AVC}.
When randomization does exist, the random seed is selected ``after'' the channel sequence was selected (mathematically, the probability over random seeds is taken after the maximum error probability over all possible sequences), and therefore prevents tuning the channel to the worst-case code.
When randomization exists, the channel inputs may be made to appear independent from the point of view of the adversary, thus limiting effective adversary strategies. Therefore the results in the current paper assume common randomness exists.

We would now like to compare the target rate $C(\overline W)$ with the randomized AVC capacity. The discrete memoryless AVC capacity without constraints may be characterized as follows: let $\mathcal{W}$ be the set of possible channels that are realized by different channel states (for example in a binary modulo-additive channel with an unknown noise sequence, there are two channels in the set -- one in which $y=x$ and another in which $y=1-x$). This set is traditionally assumed to be finite, i.e. there is a finite number of ``states'', however this constraint is immaterial for the comparison. The randomized code capacity of the AVC is \cite[Theorem 2]{Lapidoth_AVC}:
\begin{equation}\begin{split}\label{eq:1222}
\Cavc
&=
\max_Q \min_{W \in \conv(\mathcal{W})} I(Q,W)
\\&=
\min_{W \in \conv(\mathcal{W})} \max_Q  I(Q,W)
=
\min_{W \in \conv(\mathcal{W})} C(W)
,
\end{split}\end{equation}
where $\conv(\mathcal{W})$ is the convex hull of $\mathcal{W}$, which represents all channels which are realizable by a random drawing of channels from $\mathcal{W}$.\footnote{The convex hull replaces the distribution $\zeta(s)$ over channel states in \cite{Lapidoth_AVC}.} In the example, $\conv(\mathcal{W})$ would be the set of all binary symmetric channels. When input or state constraints exist, they affect \eqref{eq:1222} simply by including in the set of $Q$-s and in $\conv (\mathcal{W})$ only those priors, or channels, that satisfy the constraints (respectively). The converse of \eqref{eq:1222} is obtained by choosing the worst-case channel $W^* = \argmin{W \in \conv(\mathcal{W})} C(W)$ and implementing a discrete memoryless channel (DMC) where the channel law is $W^*$, by a random selection of channels from $\mathcal{W}$. Hence it is clear that the randomized code capacity cannot be improved by feedback. In contrast, the deterministic code AVC capacity can be improved by feedback, and in some cases made to equal to the randomized code capacity \cite{Berlkamp}\cite{Ahlswede73_AVCFB}\cite{AhlswedeCai00_AVCFB}. Therefore, most existing works on feedback in AVC deal with the deterministic case.

Since by definition $\overline W \in \conv(\mathcal{W})$, we have from \eqref{eq:1222}, $C(\overline W) \geq \Cavc$, i.e. our target rate meets or exceeds the AVC capacity. While in the traditional setting, a-priori knowledge of $\mathcal{W}$ or state constraints on the channel is necessary in order to obtain a positive rate, here we attain a rate possibly higher than the AVC capacity, without prior knowledge of $\mathcal{W}$. This is important since without such constraints, i.e. when the channel sequence is completely arbitrary, the AVC capacity is zero. This property makes the system presented here universal, with respect to the AVC parameters, a universality which also holds in an online-adversary setting (Corollary~\ref{corollary:symbolwise_random_prior_predictor1}).

We can view the difference between $\Cavc$ \eqref{eq:1222} and $C(\overline W)$ as the difference between the capacities of the worst realizable channel $W^* \in \conv(\mathcal{W})$, and the specific channel $\overline W \in \conv(\mathcal{W})$ representing the average of the sequence of channels that actually occurred. This difference is obtained by adapting the communication rate to the capacity of the average channel, and adapting the input prior to the prior that achieves this capacity, whereas in the AVC setting, the rate and the prior are determined a-priori, based on the worst-case realizable channel.

As we noted above, feedback cannot improve the randomized AVC capacity. Therefore the improvement is attained not merely by the use of feedback, but by allowing the communication rate to vary, whereas in the traditional AVC setting, one looks for a fixed rate of communication which can be guaranteed a-priori (note that the improvement is not in the worst case). In allowing the rate to vary, we have lost the formal notion of capacity (as the supremum of achievable rates), thereby making the question of setting the target rate more ambiguous, but nevertheless improved the achieved rates.

\subsection{Relation to empirical capacity and mutual information}
The capacity of the averaged channel $C(\overline W)$ is a slight generalization of the notion of \emph{empirical capacity} defined by Eswaran~\etal~\cite[\S D]{Eswaran}. The only difference is releasing the assumption made there, that the set of channel states is finite. The empirical capacity of Eswaran is in itself a generalization of the empirical capacity for modulo additive channels defined by Shayevitz and Feder~\cite{Ofer_ModuloAdditive}. Eswaran~\etal~\cite{Eswaran} assume the prior $Q$ is given a-priori and attain the empirical mutual information $I(Q,\overline W)$. The scheme used here is similar to the scheme they presented in its high level structure. We can view the current result (Theorem~\ref{theorem:C_overlineW_achievability}) as an improvement over the previous work, i.e. attaining the capacity $C(\overline W) \geq I(Q,\overline W)$, rather than the mutual information, by the addition of the universal predictor. Our result answers the question raised there \cite[\S D]{Eswaran}, whether the empirical capacity is attainable.

Another small extension is in Corollary~\ref{corollary:symbolwise_random_prior_predictor1}, showing that the result holds in an adversarial setting. This extension is outside our main focus of communicating over unknown channels, and is only used to strengthen the claim on universality with respect to the AVC parameters.

The main result (Theorem~\ref{theorem:C_overlineW_achievability}) could be derived in a conceptually simpler but crude scheme, by combining the results of Eswaran \cite{Eswaran} or our previous paper \cite{YL_individual_full} with Theorem~\ref{theorem:prior_predictor_exp}. The transmission time $n$ may be divided into multiple fixed-size blocks $i=1,\ldots,N$, and in each block, one of these schemes is operated, with an i.i.d. prior chosen by a predictor. Using Eswaran's result, for example, and ignoring some details such as finite-state assumptions, one would obtain the rate $I(\hat Q_i,\overline W_i)$ over each block, where $\overline W_i$ is the averaged channel over the block. The channel $\overline W_i$ can be well estimated (e.g. using training symbols or using the communication scheme itself). Assuming it is known, if the prediction scheme of Theorem~\ref{theorem:prior_predictor_exp} is operated over $\overline W_i$ it will guarantee the average rate over the $N$ blocks will be asymptotically at least $\frac{1}{N} \sum_{i=1}^N I(Q,\overline W_i)$ for any $Q$, and using convexity, $\frac{1}{N} \sum_{i=1}^N I(Q,\overline W_i) \geq I \left(Q,\frac{1}{N} \sum_{i=1}^N  \overline W_i \right) = I(Q,\overline W)$. Since this holds for any $Q$ this achieves the capacity of the average channel. Note that here it appears that there is no need for the uniform prior, however this is somewhat hidden in the assumption that the channel is known. Furthermore there is no need to worry about rateless blocks extending ``forever'' since the commnication scheme is re-started on each of the $N$ blocks.

\subsection{Competitive universality}
In a related paper \cite{YL_UnivModuloAdditive} we presented the concept of the iterated finite block capacity $C_\tsubs{IFB}$ of an infinite vector channel, which is similar in spirit to the finite state compressibility defined by Lempel and Ziv \cite{LZ78}. Roughly speaking, this value is the maximum rate that can be reliably attained by any block encoder and decoder, constrained to apply the same encoding and decoding rules over sub-blocks of finite length. The positive result is that $C_\tsubs{IFB}$ is universally attainable for all modulo-additive channels (i.e. over all noise sequences). The result is obtained by a system similar to the one described in Section~\ref{sec:arbitrary_var_rateless_scheme}, while the input prior is fixed to the uniform prior. The result uses two key properties of the modulo additive channel:
\begin{enumerate}
\item The channel is memoryless with respect to the input $x_i$ (i.e. current behavior is not affected by previous values of the input).
\item The capacity achieving prior is fixed for any noise sequence.
\end{enumerate}

The current work is a step toward removing the second assumption. The capacity of the averaged channel is a bound on the rate that can be obtained reliably by a transmitter and a receiver operating on a single symbol, since the channel that this system ``sees'' can be modeled as a random uniform selection of a channel out of $\{W_i\}_{i=1}^n$, which we term the ``collapsed channel'' \cite{YL_UnivModuloAdditive}. By combining $k$ symbols into a single super-symbol, we can extend the result and obtain a rate which is equal to or better from the rate obtained by block encoder and decoder operating over chunks of $k$ symbols. Therefore the current result suggests that it is possible to attain $C_\tsubs{IFB}$ for all vector channels that are memoryless in the input, i.e. that have the form defined in \eqref{eq:687}, for an arbitrary sequence of channels $W_i$ (compared to only an arbitrary noise sequence, in the previous result).

\subsection{Notes on the converse}
It is interesting to consider the converse (Theorem~\ref{theorem:C_overlineW_optimality}) from the following point of view:
Suppose a competitor is given the entire sequence of channels $W_1^n$, but is allowed to take from this sequence only the ``histogram'' (a list of channels and how many times they occurred), and devise a communication system based on this information. The rate that can be guaranteed in this case is limited by $C(\overline W)$. On the other hand, assuming common randomness exists, it is enough to know $\overline W$ in order to attain $C(\overline W)$ without feedback. To see this intuitively, we may apply a random interleaver and use the fact the interleaved channel is similar to a DMC with the channel law $\overline W$. Therefore even if one knows the entire histogram of the sequence, the average channel $\overline W$, which contains less information, contains all information necessary for communication.

To illustrate this, consider the deterministic setting, where instead of a sequence of channel laws $W_i(y|x)$ we have a sequence of deterministic functions $f_i:\mathcal{X} \to \mathcal{Y}$. This is a particular case of our problem, with $W_i(y|x) = \Ind(y = f_i(x))$. Even in this case, according to Theorem~\ref{theorem:C_overlineW_optimality}, a competitor knowing the list of functions up to order, will not be able to guarantee a rate better than $C(\overline W)$, where $\overline W = \frac{1}{n} \sum_{i=1}^n \Ind(y = f_i(x))$, i.e. a channel created by counting for each $x$, the normalized number of times a certain $y$ would appear as output.

Comparing the amount of information in the channel histogram and the averaged channel in this case, there are $|\mathcal{Y}|^{|\mathcal{X}|}$ functions, and therefore the distribution is given by $|\mathcal{Y}|^{|\mathcal{X}|}-1$ real numbers. On the other hand, the average channel is a probability distribution from $|\mathcal{X}|$ to $|\mathcal{Y}|$ and is specified by $(|\mathcal{Y}|-1)\cdot |\mathcal{X}|$ real numbers.

An interesting property revealed through the example, is that although the setting is deterministic, the result is given in terms of probability functions. These ``probabilities'' are only averages related to the deterministic function sequence, but this shows that the formulation via probabilities (or frequencies) is more natural than by specifying the function $f_i$ between the input and output.
\onlyphd{Explain more: Another interesting observation: one could consider a deterministic setting of the problem, where the channel is not probabilistic. Although the deterministic setting is more constrained, we may argue that for an unknown channel it is not trivial to assume a probability exists. }

\subsection{The required feedback rate}
We assumed the feedback channel has unlimited rate, and is free of delays and errors. This was done mainly to focus the discussion and simplify the results. It is clear from the scheme presented, that because the amount of information required to be fed back to the transmitter can be made small, the capacity of the average channel could be attained even if the feedback link has any small positive rate and a fixed delay. If the feedback channel is such that errors can be mitigated by coding with finite delay, then errors can be accommodated as well. Specifically, we show in Appendix~\ref{sec:zero_feedback_rate} that when the feedback rate is limited, or there is a fixed delay, the penalty is a gap of at most $O(\log n)$ symbols between the blocks, and that the normalized loss from this effect tends to zero. Therefore we have $\Delta_C \ntoinfty 0$ (with the notation of Theorem~\ref{theorem:C_overlineW_achievability}), with any positive feedback rate and any fixed delay. The gap may be reduced by using the time of the $i$-th block to transfer the channel information from block $i-1$ and use it only in block $i+1$ (i.e. insert a delay of one block in the prediction scheme), however this approach is not analyzed here.

\subsection{Convergence rate}
Throughout the course of this paper, as we have gradually made our assumptions more realistic, we have seen a deterioration of the rate of convergence, of the achieved rate to the target rate. We denote by $\delta_n$ the gap between the guaranteed rate and the target rate, and focus on the dominant polynomial power $p = - \lim_{n \to \infty} \frac{\ln \delta_n}{n}$, while ignoring the $\ln n$ terms. We have $p = \half$ in the synthetic problem of Section~\ref{sec:toy_problem} (assuming ``block-wise'' variation) \S\ref{sec:toy_problem}, $p = \tfrac{1}{3}$ when using the rateless scheme under assumptions of perfect average channel knowledge \S\ref{sec:arbitrary_channel_var}, and $p=\tfrac{1}{4}$ when releasing the abstract assumptions \S\ref{sec:mainproof}. The first deterioration (between $\half$ and $\tfrac{1}{3}$) is mainly attributed to the rateless coding scheme. More specifically, it stems from mixing with the uniform prior, which is necessary to bound the regret per block when the blocks have variable lengths. The second deterioration (between $\tfrac{1}{3}$ and $\tfrac{1}{4}$) can be attributed mainly to the fact that the number of bits per block $K$ has to increase in a certain rate in order to balance overheads created by the universal decoding procedure (and reduces the rate of adaptation). While the rate of convergence which was achieved deteriorates, the only upper bound we presented on the convergence rate is $p \leq \half$ (\S\ref{sec:regret_LB}), which is tight only for the first case. We do not know whether better convergence rates can be attained in Theorems~\ref{lemma:C_overlineW_achievability_w_side_info},\ref{theorem:C_overlineW_achievability}.

\begin{table*}
  \centering
\begin{tabular}{|p{2cm}|p{2.5cm}|p{3.5cm}|p{2.5cm}|p{5cm}|}
\hline
  &  Synthetic problem (``Block-wise variation'') & Arbitrarily varying channel, with side information on average channel and without communication overheads  & Arbitrarily varying channel & Notes \\ \hline
  Reference &  \S\ref{sec:toy_problem}, Theorem~\ref{theorem:prior_predictor_exp} & \S\ref{sec:arbitrary_var_exp_prior_predictor}, Lemma~\ref{lemma:C_overlineW_achievability_w_side_info} & \S\ref{sec:problem_setting},\S\ref{sec:mainproof}, Theorem~\ref{theorem:C_overlineW_achievability} & \\ \hline
  $C_1$ Attainability & No & No & No & $C_1$ = Capacity of $\{W_i\}_{1}^n$ = Mean capacity $\frac{1}{n} \sum_i C(W_i)$ \\ \hline
  $C_2$ Attainability & Yes & No & No & $C_2$ = Mean mutual information with fixed prior $\max_Q \frac{1}{n} \sum_i I(Q,W_i)$ \\  \hline
  $C_3$ Attainability & Yes & Yes & Yes & $C_3$ = Capacity of the time-averaged channel $C(\overline W) = C \left( \frac{1}{n} \sum_i W_i \right)$
\begin{enumerate}
\item Best attainable rate not using time structure (Theorem~\ref{theorem:C_overlineW_optimality}).
\item $C_3 \geq \Cavc$ (Section~\ref{sec:discussion_AVC})
\end{enumerate} \\ \hline
  Normalized regret lower bound & $O \left(\frac{1}{n} \right)^{\half}$ & $O \left(\frac{1}{n} \right)^{\half}$ & $O \left(\frac{1}{n} \right)^{\half}$ & \\ \hline
  Normalized regret attained & $O \left(\frac{\ln n}{n} \right)^{\half}$ & $O \left(\frac{\ln n}{n} \right)^{\tfrac{1}{3}}$ & $O \left(\frac{\ln^2 n}{n} \right)^{\tfrac{1}{4}}$ &  \\ \hline
 \end{tabular}  \caption{Summary of the results}\label{tbl:prior_pred_summary}
\end{table*}

\subsection{Comments on the prediction scheme}\label{sec:discussion_prediction_scheme}
The results in this paper were obtained by exponential weighting. This scheme was selected mainly due to its simplicity and elegance. Unfortunately, the exponential weighting is performed over a continuous domain (of probabilities), and therefore it is not immediately implementable. Of course, the simplest practical solution could be discrete sampling of the unit simplex and replacement of the integrals by sums. Since the mutual information is continuous, it is possible to bound the error resulting from this discretization. An alternative way is to quantize the set of priors. Instead of competing against a continuum of reference schemes, we first reduce the number of reference schemes to a finite one, by creating a ``codebook'' of priors $\{Q_m\}$. This codebook is designed so that the penalty in the mutual information resulting from rounding to the nearest codeword, is small. This quantization is useful in terms of the feedback link, which now only has to convey the index $m$. Having quantized the priors, we may replace the predictors shown here by standard schemes used for competition against a finite set of references \cite[\S 2]{Nicolo},\cite{Vovk97}. See a rough analysis of this approach in Appendix~\ref{sec:prior_quantization_analysis}. An alternative approach is to bypass the explicit calculation of the predictor $\hat Q_i$ and use a rejection-sampling based algorithm to generate a random variable $X \sim \hat Q_i$. This approach is demonstrated in Appendix~\ref{sec:rejection_sampling_predictor}.

Zinkevich \cite{Zinkevich03} proposed a computationally efficient online algorithm, based on gradient descent, to solve a problem of minimizing the sum of convex functions, each revealed to the forecaster after the decision was made (a similar setting to that of Lemma~\ref{lemma:rateless_prior_predictor_lemma}). To apply Zinkevich's results to our problem, some modifications are required. The mutual information does not have a bounded gradient (which is required by \cite{Zinkevich03}), but this could be bypassed by keeping away from the boundary of $\Delta_{\mathcal{X}}$, i.e. from these points for which one of the elements of $Q$ is $0$ or $1$. One way to accomplish this is by mixing with the uniform prior when defining the target rate, and use $\max_Q \sum_i I((1-\lambda)Q + \lambda U, W_i)$ as a target, and then bounding the loss induced by this mixture. In the rateless scheme, a bound on the maximum value of $m_i F_i(Q)$ (of Lemma~\ref{lemma:rateless_prior_predictor_lemma}) is required and can be obtained using the same methods presented here.

Another application of sequential algorithms to solve problems related to AVC's was proposed by Buchbinder~\etal~\cite{Buchbinder09} who used a sequential algorithm to solve a problem of dynamic transmit power allocation, where the current channel state is known but future states are arbitrary.

\subsection{The combination of the communication scheme with the predictor}
In the communication scheme proposed in Section~\ref{sec:arbitrary_var_rateless_scheme} we chose to use an i.i.d. prior during each block, and update the prior only at the end of the block. This choice is motivated by the following considerations:
\begin{itemize}
\item Assuming no explicit training symbols are transmitted, the estimation of the channel $\overline{W}$ is done based on the encoded sequence, which is known to the receiver only after decoding (at the end of the block).
\item Varying the prior throughout the block inserts memory into the channel input, which complicates the analysis.
\end{itemize}
The result of this is a relatively slow update of the prior, essentially limited by the block size, which is determined based on communication related considerations (overheads and error probabilities). An alterative would be learning the channel through random training symbols (see for example \cite{Ofer_ModuloAdditive}), and updating the prior from time to time, without relation to the rateless blocks.

\subsection{The behavior of the regret for binary channels}
In Section~\ref{sec:regret_LB} we have shown a lower bound on the redundancy in attaining $C_2$ by using a counter example with $|\mathcal{X}|=4, |\mathcal{Y}|=2$. It is worth mentioning that for the set of binary channels $|\mathcal{X}|=|\mathcal{Y}|=2$, the normalized regret is not necessarily $O \left(\sqrt{\frac{1}{n}} \right)$. For this set of channels, the optimal prior does not reach the boundaries of $[0,1]$: the two input probabilities $\Pr(X=x)$ are always in $[e^{-1}, 1-e^{-1}]$ \cite{Shulman_Prior}. It is possible to show that the loss function $l(Q,W) = 1 - \frac{I(Q,W)}{I_{\max}}$ satisfies conditions 1,2,4 in Cesa-Bianchi and Lugosi's book \cite{Nicolo} Theorem 3.1 (but not condition 3). This fact together with experimental results showing convergence of the FL predictor, suggests that the normalized minimax regret in this case may converge like $O \left(\frac{\log n}{n} \right)$.

\subsection{The uniform component in prior predictor}
In the prediction scheme of Theorems~\ref{lemma:C_overlineW_achievability_w_side_info},\ref{theorem:C_overlineW_achievability}, we mixed a uniform prior with an exponentially weighted predictor \eqref{eq:Qi3382}. This mixing has two advantages:
\begin{enumerate}
\item Enabling to bound the instantaneous regret caused by a large block due to a low mutual information
\item Enabling channel estimation by making sure all input symbols have a non zero probability.
\end{enumerate}
Note that alternative solutions are use of training symbols at random locations and termination and re-transmission of blocks whose length exceeds a threshold.

Mixing the exponentially weighted predictor with a uniform distribution is a technique used in prediction problems with partial monitoring, where the predictor only has access to its own loss (or a function of it) and not to the loss of the competitors \cite[\S 6]{Nicolo}, and effectively assigns some time instances for sampling the range of strategies. In our problem the uniform prior plays two roles. One, is related to the rateless communication scheme, which required to relate the gains of the predictor to the gain of any alternative prior $Q$ \eqref{eq:3402} in order to have an upper bound on the latter \eqref{eq:850n}. The second role is in the convergence of the estimated channel (Proposition~\ref{prop:symbolwise_scheme_channel_convergence2}). The second role is similar to the role of uniform distribution in partial monitoring problems: the channel $W(y|x)$ cannot be estimated for input values $x$ that occur with zero probability.

Note that even without the explicit uniform component $\lambda U$, the exponential weighting element $\int w_i(Q) Q d Q$ in \eqref{eq:Qi3382} includes a small uniform component. Particularly, since referring to \eqref{eq:8385}, $1 \leq e^{\eta \sum_{j=1}^{i-1} m_j \cdot I(Q, \overline W_j)} \leq e^{\eta n I_{\max}}$, $w_i(Q) \geq \frac{1}{\vol(\Delta_{\mathcal{X}})} e^{-\eta n I_{\max}}$ and
\begin{equation}\begin{split}\label{eq:722}
\int_{\Delta_{\mathcal{X}}} w_i(Q) Q d Q
& \geq
e^{-\eta n I_{\max}} \underbrace{\frac{1}{\vol(\Delta_{\mathcal{X}})}  \int_{\Delta_{\mathcal{X}}} Q d Q}_{U}
\\& =
e^{-\eta n I_{\max}} \cdot U
.
\end{split}\end{equation}
However this value is too small for our purpose.

\subsection{Continuous channels}\label{sec:continuous_channels}
In the current paper we assumed the input and output alphabets are finite. In general it is not possible to universally attain $C_2$ or $C_3$, even in the context of the synthetic problem of Section~\ref{sec:toy_problem}, when the alphabet size $\mathcal{X}$ is infinite. This is since in the continuous case one is trying to assign a probability $Q$ to an infinite set of values, where the values producing the capacity may be a small subgroup. Consider the following example:
\begin{example}
Let the channel $W_{a}$, with input $x$ and output $y$ ($x,y \in \mathbb{R}$) be defined by the arbitrary sequence $\{a_k\}_1^{\infty}$, $a_k \in \mathbb{R}$, with all $a_i \neq a_k (i \neq k)$. The channel rule is defined by:
\begin{equation}\label{eq:703}
y = \left\{ \begin{array}{cc} k & x = a_k \\ 0 & o.w.  \end{array} \right.
.
\end{equation}
For any sequential predictor (even randomized) we can find a sequence of channels $\{W_a\}$ such that the values of the sequence $\{a_i\}$ at each step have total probability zero (since the input distribution may have at most a countable group of discrete values with non zero probability). Therefore we can always find a sequence of channels where the rate obtained by the predictor would be zero. On the other hand, each channel $W_a$ has infinite capacity (since it can transmit noiselessly any integer number). Therefore the value of $C_2$ is infinite (it is enough to choose a prior suitable for one of the channels in the sum \eqref{eq:189}).
\end{example}

It stands to reason that under suitable continuity conditions on $W(y|x)$ and input constraints on $Q(x)$, we may convert the problem to a discrete one, while bounding the loss in this conversion, by discretization of the input -- i.e. by selecting the input from a finite grid, or alternatively assuming a parametrization of the channel.

\section{Conclusion}\label{sec:conclusion}
We considered the problem of adapting an input prior for communication over an unknown and arbitrarily varying channel, comprised of an arbitrary sequence of memoryless channels, using feedback from the receiver. We showed that it is possible to asymptotically approach the capacity of the time-averaged channel universally for every sequence of channels. This  rate equals or exceeds the randomized AVC capacity of any memoryless channel with the same inputs, and thus the system is universal with respect to the AVC model. The result holds also when the channel sequence is determined adversatively. We also presented negative results showing which communication rates or minimax regret convergence rates cannot be attained universally (see a summary in Table~\ref{tbl:prior_pred_summary}), and presented a simplified synthetic problem relating to prediction of the communication prior, which may have applications for block-fading channels.

When examining the role of feedback in combating unknown channel, previous works mainly focused on the gains of rate adaptation, and here we have seen an additional aspect, namely selection of the communication prior, in which feedback improves the communication rate. The results have implications on competitive universality in communication, and suggest that with feedback, it would be possible for any memoryless AVC, to universally achieve a rate comparable to that of any finite block system, without knowing the channel sequence.

When comparing the results to the traditional AVC results, the former setting was prevailed by the notion of capacity, and thus, even when feedback was assumed, it was not used for adapting the communication rate. Here we have shown for the first time, that rates equal to or better from the AVC capacity can be attained universally, when releasing the constraint of an a-priori guaranteed rate. This demonstrates the validity of the alternative ``opportunistic'' problem setting that has been considered in the last decade for feedback communication over unknown channels, a setting which does not focus on capacity.

\section*{Acknowledgment}
We thank Yishay Mansour for helpful discussions on the universal prediction problem.

\appendix{}
\subsection{Proof of Lemma~\ref{lemma:F_exp_weight_UB}}\label{sec:proof_of_lemma1}
Lemma~\ref{lemma:F_exp_weight_UB} relates the exponential weighting of a bounded and concave real function $a \leq F(\vr x) \leq b$ over a convex vector region $\vr x \in S \subset \mathbb{R}^d$ to its maximum.

\textit{Proof:}
Let $\vr x^*$ denote a global maximum of $F(\vr x)$ in $S$ (which exists since $F$ is concave and $S$ is closed). Then from the concavity of $F$ for any $\lambda \in [0,1]$ we have:
\begin{equation}\label{eq:2600}
F(\lambda \vr x + (1-\lambda) \vr x^*) \geq \lambda F(\vr x) + (1-\lambda) F(\vr x^*) \geq \lambda a + (1-\lambda) F(\vr x^*)
.
\end{equation}
Note that the RHS is a constant. Denote $S_{\lambda} \defeq \{\lambda \vr x + (1-\lambda) \vr x^* : \vr x \in S\} = \lambda S + (1-\lambda) \vr x^*$. Then due to convexity $S_{\lambda} \subset S$ and due to the shrinkage $\vol(S_{\lambda}) = \lambda^d \vol(S)$. Furthermore by (\ref{eq:2600}), $\forall \vr x \in S_{\lambda}: F(\vr x) \geq \lambda a + (1-\lambda) F(\vr x^*)$. We have:
\begin{equation}\begin{split}\label{eq:977}
\int_S e^{\eta F(\vr x)} d \vr x
& \geq
\int_{S_{\lambda}} e^{\eta F(\vr x)} d \vr x
=
\int_{S_{\lambda}} e^{\eta (\lambda a + (1-\lambda) F(\vr x^*))} d \vr x
\\ & =
e^{\eta (\lambda a + (1-\lambda) F(\vr x^*))} \vol(S_{\lambda})
\\ & =
e^{\eta F(\vr x^*)} \cdot e^{- \eta \lambda (F(\vr x^*) - a)} \lambda^d \vol(S)
\\ & \geq
e^{\eta F(\vr x^*)} \cdot e^{- \eta \lambda (b - a)} \lambda^d \vol(S)
.
\end{split}\end{equation}
Therefore,
\begin{equation}
\overline{F}
\defeq
\frac{1}{\eta} \ln \left[ \frac{1}{\vol(S)} \int_S e^{\eta F(\vr x)} d \vr x \right]
\geq
F(\vr x^*) -\lambda (b-a) + \frac{d \ln \lambda}{\eta}
.
\end{equation}
Maximizing the RHS with respect to $\lambda$ we obtain
\begin{equation}\label{eq:2623}
\lambda = \frac{d}{\eta (b-a)}
,
\end{equation}
where $\lambda \leq 1$ by the assumptions of the lemma, and substituting $\lambda$ we have:
\begin{equation}
\overline{F}
\geq
F(\vr x^*) - \frac{d}{\eta} \left( 1 + \ln \frac{\eta  (b-a)}{d} \right)
=
F(\vr x^*) - \frac{d}{\eta} \ln \frac{\eta  e  (b-a)}{d}
.
\end{equation}
Rearranging yields the desired result.
\endofproof

\subsection{Proof of Lemma~\ref{lemma:rateless_prior_predictor_lemma}}\label{sec:proof_rateless_prior_predictor_lemma}
During the course of the derivation below we attempt to optimize the asymptotical form of the loss (up to constant factors), and thus we make simplifying assumptions on the parameters, which hold asymptotically for large enough $n$. For finite $n$ these assumptions might lead to suboptimal results. We do not discuss the assumptions during the course of the derivation and we collect them at the end. All integrals below are by default over the unit simplex $Q \in \Delta_{\mathcal{X}}$.

In the block-wise variation setting (Section~\ref{sec:toy_problem}), our target was to control the growth rate of the regret. Here, at each block $i$, by \eqref{eq:683t} the gain of the competitor using prior $Q$ is $m_i F_i(Q)$ (bits), while the universal scheme sends a fixed number of bits $K$. Therefore the gain of the competitor $m_i F_i(Q)$ and the instantaneous regret $m_i F_i(Q) - K$ are related by a constant, and it is more convenient to base the derivation on the gain rather than the regret. The potential function $\Phi$ will be used as an approximation of the $\max$ in \eqref{eq:683t}.

Denote the cumulative gain of the competitor with prior $Q$ as:
\begin{equation}\label{eq:879ph}
G_i(Q) \defeq \sum_{j=1}^i m_j F_j(Q)
,
\end{equation}
And the potential function of $G_i(Q)$ as:
\begin{equation}\label{eq:868}
\Phi_i \defeq \Phi(G_i(Q))
.
\end{equation}
Note that $\Phi_i$ is not a function of $Q$ due to the integration over $Q$ performed by $\Phi(\cdot)$. We can now write $w_i(Q)$ as:
\begin{equation}\label{eq:841}
w_i(Q) = \frac{e^{\eta G_{i-1}(Q)}}{\Phi_{i-1}}
.
\end{equation}
The growth of the potential is bounded by:
\begin{equation}\begin{split}\label{eq:3410}
\Phi_i
& =
\int e^{\eta G_i(Q)} d Q
\\& =
\int e^{\eta G_{i-1}(Q)} e^{\eta m_i F_i(Q)} d Q
\\ & \stackrel{\eqref{eq:841}}{=}
\int \Phi_{i-1} w_i(Q) e^{\eta m_i F_i(Q)} d Q
\\ & \stackrel{\eqref{eq:ex483}}{\leq}
\Phi_{i-1} \int w_i \left[ 1 + \eta m_i F_i(Q) + \eta^2 m_i^2 F_i(Q)^2 \right] d Q
\\ & =
\Phi_{i-1} \left[ 1 + \eta \int w_i m_i F_{i} d Q + \eta^2 \int w_i m_i^2 F_{i}^2 d Q \right]
,
\end{split}\end{equation}
where in the last inequality we used Lemma~\ref{lemma:exp_second_order_bound} and assumed $\eta m_i F_i(Q) \leq 1$. The dependence of $F_i$ and $w_i$ on $Q$ is suppressed for brevity. We now bound the integrals $\int w_i m_i F_{i} d Q$ and $\int w_i m_i^2 F_{i}^2 d Q$. The property that a badly chosen prior may cause the iterative system to get stuck (not transmitting any block) translates into the fact that without placing any limitations on $\hat Q_i$, the competitor's gain, $m_i F_i(Q)$ may be unbounded, since $m_i$ might be indefinitely large while $F_i(Q)$ can be any positive value. This is prevented by mixing with the uniform prior, which enables us to link $F_i(Q)$ with $F_i(\hat Q_i)$. Since in the context of the lemma we do not assume $F_i(Q)$ is the mutual information, we use a bound which is slightly looser than Shulman and Feder's \eqref{eq:3397}, but is based on the same technique \cite{Shulman_Prior}, and only assumes concavity.

Define $x+z$ as modulo-addition over the set $\mathcal{X}$, and write $U(x) = \frac{1}{|\mathcal{X}|} \sum_{z \in \mathcal{X}} Q(x+z)$ for any $Q$, i.e. express the uniform prior as the mean of all cyclic rotations of $Q$. Using concavity and non-negativity of $F$:
\begin{equation}\begin{split}\label{eq:758}
F_i(U)
&=
F_i \left( \frac{1}{|\mathcal{X}|} \sum_{z \in \mathcal{X}} Q(x+z) \right)
\\ & \geq
\frac{1}{|\mathcal{X}|} \sum_{z \in \mathcal{X}} F_i \left(Q(x+z) \right)
\geq
\frac{F_i \left(Q \right)}{|\mathcal{X}|}
.
\end{split}\end{equation}
Because the prior \eqref{eq:Qi3382} has the structure $\hat Q_i = (1-\lambda) Q' + \lambda U$, by the concavity of $F_i$:
\begin{equation}\begin{split}\label{eq:3402}
\forall Q,i: F_i(\hat Q_i)
& \geq
(1-\lambda) F_i(\hat Q') + \lambda F_i(U)
\\ & \geq
\lambda F_i(U)
\stackrel{\eqref{eq:758}}{\geq}
\frac{\lambda}{|\mathcal{X}|} F_i(Q)    ,
\end{split}\end{equation}

Using \eqref{eq:3402} in conjunction with \eqref{eq:mi728} we have
\begin{equation}\begin{split}\label{eq:850n}
m_i F_i(Q)
& \stackrel{\eqref{eq:3402}}{\leq}
\frac{|\mathcal{X}|}{\lambda} m_i F_i(\hat Q_i)
\\& \stackrel{\eqref{eq:mi728}}{\leq}
\frac{|\mathcal{X}|}{\lambda} K
,
\end{split}\end{equation}
which yields a bound on the competitor gain in each block. We now bound the two integrals appearing in \eqref{eq:3410}. Starting with the first integral, using the concavity of $F_i$:
\begin{equation}\begin{split}\label{eq:832}
K
& \stackrel{\eqref{eq:mi728}}{\geq}
m_i F_i(\hat Q_i)
\stackrel{\eqref{eq:Qi3382}}{=}
m_i F_i \left((1-\lambda) \int  w_i(Q) Q d Q + \lambda U \right)
\\ & \geq
m_i (1-\lambda) \int w_i(Q) F_i(Q) d Q + \lambda m_i F_i(U)
\\ & \geq
m_i (1-\lambda) \int w_i(Q) F_i(Q) d Q
,
\end{split}\end{equation}
from which we obtain
\begin{equation}\label{eq:845}
\int w_i(Q) m_i F_i(Q) d Q  \leq \frac{K}{1-\lambda}    ,
\end{equation}

The second order term is bounded as follows:
\begin{equation}\begin{split}\label{eq:850b}
\int w_i(Q) m_i^2 F_i^2(Q) dQ
& =
\int w_i(Q) (m_i F_i(Q)) (m_i F_i(Q)) dQ
\\ & \stackrel{\eqref{eq:850n}}{\leq}
\frac{|\mathcal{X}|}{\lambda} K \cdot \int w_i(Q) m_i F_i(Q) dQ
\\ & \stackrel{\eqref{eq:845}}{\leq}
\frac{|\mathcal{X}|}{\lambda} K \cdot \frac{K}{1-\lambda}
.
\end{split}\end{equation}

Recall that in the classical weighted average predictor \cite{Nicolo}, the product of the instantaneous regret and the weighting function is guaranteed to be non positive (Blackwell condition). Similarly in the previous section we obtained $\int w(Q) r_i(Q) dQ \leq 0$ (see \eqref{eq:w467}). In the present case, if we define $r_i(Q) = m_i F_i(Q) - K$, then by \eqref{eq:845} we have $\int w(Q) r_i(Q) dQ \leq \frac{K}{1-\lambda} - K =  \frac{K \cdot \lambda}{1-\lambda}$, i.e. due to the inclusion of the uniform prior (which is needed for $m_i F_i$ to be bounded), this integral may be positive, although arbitrarily small. Thus, we pay a price in the first order term in order to be able to bound the second order term.

Plugging the bounds \eqref{eq:845}, \eqref{eq:850b} into \eqref{eq:3410} we have:
\begin{equation}\begin{split}\label{eq:3412}
\Phi_i
& \stackrel{\eqref{eq:3410}}{\leq}
\Phi_{i-1} \left[ 1 + \eta \int w_i m_i F_{i} d Q + \eta^2 \int w_i m_i^2 F_{i}^2 d Q \right]
\\ & \stackrel{(\ref{eq:845}),(\ref{eq:850b})}{\leq}
\Phi_{i-1} \Bigg[ 1 + \eta \cdot \frac{K}{1-\lambda}
+ \eta^2 \frac{|\mathcal{X}|}{\lambda} K \cdot \frac{K}{1-\lambda} \Bigg]
\\ & \stackrel{\eqref{eq:ex483}}{\leq}
\Phi_{i-1} e^{\eta \cdot \frac{K}{1-\lambda} \left( 1 + \frac{ \eta \cdot  K \cdot |\mathcal{X}|}{\lambda}  \right)}
\\ & \leq \ldots \leq
\Phi_0 e^{\eta \cdot \frac{K \cdot i}{1-\lambda} \left( 1 + \frac{ \eta \cdot  K \cdot |\mathcal{X}|}{\lambda}  \right)}
.
\end{split}\end{equation}
In the last step we applied the same relation inductively. Using \eqref{eq:3412} we can obtain a bound on $\Phi(G_{B+1}(Q))$, and we now use Lemma~\ref{lemma:F_exp_weight_UB} to relate this bound to $G_{B+1}(Q)$ and to the target rate. $\Rtarget = \frac{1}{n} \max_Q G_{B+1}(Q)$. The dimension is $d=\dim(\Delta_{\mathcal{X}}) = |\mathcal{X}|-1$. By \eqref{eq:683t} we have
\begin{equation}\label{eq:580b}
\underbrace{0}_{\defeq a} \leq G_{B+1}(Q) \leq \underbrace{n \cdot \max(\Rtarget, I_{\max})}_{\defeq b}.
\end{equation}
The reason for setting the upper bound as $b = n \cdot \max(\Rtarget, I_{\max})$ rather than just $n \cdot \Rtarget$, is technical, as this simplifies the conditions required to meet the requirements of the lemma. To satisfy $\eta (b-a) \geq d$ we only require $\eta \geq \frac{|\mathcal{X}|-1}{n I_{\max}}$. By Lemma~\ref{lemma:F_exp_weight_UB} and \eqref{eq:3412} we have:
\begin{equation}\begin{split}\label{eq:3499}
G_{B+1}(Q)
& \stackrel{\eqref{eq:exp_weight_UB_2603}}{\leq}
\frac{1}{\eta} \ln \frac{\Phi(G_{B+1}(\tilde Q))}{\Phi(0)} + n \delta_1(\Rtarget)
\\ & =
\frac{1}{\eta} \ln \frac{\Phi_{B+1}}{\Phi_0} + n \delta_1(\Rtarget)
\\ & \stackrel{\eqref{eq:3412}} \leq
\frac{K \cdot (B+1)}{1-\lambda} \left( 1 + \frac{ \eta \cdot  K \cdot |\mathcal{X}|}{\lambda}  \right) + n \delta_1(\Rtarget)
,
\end{split}\end{equation}
where
\begin{equation}\label{eq:979}
\delta_1(\Rtarget) \defeq \frac{|\mathcal{X}|-1}{n \eta} \cdot \ln \left( \frac{\eta e n \max(\Rtarget, I_{\max})}{|\mathcal{X}|-1}  \right)
,
\end{equation}
is the redundancy term introduced by Lemma~\ref{lemma:F_exp_weight_UB}.
Bounding $\Rtarget$ using \eqref{eq:3499}, while substituting $K(B+1) = KB + K = nR + K$, we obtain:
\begin{equation}\begin{split}\label{eq:3499b}
\Rtarget
&=
\frac{1}{n} \max_Q G_{B+1}(Q)
\\ & \leq
\left( R + \frac{K}{n} \right) \frac{1}{1-\lambda} \left( 1 + \frac{ \eta \cdot  K \cdot |\mathcal{X}|}{\lambda}  \right) + \delta_1(\Rtarget)
.
\end{split}\end{equation}
After rearrangement we have the following bound on $R$:
\begin{equation}\label{eq:3499c}
R \geq \left( \Rtarget -  \delta_1(\Rtarget) \right) \cdot (1 - \delta_2)  - \delta_3
,
\end{equation}
where
\begin{eqnarray}
1 - \delta_2 & \defeq & \left( 1 + \frac{ \eta \cdot  K \cdot |\mathcal{X}|}{\lambda}  \right)^{-1} \cdot (1-\lambda)                               \\
\delta_3 & \defeq & \frac{K}{n}
.
\end{eqnarray}
The rest of the proof of Lemma~\ref{lemma:rateless_prior_predictor_lemma} is an algebraic derivation focused on simplifying and optimizing the bound above.
The lower bound on $R$ in the RHS of \eqref{eq:3499c} is increasing with respect to $\Rtarget$. This is since $\frac{\partial}{\partial \Rtarget} \delta_1$ is zero for $\Rtarget \leq I_{\max}$ and for $\Rtarget \geq I_{\max}$ the derivative $\frac{\partial}{\partial \Rtarget} \delta_1$ is $\frac{|\mathcal{X}|-1}{n \eta \Rtarget}$, which by our assumption $\eta \geq \frac{|\mathcal{X}|-1}{n I_{\max}}$ is smaller than $1$. Therefore $\frac{\partial}{\partial \Rtarget} (\Rtarget - \delta_1(\Rtarget)) \geq 0$. In order to optimize the parameters, we assume for now that $\Rtarget \leq I_{\max}$ and bound the difference $R - \Rtarget$. Using $\frac{1}{1 + t} \geq 1 - t$ we have
\begin{equation}\label{eq:1006}
1 - \delta_2 \geq \left( 1 - \frac{ \eta \cdot  K \cdot |\mathcal{X}|}{\lambda}  \right) \cdot (1-\lambda) \geq 1 - \frac{ \eta \cdot  K \cdot |\mathcal{X}|}{\lambda} - \lambda
.
\end{equation}
From \eqref{eq:3499c}, under the assumption $\Rtarget \leq I_{\max}$ we have:
\begin{equation}\begin{split}\label{eq:3499d}
R
& \geq
\left( \Rtarget -  \delta_1(I_{\max}) \right) \cdot (1 - \delta_2)  - \delta_3
\\ & \geq
\Rtarget -  \delta_1(I_{\max}) - \delta_2 \cdot I_{\max}  - \delta_3
\\& \geq
\Rtarget -  \delta_1(I_{\max}) - \frac{ \eta \cdot  K \cdot |\mathcal{X}|  \cdot I_{\max}}{\lambda} - \lambda \cdot I_{\max}  - \delta_3
,
\end{split}\end{equation}
We further simplify $\delta_1(I_{\max})$ by making the assumption that $\eta \leq \frac{|\mathcal{X}|-1}{e I_{\max}}$ and therefore $\ln \left( \frac{\eta e \max(\Rtarget, I_{\max})}{|\mathcal{X}|-1}  \right) \leq 0$, and $\delta_1(I_{\max}) \leq \frac{|\mathcal{X}|-1}{n \eta} \cdot \ln \left( n  \right)$. Using these simplifications we further bound the RHS of \eqref{eq:3499d} by $\Rtarget - \Dpred$ where
\begin{equation}\label{eq:939}
\Dpred = I_{\max} \cdot \lambda +  \underbrace{\frac{c_0}{\lambda}}_{\defeq a_1} \cdot \eta + \underbrace{(|\mathcal{X}|-1) \cdot \frac{\ln (n)}{n}}_{\defeq b_1} \cdot \frac{1}{\eta} + \delta_3
,
\end{equation}
and $c_0 = K \cdot |\mathcal{X}| \cdot I_{\max}$.

Applying Lemma~\ref{lemma:ab_alphabeta} to the optimization of the two terms depending on $\eta$ in \eqref{eq:939} (marked $a_1,b_1$, with powers $\alpha=1, \beta=1$) we have:
\begin{equation}\label{eq:962}
\eta^* = \sqrt{\frac{b_1}{a_1}} = \sqrt{\frac{|\mathcal{X}|-1}{c_0} \cdot \frac{\ln (n) \cdot \lambda}{n}}
,
\end{equation}
and
\begin{equation}\begin{split}\label{eq:966}
\Dpred \Big|_{\eta=\eta^*}
&=
I_{\max} \cdot \lambda + 2 \sqrt{a_1 b_1}   + \delta_3
\\& =
I_{\max} \cdot \lambda + 2 \sqrt{\frac{c_0 (|\mathcal{X}|-1) \cdot \ln (n)}{n \lambda}} + \delta_3
.
\end{split}\end{equation}
Substituting $c_0$ yields $\Dpred$ and $\eta$ stated in the Lemma. Now, the derivation involving equations \eqref{eq:3499d} -- \eqref{eq:966} assumes $\Rtarget \leq I_{\max}$. Since the lower bound \eqref{eq:3499c} on $R$ is increasing with respect to $\Rtarget$, in the case that $\Rtarget > I_{\max}$ we are guaranteed to obtain a better lower bound on $R$ than the lower bound $R \geq I_{\max} - \Dpred$ attained for $\Rtarget = I_{\max}$ (in other words, the RHS of \eqref{eq:3499c} for $\Rtarget = I_{\max}$ is at least $I_{\max} - \Dpred$). Therefore the bound can be stated as $R \geq \min(\Rtarget, I_{\max}) - \Dpred$.

We now collect the various assumptions we have made along the way. We use the same technique used in the proof of Theorem~\ref{theorem:prior_predictor_exp}, of showing that if the assumptions do not hold then (possibly under some simple conditions), $\Dpred \geq I_{\max}$ and therefore the lemma holds in a void way (since the RHS of \eqref{eq:688} becomes non-positive).

In \eqref{eq:3410} we assumed $\eta m_i F_i(Q) \leq 1$. Using the upper bound of \eqref{eq:850n} we have the sufficient condition $\eta \frac{|\mathcal{X}|}{\lambda} K \leq 1$. If this condition doesn't hold, i.e. $\eta \frac{|\mathcal{X}|}{\lambda} K > 1$, then the second term in \eqref{eq:939} satisfies $\frac{c_0}{\lambda} \eta = \frac{K \cdot |\mathcal{X}| \eta}{\lambda} \cdot I_{\max} >  I_{\max}$, so $\Dpred > I_{\max}$ and the lemma holds in a void way. Before \eqref{eq:939} we assumed $\eta \leq \frac{|\mathcal{X}|-1}{e I_{\max}}$. When the opposite is true, then second term in \eqref{eq:939} satisfies the $\frac{c_0}{\lambda} \eta = \frac{K \cdot |\mathcal{X}| \eta}{\lambda} \cdot I_{\max} > \frac{K \cdot |\mathcal{X}| (|\mathcal{X}|-1)}{e \cdot \lambda} > \frac{2}{e} \cdot K > \frac{K}{2}$. By requiring $K \geq 2 I_{\max}$ we have that in this case the Lemma will also be true in a void way. To use Lemma~\ref{lemma:F_exp_weight_UB} we required $\eta \geq \frac{d}{b-a} = \frac{|\mathcal{X}|-1}{n I_{\max}}$. If the opposite is true, then the third term in \eqref{eq:939} satisfies $\frac{(|\mathcal{X}|-1) \cdot \ln (n)}{\eta n} > I_{\max} \ln (n)$, and thus if $n \geq e$, $\Dpred > I_{\max}$. Therefore, by requiring $n > e$ and $K \geq 2 I_{\max}$, we have that if any of the assumptions we made does not hold, the lemma is true in a void way.
This concludes the proof of Lemma~\ref{lemma:rateless_prior_predictor_lemma}. \endofproof

\subsection{Proof of Lemma~\ref{lemma:fMI_properties}}\label{sec:proof_of_lemma_fMI_properties}
In this proof we use nats ($\log$-s are natural base). This does not change the results since all values scale according to the base of the $\log$-s. Also, we assume all probabilities and false probabilities are non-zero. It is easy to check that the results for zero probabilities follow by replacing zeros with small probabilities and taking the limit using $p \log p \arrowexpl{p \to 0} 0$.

\textbf{Non negativity}
Define $\fpr{p}(y) = \sum_x Q(x) \fpr{W}(y|x)$ and write:
\begin{equation}\begin{split}\label{eq:3398}
- \fpr{I}(Q,\fpr{W})
&=
\sum_{x,y} Q(x) \fpr{W}(y|x) \log \left( \frac{\fpr{p}(y)}{\fpr{W}(y|x)} \right)
\\&\stackrel{\log t \leq t-1}{\leq}
\sum_{x,y} Q(x) \fpr{W}(y|x) \left( \frac{\fpr{p}(y)}{\fpr{W}(y|x)} - 1 \right)
\\&=
\sum_{x} Q(x) \cdot \sum_y \fpr{p}(y) -  \sum_{x,y} Q(x) \fpr{W}(y|x)
\\&=
1-1
=
0
.
\end{split}\end{equation}

\textbf{Concavity with respect to $Q$}:
Denote as above $\fpr{p}(y) = \sum_x Q(x) \fpr{W}(y|x)$ and write:
\begin{equation}\begin{split}\label{eq:3411}
\fpr{I}(Q,\fpr{W})
&=
\sum_{x,y} Q(x) \fpr{W}(y|x) \log \frac{\fpr{W}(y|x)}{\fpr{p}(y)}
\\&=
\sum_{x,y} Q(x) \fpr{W}(y|x) \log \fpr{W}(y|x) -  \sum_{y} \fpr{p}(y) \log  \fpr{p}(y)
.
\end{split}\end{equation}
The left hand term is linear with respect to $Q$. The function $t \log t$ is convex in $t$ (for all $t \geq 0$), and $\fpr{p}(y)$ is linear in $Q$, therefore the right hand term is convex in $Q$, and so $\fpr{I}$ is concave with respect to $Q$.

\textbf{Convexity with respect to $\fpr{W}$}:
Let $\lambda_i \geq 0, \sum \lambda_i = 1$, and $\fpr{W}(y|x) = \sum_i \lambda_i \fpr{W}_i(y|x)$. We prove that $\Delta \defeq \fpr{I}(Q, \fpr{W}) - \sum_i \lambda_i \fpr{I}(Q,\fpr{W}_i) \leq 0$. Define the respective output distributions as $\fpr{p}_i(y) = \sum_x Q(x) \fpr{W}_i(y|x)$ and $\fpr{p}(y) = \sum_x Q(x) \fpr{W}(y|x) = \sum_i \lambda_i \fpr{p}_i(y)$, then we have

\begin{equation}\begin{split}\label{eq:3422}
\Delta
&=
\fpr{I}(Q, \fpr{W}) - \sum_i \lambda_i \fpr{I}(Q,\fpr{W}_i)
\\&=
\sum_{x,y} Q(x) \underbrace{\fpr{W}(y|x)}_{\sum_i \lambda_i \fpr{W}_i(y|x)} \log \left( \frac{\fpr{W}(y|x)}{\fpr{p}(y)} \right)
\\& \qquad -
\sum_{x,y,i} \lambda_i Q(x) \fpr{W}_i(y|x) \log \left( \frac{\fpr{W}_i(y|x)}{\fpr{p}_i(y)} \right)
\\&=
\sum_{x,y,i} \lambda_i Q(x) \fpr{W}_i(y|x) \log \left( \frac{\fpr{W}(y|x) \cdot \fpr{p}_i(y)}{\fpr{W}_i(y|x) \cdot \fpr{p}(y)} \right)
\\& \leq
\sum_{x,y,i} \lambda_i Q(x) \fpr{W}_i(y|x) \left( \frac{\fpr{W}(y|x) \cdot \fpr{p}_i(y)}{\fpr{W}_i(y|x) \cdot \fpr{p}(y)} - 1 \right)
\\& =
\sum_{x,y,i} \lambda_i Q(x) \cdot \frac{\fpr{W}(y|x) \cdot \fpr{p}_i(y)}{\fpr{p}(y)} - \sum_{x,y,i} \lambda_i Q(x) \fpr{W}_i(y|x)
\\& =
\sum_{x,y} Q(x) \fpr{W}(y|x) - \sum_{x,y} Q(x) \fpr{W}(y|x)
=
0
.
\end{split}\end{equation}

\textbf{Boundness:}
Since $\sum_{x'} Q(x') \fpr{W}(y|x') \geq \fpr{W}(y|x) Q(x)$ we have $\log \left( \frac{\fpr{W}(y|x)}{\sum_{x'} Q(x') \fpr{W}(y|x')} \right) \leq \log \left( \frac{\fpr{W}(y|x)}{Q(x) \fpr{W}(y|x)} \right) = \log \left( \frac{1}{Q(x)} \right)$. Now write:
\begin{equation}\begin{split}\label{eq:3509}
\fpr{I}(Q,\fpr{W})
& \defeq
\sum_{x,y} Q(x) \fpr{W}(y|x) \log \left( \frac{\fpr{W}(y|x)}{\sum_{x'} Q(x') \fpr{W}(y|x')} \right)
\\& \leq
\sum_{x,y} Q(x) \fpr{W}(y|x) \log \left( \frac{1}{Q(x)} \right)
\\& \leq
\sum_{x} \sigma Q(x) \log \left( \frac{1}{Q(x)} \right)
\\& =
\sigma \cdot H(Q)
\leq
\sigma \cdot \log |\mathcal{X}|
.
\end{split}\end{equation}

\subsection{Proof of Lemma~\ref{lemma:fMI_Lp_bound} and $L_p$ bounds on differences of entropies and capacities}\label{sec:Lp_bound_on_capacity}
In this section we prove Lemma~\ref{lemma:fMI_Lp_bound}, relating the $L_p$ norm difference of two channels (one of which may be a false distribution) to the difference in capacities. Two intermediate results that are captured in Lemmas~\ref{lemma:false_entropy_L1_bound},\ref{lemma:false_entropy_Lp_bound} are an extension of the $L_1$ bound of Cover \& Thomas to false distributions and a trivial extension of the same bound to $L_p$ norms.

We begin with the following $L_1$ bound on entropy from Cover \& Thomas \cite{CoverThomas_InfoTheoryBook}:
\begin{lemma}[$L_1$ bound on entropy, Theorem 7.3.3 of \cite{CoverThomas_InfoTheoryBook}]\label{lemma:CoverThomas_L1_bound}
Let $Q,P$ be two distributions on the finite alphabet $\mathcal{Y}$ with $\| Q - P \|_1 \leq \half$, then
\begin{equation}\label{eq:1437a}
|H(Q)-H(P)| \leq -\|Q-P\|_1 \cdot \log \left( \frac{\|Q-P\|_1}{|\mathcal{Y}|} \right)
.
\end{equation}
\end{lemma}
Also note that the function $-t \log \frac{t}{|\mathcal{Y}|}$ is monotonous non decreasing for $t \leq e^{-1}  |\mathcal{Y}|$, as can be verified by differentiation. Our first step is to extend the lemma to a case where one of $P,Q$ is a false distribution. In Cover and Thomas' proof, the first step is to write entropy as $H(P) = \sum_y f(P(y))$ where $f = -t \log t$ and to show that for all $0 \leq v \leq \half$ and $0 \leq t \leq 1-v$, the difference in $f$ is bounded by $|f(t+v)  - f(t)| \leq v \log v$. Here $t$ represents the minimum of $P(y), Q(y)$ for a certain $y$, $v$ the absolute difference, and $t+v$ the maximum of $P(y), Q(y)$. Then, the difference in entropy is bounded by the sum of the absolute values, this bound is substituted in the summation, and convexity arguments are use to bring it to the desired form. The only step that needs to be modified is showing that $|f(t+v)  - f(t)| \leq v \log v$, where now $t$ is no longer bounded to $t \leq 1-v$. It can be verified by differentiating the function $g(t) = f(t+v)  - f(t)$ with respect to $t$ that the derivative is always negative for $v > 0$. In addition, $g(0) > 0$, therefore the maximum absolute of this function, which is the absolute value of either the the maximum or the minimum, occurs at either end of the region to which $t$ is limited. In the original proof this yields $|f(t+v)  - f(t)| = |g(t)| \leq \max (|g(0)|, |g(1-v)|) = \max(f(v), f(1-v)) = -v \log v$ (notice that $f(0)=f(1)=0$). Here, since one of $P,Q$ is a legitimate distribution, $t \leq 1$ (as the minimum of the two) and we have instead: $|f(t+v)  - f(t)| = |g(t)| \leq \max (|g(0)|, |g(1)|) = \max(f(v), -f(1+v))$. As we will show below, if we limit $v \leq \tfrac{1}{4}$ we have $f(v) \geq -f(1+v)$, and therefore the bound $|f(t+v)  - f(t)| = |g(t)| \leq f(v)$ applies as in the original proof and Cover \& Thomas' result holds. To show this, consider the function $g(v) = -v \ln v - (v+1) \ln (v+1)$. This function is $0$ for $v=0$, and the derivative is $g'(v) = -\ln v - 1 -\ln (v+1) - 1 = -\ln (v(v+1)e^2)$, it is positive in a certain interval $(0,v_1)$ and negative for $v > v_1$, and therefore it crosses $0$ only once. Calculating this function for $v=\tfrac{1}{4}$ yields a positive value, therefore it is positive for all $v \leq \tfrac{1}{4}$. We capture this variation of Cover \& Thomas result in the following lemma:

\begin{lemma}[$L_1$ bound on false entropy difference]\label{lemma:false_entropy_L1_bound}
Let $P$ be a distribution on the finite alphabet $\mathcal{Y}$ and $\fpr{P}$ be a false distribution on the same alphabet, with $\| \fpr{P} - P \|_1 \leq \tfrac{1}{4}$, then
\begin{equation}\label{eq:1437}
|\fpr{H}(\fpr{P})-H(P)| \leq -\|\fpr{P}-P\|_1 \log \left( \frac{\|\fpr{P}-P\|_1}{|\mathcal{Y}|} \right)
,
\end{equation}
where the false entropy $\fpr{H}$ is defined as
\begin{equation}\label{eq:f_entropy_def1459}
\fpr{H}(\fpr{P}) \defeq - \sum_{y \in \mathcal{Y}} \fpr{P}(y) \log \fpr{P}(y)
.
\end{equation}
\end{lemma}

We first convert the bound to the $L_p$ norm ($p \geq 1$). To relate the norms we use \Holder's inequality: for two vectors $\vr a, \vr b$, $\sum_i | a_i b_i | \leq \| a \|_p \cdot \| a \|_{\overline p}$, where $\overline p^{-1} = 1 - p^{-1}$ is the \Holder~conjugate of $p$ and by convention for $p=\infty$ we define $1/p=0$ (note that $\overline p \geq 1$ and the conjugate of $p=\infty$ is $\overline p = 1$). We have
\begin{equation}\begin{split}\label{eq:1443}
\| \fpr{P}  - P \|_1
&=
\sum_y 1 \cdot |\fpr{P}(y)-P(y)|
\leq
\|\fpr{P}-P \|_p \cdot \| \vr 1 \|_{\overline p}
\\& =
\|\fpr{P}-P \|_p \cdot (\sum_{y \in \mathcal{Y}} 1^{\overline p})^{1/{\overline p}}
=
\|\fpr{P}-P \|_p \cdot |\mathcal{Y}|^{1/{\overline p}}
\\& =
\|\fpr{P}-P \|_p \cdot |\mathcal{Y}|^{1 - 1/p}
.
\end{split}\end{equation}
Assuming $\|\fpr{P}-P \|_p \cdot |\mathcal{Y}|^{1 - 1/p} \leq e^{-1}  |\mathcal{Y}|$ we can use the monotonicity of the bound of Lemma~\ref{lemma:false_entropy_L1_bound}, and write:
\begin{equation}\begin{split}\label{eq:1444}
|\fpr{H}(\fpr{P})-H(P)|
& \leq
-\|\fpr{P}-P\|_1 \log \left( \frac{\|\fpr{P}-P\|_1}{|\mathcal{Y}|} \right)
\\& \leq
-\|\fpr{P}-P \|_p \cdot |\mathcal{Y}|^{1 - 1/p} \log \left( \frac{\|\fpr{P}-P \|_p}{|\mathcal{Y}|^{1/p}} \right)
\\ & \defeq
f_p \left(\|\fpr{P}-P \|_p \right)
,
\end{split}\end{equation}
where we defined
\begin{equation}\label{eq:1469}
f_p(t) = -t \cdot |\mathcal{Y}|^{1 - 1/p} \log \left( \frac{t}{|\mathcal{Y}|^{1/p}} \right)
.
\end{equation}
$f_p(t)$ is concave with respect to $t$ (because $-t \ln t$ is concave in $t \geq 0$), and is monotonically non decreasing for $t \leq e^{-1} |\mathcal{Y}|^{1/p}$, as can be verified by differentiation. Furthermore, to meet the requirement $\|\fpr{P}-P \|_1 \leq \tfrac{1}{4}$ of Lemma~\ref{lemma:false_entropy_L1_bound}, it is sufficient that $\|\fpr{P}-P \|_p \cdot |\mathcal{Y}|^{1 - 1/p} \leq \tfrac{1}{4}$ (by \eqref{eq:1443}), and in addition prior to \eqref{eq:1444} we have assumed $\|\fpr{P}-P \|_p \leq e^{-1}  |\mathcal{Y}|^{1/p}$, however it is easy to see that this condition is dominated by the previous one. Since $1-1/p \geq 1$, and $|\mathcal{Y}| > 1$, it is sufficient to require $\|\fpr{P}-P \|_p \leq \tfrac{1}{4}$. In summary, we have the following result:

\begin{lemma}[$L_p$ bound on false entropy difference]\label{lemma:false_entropy_Lp_bound}
Let $p \geq 1$, $P$ be a distribution on the finite alphabet $\mathcal{Y}$, and $\fpr{P}$ be a false distribution on the same alphabet with $\| \fpr{P} - P \|_p \leq \tfrac{1}{4}$, then:
\begin{equation}\label{eq:1437p}
|\fpr{H}(\fpr{P})-H(P)| \leq f_p \left(\|\fpr{P}-P \|_p \right)
,
\end{equation}
where $f_p$ is defined in \eqref{eq:1469}, and it is concave and monotonically non-decreasing for $t \leq \tfrac{1}{4}$.
\end{lemma}

We now write the false mutual information \eqref{eq:fMI_def3386} as a difference of false entropies \eqref{eq:f_entropy_def1459}:
\begin{equation}\label{eq:1443f}
\fpr{I}(Q,\fpr{W}) = \fpr{H} \left(\sum_x \fpr{W}(y|x) Q(x) \right) - \sum_x Q(x) \fpr{H}(\fpr{W}(y|x))
.
\end{equation}
The above is analogous to the equality $I(X;Y) = H(Y) - H(Y|X)$. For the channels $W,\fpr{W}$ define the difference as $\delta_{xy} = W(y|x) - \fpr{W}(y|x)$ and define the output distributions as $P_Y(y) =\sum_x W(y|x) Q(x)$ and $\fpr{P}_Y(y) =\sum_x \fpr{W}(y|x) Q(x)$, then by the triangle inequality:
\begin{equation}\begin{split}\label{eq:1449}
& |\fpr{I}(Q,\fpr{W}) - I(Q,W) |
\leq
| H(P_Y) - H(\fpr{P}_Y) |
\\ & \qquad +
\sum_x Q(x) \left| \fpr{H}(\fpr{W}(y|x)) - H(W(y|x)) \right|
.
\end{split}\end{equation}
We begin with the difference $H(P_Y) - \fpr{H}(\fpr{P}_Y)$. By the $L_p$ bound of Lemma~\ref{lemma:false_entropy_Lp_bound} we have:
\begin{equation}\label{eq:1468}
H(P_Y) - H(\fpr{P}_Y) \leq f_p( \|P_Y-\fpr{P}_Y\|_p )
.
\end{equation}
Using the triangle inequality:
\begin{equation}\begin{split}\label{eq:1473}
\|P_Y-\fpr{P}_Y\|_p
&=
\| \sum_x Q(x) (W(y|x) - \fpr{W}(y|x)) \|_p
\\&=
\| \sum_x Q(x) \delta_{xy} \|_p
\\&\leq
\sum_x \|  Q(x) \delta_{xy} \|_{p,y}
\\&=
\sum_x Q(x) \| \delta_{xy} \|_{p,y}
,
\end{split}\end{equation}
where the notation $\| \Box \|_{p,y}$ is used to emphasize that the norm operation is with respect to $y$ only. Using \Holder's inequality,
\begin{equation}\begin{split}\label{eq:1473b}
\sum_x Q(x) \| \delta_{xy} \|_{p,y}
& \leq
\| Q(x) \|_{\overline p} \cdot \Big\| \| \delta_{xy} \|_{p,y} \Big\|_p
\\ & =
\left( \sum_x Q(x)^{\overline p} \right)^{1/{\overline p}} \cdot \| \delta_{xy} \|_p
\\& \stackrel{\overline p \geq 1}{\leq}
\left( \sum_x Q(x) \right)^{1/{\overline p}} \cdot \| \delta_{xy} \|_p
\\& =
\| \delta_{xy} \|_p
.
\end{split}\end{equation}
Assuming $\| \delta_{xy} \|_{p} \leq \tfrac{1}{4}$, $f_p$ is monotonously increasing, and combining the inequalities above we have:
\begin{equation}\label{eq:1506}
H(P_Y) - \fpr{H}(\fpr{P}_Y) \leq f_p \left( \| \delta_{xy} \|_p \right)
.
\end{equation}

For the second part of \eqref{eq:1449}, by the $L_p$ bound we have:
\begin{equation}\label{eq:1468a}
\left| \fpr{H}(\fpr{W}(y|x)) - H(W(y|x)) \right| \leq f_p (\|\delta_{xy}\|_{p,y})
.
\end{equation}
Using the concavity and monotonicity of $f_p$:
\begin{equation}\begin{split}\label{eq:1496}
&
\sum_x Q(x) \left| \fpr{H}(\fpr{W}(y|x)) - H(W(y|x)) \right|
\\ & \leq
\sum_x Q(x)  f_p (\|\delta_{xy}\|_{p,y})
\\& \leq
f_p \left( \sum_x Q(x) \|\delta_{xy}\|_{p,y} \right)
\\& \stackrel{\eqref{eq:1473b}}{\leq}
f_p (\| \delta_{xy} \|_{p})
,
\end{split}\end{equation}
where the monotonicity of $f_p$ is again guaranteed by the condition $\| \delta_{xy} \|_{p} \leq \tfrac{1}{4}$. Plugging \eqref{eq:1506} and \eqref{eq:1496} into \eqref{eq:1449} we have:
\begin{equation}\label{eq:1526}
\left| \fpr{I}(Q,\fpr{W}) - I(Q,W) \right| \leq 2 f_p \left(\| \delta_{xy} \|_{p} \right)
.
\end{equation}
which proves the bound on mutual information. The bound on capacity is trivially obtained from \eqref{eq:1526} above by writing $\fpr{I}(Q,\fpr{W}) \geq I(Q,W) - 2 f_p \left(\| \delta_{xy} \|_{p} \right)$ and maximizing both sides with respect to $Q$ (and similarly for the other direction).
\endofproof

\subsection{Proofs of small Lemmas}\label{sec:proofs_of_small_lemmas}
\textit{Proof of Lemma~\ref{lemma:exp_second_order_bound}: }
We would like to prove that $1 + x \leq e^x \leq 1 + x + x^2$. Using a finite tailor series we have:
\begin{equation}\label{eq:425}
e^x = 1 + e^0 \cdot x + \half e^t x^2
,
\end{equation}
where $t \in [0,x] \cup [x, 0]$ is a point between $0$ and $x$. This proves the lower bound. Also, for $x \leq 0$ since $e^t \leq 1$ this also proves the upper bound. For $0 < x \leq 1$, the right inequality can be made tighter, by writing the full Tailor expansion:
\begin{equation}\begin{split}
e^x
&=
\sum_{m=0}^{\infty} \frac{1}{m!} x^m
=
1 + x + \sum_{m=2}^{\infty} \frac{1}{m!} x^m
\\ & \leq
1 + x + x^2 \sum_{m=2}^{\infty} \frac{1}{m!}
=
1 + x + x^2 (e^{1} - 1 - 1)
\\ & =
1 + x + (e - 2) x^2
\leq
1 + x + x^2
.
\end{split}\end{equation}
\endofproof

\textit{Proof of Lemma~\ref{lemma:ab_alphabeta}:}
$f(t)$ is continuous and differentiable therefore $f'(t)=0$ at the maximum. Derivation yields $f'(t)= \alpha a  \cdot t^{\alpha-1} - \beta b \cdot t^{-\beta-1}$, and $f'(t)=0$ yields the single solution $t^*$ stated in the Lemma. This is a single maximum since $f'(t)$ is positive for $t < t^*$ and negative for $t > t^*$.
\endofproof

\subsection{Proof of Theorem~\ref{theorem:C_overlineW_optimality}: the optimality of averaged channel capacity}\label{sec:proof_CW_max_rate}
In this section we prove Theorem~\ref{theorem:C_overlineW_optimality} presented in \S\ref{sec:arbitrary_var_target_rate} (regarding the optimality of $C(\overline W)$). For a given sequence $W_1^n$, consider the ``permutation'' channel generated by uniformly selecting a random permutation $\Pi$ of the indices $i=1,\ldots,n$, rearranging the sequence $W_1^n$ to a permuted sequence $T_i = W_{\pi_i}$, and
applying the channel $\Pr(\vr Y | \vr X, \pi) = \prod_i T_i(Y_i | X_i)$ to the input (i.e. using the channels $W_i$ in permuted order). Suppose there is a system achieving the rate $R(W_1^n) - \Delta$ with probability $1-\delta$ and error probability $\epsilon$. Since this rate is fixed for all drawing of $\Pi$, the system can guarantee the rate $R(W_1^n)-\Delta$ a-priori (with probability $1-\delta$), and we can convert the rate-adaptive system to a fixed-rate system, delivering a message $\msg$ of $n (R(W_1^n) - \Delta)$ bits, with probability of error at most $\epsilon + \delta$. Once we constrain the discussion to the permutation channel induced by the deterministic sequence $W_1^n$, we can assume this sequence is known to the transmitter and the receiver.

By a standard application of Fano's inequality \cite[Theorem 2.10.1]{CoverThomas_InfoTheoryBook}, we have:
\begin{equation}\begin{split}\label{eq:2514}
I(\msg; \vr Y)
&=
H(\msg) - H(\msg | \vr Y)
\\& \geq
n (R(W_1^n) - \Delta) (1 - (\epsilon+\delta)) - h_b(\epsilon + \delta)
.
\end{split}\end{equation}
Rearranging and using $h_b(p) \leq 1$ we have:
\begin{equation}\label{eq:2523}
R(W_1^n) \leq \frac{\frac{1}{n} I(\msg; \vr Y) +  \frac{1}{n}}{1 - \epsilon - \delta} + \Delta
.
\end{equation}

In the main part of the proof we will show that approximately, $\frac{1}{n} I(\msg; \vr Y) \leq C(\overline W)$.  Note that because of feedback, $X_i$ may be a function of $\msg$ and $\vr Y^{i-1}$, and therefore $I(\vr X^n; \vr Y^n)$ does not give a tight bound on the rate. As noted in the outline presented in Section~\ref{sec:arbitrary_var_target_rate}, if the channels $T_i$ were selected from $W_1^n$ with replacement, this result would be obvious, since feedback would not be helpful. In the permuted channel, a system with feedback can use past channel outputs to gain some knowledge about the future behavior of the channel. The point of the proof is to show that there is no considerable gain from this knowledge, and even a knowledge of the actual list of channels that were already picked does not change the mutual information considerably.

\begin{figure}
\centering
\ifpdf
  \setlength{\unitlength}{1bp}%
  \begin{picture}(220.24, 130.39)(0,0)
  \put(0,0){\includegraphics{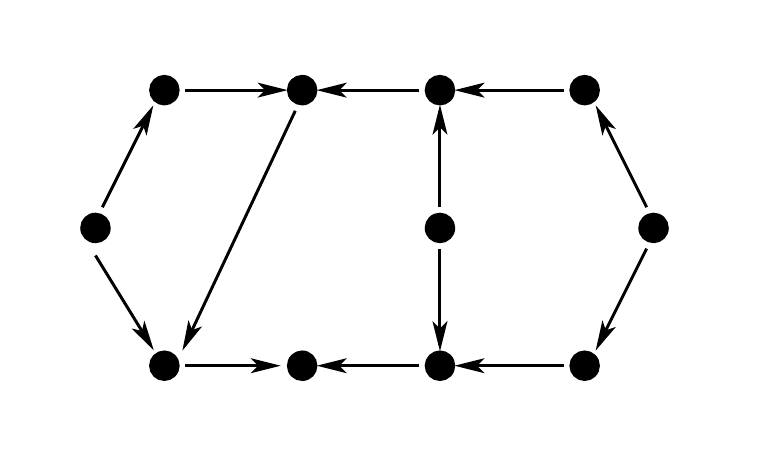}}
  \put(5.67,63.37){\fontsize{9.96}{11.95}\selectfont $m$}
  \put(43.37,7.81){\fontsize{9.96}{11.95}\selectfont $X_i$}
  \put(41.39,116.95){\fontsize{9.96}{11.95}\selectfont $\vr X^{i-1}$}
  \put(194.17,63.37){\fontsize{9.96}{11.95}\selectfont $\pi$}
  \put(118.77,116.95){\fontsize{9.96}{11.95}\selectfont $T^{i-1}$}
  \put(79.09,116.95){\fontsize{9.96}{11.95}\selectfont $\vr Y^{i-1}$}
  \put(81.07,7.81){\fontsize{9.96}{11.95}\selectfont $Y_i$}
  \put(122.74,7.81){\fontsize{9.96}{11.95}\selectfont $T_i$}
  \put(162.43,116.95){\fontsize{9.96}{11.95}\selectfont $\pi^{i-1}$}
  \put(164.41,7.81){\fontsize{9.96}{11.95}\selectfont $\pi_i$}
  \put(134.65,61.39){\fontsize{9.96}{11.95}\selectfont $W_1^n$}
  \end{picture}%
\else
  \setlength{\unitlength}{1bp}%
  \begin{picture}(220.24, 130.39)(0,0)
  \put(0,0){\includegraphics{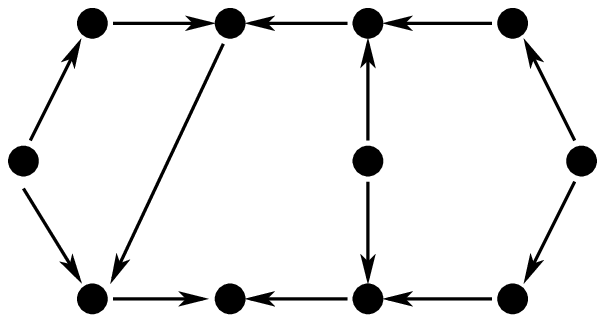}}
  \put(5.67,63.37){\fontsize{9.96}{11.95}\selectfont $m$}
  \put(43.37,7.81){\fontsize{9.96}{11.95}\selectfont $X_i$}
  \put(41.39,116.95){\fontsize{9.96}{11.95}\selectfont $\vr X^{i-1}$}
  \put(194.17,63.37){\fontsize{9.96}{11.95}\selectfont $\pi$}
  \put(118.77,116.95){\fontsize{9.96}{11.95}\selectfont $T^{i-1}$}
  \put(79.09,116.95){\fontsize{9.96}{11.95}\selectfont $\vr Y^{i-1}$}
  \put(81.07,7.81){\fontsize{9.96}{11.95}\selectfont $Y_i$}
  \put(122.74,7.81){\fontsize{9.96}{11.95}\selectfont $T_i$}
  \put(162.43,116.95){\fontsize{9.96}{11.95}\selectfont $\pi^{i-1}$}
  \put(164.41,7.81){\fontsize{9.96}{11.95}\selectfont $\pi_i$}
  \put(134.65,61.39){\fontsize{9.96}{11.95}\selectfont $W_1^n$}
  \end{picture}%
\fi
\caption{\label{fig:dependence_graph_for_converse_channel}%
 A dependence graph for the variables of the permutation channel in Appendix~\ref{sec:proof_CW_max_rate}. Each node is a (potentially random) function of the nodes with arrows pointing toward it.}
\end{figure}

We denote by $\Pi$ the random permutation and by $\pi$ a specific instance of the permutation. We bound the mutual information as follows:
\begin{equation}\begin{split}\label{eq:2507}
I(\vr Y^n ; \msg)
&=
\sum_{i=1}^n I(Y_i ; \msg | \vr Y^{i-1})
\\& =
\sum_{i=1}^n \left( H(Y_i | \vr Y^{i-1}) - H(Y_i | \vr Y^{i-1}, \msg) \right)
\\& \stackrel{(a)}{\leq}
\sum_{i=1}^n \left( H(Y_i) - H(Y_i | \vr Y^{i-1}, \msg, \Pi^{i-1}, X_i) \right)
\\& \stackrel{(b)}{=}
\sum_{i=1}^n \left( H(Y_i) - H(Y_i | \Pi^{i-1}, X_i) \right)
,
\end{split}\end{equation}
where (a) is because conditioning reduces entropy (used twice), and (b) is since $\vr Y^{i-1}, \msg \leftrightarrow T^{i-1}, X_i \leftrightarrow Y_i$ (in other words,  $\pi^{i-1}, X_i$ gives all relevant information on $Y_i$). This can be seen from the functional dependence graph in Fig.\ref{fig:dependence_graph_for_converse_channel}. Let $Z_i$ be a random variable generated by passing $X_i$ through the channel $\overline W$ (i.e. $\Pr(Z_1^n | X_1^n) = \prod_{i=1}^n \overline W(Z_i | X_i)$). Next we show that $H(Y_i) \approx H(Z_i)$ and $H(Y_i | \Pi^{i-1}, X_i) \approx H(Z_i | X_i)$.

Given $\Pi^{i-1}$, the channel law between $X_i$ and $Y_i$ is a random pick from the group of $n-i+1$ channels that are not included in $\{\Pi_j\}_{j=1}^{i-1}$:
\begin{equation}\begin{split}\label{eq:2528}
& \Pr(Y_i=y | \Pi^{i-1}, X_i=x)
\\& =
\sum_{k=1}^n \Pr(Y_i=y | \Pi^{i-1}, \Pi_i = k, X_i=x)
\\ & \qquad \cdot \Pr(\Pi_i = k | \Pi^{i-1}, X_i=x)
\\&=
\sum_{k \not\in \{\Pi^{i-1}\}} W_k(y | x) \cdot \frac{1}{n-i+1}
\defeq
\overline W_{\Pi^{i-1}}(y|x)
.
\end{split}\end{equation}
The average channel given the past indices $W_{\pi^{i-1}}(y|x)$ is an average of $n-i+1$ values $0 \leq W_k(y | x) \leq 1$. Note that the indices $k$ belong to $\Pi_{i}^n$, so the notation may be confusing, but it is used to stress the causal dependence on $\Pi^{i-1}$.

Considering the random variable $\overline W_{\Pi^{i-1}}(y|x)$ generated by calculating this channel over all drawings of $\Pi$, the set $k \not\in \{\Pi^{i-1}\}$ becomes a random set of $n-i+1$ distinct indices from $1,\ldots,n$, chosen uniformly from all such sets. $\overline W_{\Pi^{i-1}}(y|x)$ is an average of $n-i+1$ values $0 \leq W_k(y | x) \leq 1$, sampled uniformly without replacement from the set $\{ W_k(y|x) \}_{k=1}^n$ (for any specific $x,y$). It was shown by Hoeffding \cite[\S 6]{Hoeffding} that averages of variables sampled without replacement obey the same bounds \cite[Theorem~1]{Hoeffding} with respect to the probability to deviate from their mean, as independent random variables. Specifically, applying Hoeffding's bounds (combining Theorem~1 with Section~6 in \cite{Hoeffding}), and since $\E \left[ \overline W_{\Pi^{i-1}}(y|x) \right] = \overline W$, we have:
\begin{equation}\label{eq:2543}
\Pr \{ | \overline W_{\Pi^{i-1}}(y|x) - \overline W | \geq t \} \leq 2 e^{-2 (n - i + 1) t^2}
.
\end{equation}
Using the union bound over all $|\mathcal{X}| \cdot |\mathcal{Y}|$ values of $x,y$ (see the proof of Proposition~\ref{prop:symbolwise_scheme_channel_convergence2}), we have:
\begin{equation}\label{eq:2543a}
\Pr \{ \| \overline W_{\Pi^{i-1}} - \overline W \|_{\infty} \geq t \} \leq 2 |\mathcal{X}| \cdot |\mathcal{Y}| e^{-2 (n - i + 1) t^2}
,
\end{equation}
where the $L_{\infty}$ norm is over $x,y$. To further simplify, we pick a small value $\epsilon_0$, and from now on we assume $i \leq (1-\epsilon_0) n$. Substituting in \eqref{eq:2543a}, we have:
\begin{equation}\label{eq:2543b}
\Pr \{ \| \overline W_{\Pi^{i-1}} - \overline W \|_{\infty} \geq t \} \leq 2 |\mathcal{X}| \cdot |\mathcal{Y}| e^{-2 \epsilon_0 n t^2} \defeq p
,
\end{equation}
Since $H(\cdot)$ is uniformly continuous (see Lemma~\ref{lemma:false_entropy_Lp_bound}), for any $\epsilon_0$ there is a $t$ such that if $\| P_1(y) - P_2(y) \|_{\infty} \leq 2 t$ then $| H(P_1) - H(P_2) | \leq \epsilon_0$. For a given $\epsilon_0$ we choose the value of $t$ such that this requirement is satisfied, so that together with \eqref{eq:2543b} we have:
\begin{equation}\label{eq:2580}
\forall x: \Pr \{ | H(\overline W_{\pi^{i-1}}(\cdot|x)) -  H(\overline W)| \leq \epsilon_0 \} \geq 1 - p
.
\end{equation}

We use the following relation to translate proximity in probability to proximity of the expected values: if $A,B \in [0,A_{\max}]$ are two random variables satisfying $\Pr\{ |A - B| \leq \epsilon \} \geq 1-p$ (for some $\epsilon,p \in [0,1]$), then
\begin{equation}\begin{split}\label{eq:2592}
\big| \E [A] - \E [B] \big|
&=
\Big| \E [(A-B) \cdot \Ind(|A - B| \leq \epsilon)]
\\& \qquad + \E [(A-B) \cdot \Ind(|A - B| > \epsilon)] \Big|
\\ & \leq
\E [|A-B| \cdot \Ind(|A - B| \leq \epsilon)]
\\ & \qquad + \E [|A-B| \cdot \Ind(|A - B| > \epsilon)]
\\& \leq
\epsilon + \E [A_{\max} \cdot \Ind(|A - B| > \epsilon)]
\\ & \leq
\epsilon + A_{\max} \cdot p
.
\end{split}\end{equation}
Applying this inequality to bound $H(Y_i | \Pi^{i-1}, X_i)$ we have:
\begin{equation}\begin{split}\label{eq:2572}
H(Y_i | \Pi^{i-1}, X_i)
&=
\sum_{x, \pi} H(Y_i | \Pi^{i-1} = \pi^{i-1}, X_i=x)
\\ & \qquad \cdot \Pr(\Pi = \pi, X_i=x)
\\&\stackrel{\eqref{eq:2528}}{=}
\sum_{x, \pi} H(\overline W_{\pi^{i-1}}(\cdot|x)) \cdot \Pr(\Pi = \pi, X_i=x)
\\ & =
\E \left[ H(\overline W_{\Pi^{i-1}}(\cdot|X_i)) \right]
\\& \stackrel{\eqref{eq:2580}, \eqref{eq:2592}}{\geq}
\E \left[ H(\overline W(\cdot|X_i)) \right] - \epsilon_0 - \log |\mathcal{Y}| \cdot p
\\& =
H(Z_i|X_i) - \epsilon_0 - \log |\mathcal{Y}| \cdot p
.
\end{split}\end{equation}

We now show that the distributions of $Y_i$ and $Z_i$ are similar (note that they are not equal, due to the possible dependence of $X_i$ on $\Pi^{i-1}$).
\begin{equation}\begin{split}\label{eq:2574}
&\left| \Pr(Y_i=y) - \Pr(Z_i=y) \right|
\\& =
\left| \E \left[ \Pr(Y_i=y | \Pi^{i-1}, X_i) \right] - \E \left[ \Pr(Z_i=y | X_i) \right] \right|
\\&\stackrel{\eqref{eq:2528}}{=}
\left| \E \left[ \overline W_{\Pi^{i-1}}(y|X_i) \right] - \E \left[ \overline W(y | X_i) \right] \right|
\stackrel{\eqref{eq:2543b}, \eqref{eq:2592}}{\leq}
t + p
.
\end{split}\end{equation}
Since for any $\epsilon_0, t$ we have $p \ntoinfty 0$ \eqref{eq:2543b}, we can choose $n$ large enough such that $p \leq t$ and we have $\left| \Pr(Y_i=y) - \Pr(Z_i=y) \right| \leq 2t$. Then, by our selection of $t$ (before \eqref{eq:2580}), we shall have:
\begin{equation}\label{eq:2628}
\left| H(Y_i) - H(Z_i) \right| \leq \epsilon_0
.
\end{equation}
Returning to \eqref{eq:2507}, and treating the first $(1-\epsilon_0)n$ and the last $\epsilon_0 n$ symbols separately, we have:
\begin{equation}\begin{split}\label{eq:2507b}
I(\vr Y^n ; \msg)
& \leq
\sum_{i=1}^n \left( H(Y_i) - H(Y_i | \Pi^{i-1}, X_i) \right)
\\ & \stackrel{\eqref{eq:2572}, \eqref{eq:2628}}{\leq}
\sum_{i=1}^{(1-\epsilon_0) n} \Big[ (H(Z_i) + \epsilon_0)
\\ & \qquad - ( H(Z_i|X_i) - \epsilon_0 - \log |\mathcal{Y}| \cdot p)  \Big]
\\ & \qquad +  \epsilon_0 \cdot n \cdot \log |\mathcal{Y}|
\\& \leq
\sum_{i=1}^{n} I(Z_i; X_i) + n \underbrace{( 2 \epsilon_0 + (\epsilon_0 + p) \cdot \log |\mathcal{Y}|)}_{\delta_0}
\\& \leq
n \cdot C(\overline W) + n \delta_0
.
\end{split}\end{equation}
Because $\epsilon_0$ is a parameter of choice, and for any $\epsilon_0, t$ we have $p \ntoinfty 0$ \eqref{eq:2543b}, we can make $\delta_0$ a small as desired for $n$ large enough. Returning to \eqref{eq:2523} we have:
\begin{equation}\begin{split}\label{eq:2696}
R(W_1^n)
& \leq
\frac{C(\overline W) + \delta_0  + 1/n}{(1-\epsilon -\delta)} + \Delta
\\& \leq
(C(\overline W) + \delta_0  + 1/n)(1 + \epsilon + \delta) + \Delta
\\& \leq
C(\overline W) + \underbrace{(\delta_0  + 1/n)(1 + \epsilon + \delta) + (\epsilon + \delta) I_{\max} + \Delta}_{\delta_1}
,
\end{split}\end{equation}
where $I_{\max}$ is defined in \eqref{eq:Imax_def}. Since by Definition~\ref{def:attainability_of_RW}, the above must hold for every $\epsilon, \delta, \Delta$, for $n$ large enough, and $\delta_0 \ntoinfty 0$ (see \eqref{eq:2507b} and the discussion following it), we can make $\delta_1$ as small as desired by taking $n \to \infty$. This concludes the proof of Theorem~\ref{theorem:C_overlineW_optimality}.
\endofproof

\subsection{Proof of Lemma~\ref{lemma:C_overlineW_achievability_w_side_info}}\label{sec:proof_C_overlineW_achievability_w_side_info}
Clearly, using the assumptions of this section, $F_i(Q) = I(Q,\overline W_i)$ satisfies the conditions of the lemma. The lemma assumes there are $B+1$ blocks and the rate is $\frac{KB}{n}$, which corresponds to a case where the last block was not decoded, however it holds as a lower bound even if the last block was decoded. We now optimize the value of $\lambda$. Starting from \eqref{eq:743}:
\begin{equation}\label{eq:990}
\Dpredsup{*}(\lambda) = \frac{K}{n} + I_{\max} \cdot \lambda + \underbrace{c_1 \sqrt{\frac{\ln (n)}{n}}}_{b_2} \lambda^{-\half}
,
\end{equation}
we determine $\lambda$ using Lemma~\ref{lemma:ab_alphabeta} (with $\alpha=1, \beta=\half$) and obtain:
\begin{equation}\begin{split}\label{eq:993}
\Dpredsup{*} (\lambda^*)
&=
\left( \frac{\beta}{\alpha} \right)^{\frac{\alpha}{\alpha+\beta}} \left[ 1
+ \frac{\alpha}{\beta} \right] \cdot I_{\max}^{\frac{\beta}{\alpha+\beta}} \cdot b_2^{\frac{\alpha}{\alpha+\beta}} + \frac{K}{n}
\\&=
3 \cdot 2^{-\frac{2}{3}} \cdot I_{\max}^{\frac{1}{3}} \cdot \left( c_1 \sqrt{\frac{\ln (n)}{n}} \right)^{\frac{2}{3}} + \frac{K}{n}
\\& \stackrel{\eqref{eq:795}}{=}
3 \left( K \mathcal{X}| (|\mathcal{X}|-1)  \right)^{\frac{1}{3}} I_{\max}^{\frac{2}{3}} \cdot \left( \frac{\ln (n)}{n} \right)^{\frac{1}{3}} + \frac{K}{n}
\\ & \leq
\left(\frac{K}{n}\right)^{\tfrac{1}{3}} \cdot \left[ 3  \left( |\mathcal{X}|  \cdot I_{\max} \right)^{\tfrac{2}{3}} \ln^{\tfrac{1}{3}}(n) + \left(\frac{K}{n}\right)^{\tfrac{2}{3}} \right]
\\ & \stackrel{(a)}{\leq}
\left(\frac{K}{n}\right)^{\tfrac{1}{3}} \cdot  4 \cdot \left( |\mathcal{X}|  \cdot I_{\max} \right)^{\tfrac{2}{3}} \ln^{\tfrac{1}{3}}(n)
\\ & =
4 \cdot K^{\tfrac{1}{3}} \cdot |\mathcal{X}|^{\tfrac{2}{3}} \cdot I_{\max}^{\tfrac{2}{3}} \cdot \left( \frac{\ln (n)}{n} \right)^{\frac{1}{3}}
\defeq \Dpred
,
\end{split}\end{equation}
where in (a) we assumed $K \leq |\mathcal{X}|  \cdot n \cdot I_{\max}$. If the contrary is true, the first term in \eqref{eq:990} yields $\Dpred > \frac{K}{n} > I_{\max}$ and the theorem is true in a void way. Similarly, we do not have to worry about the case $\lambda^* > 1$ since also in this case, due to the second term in \eqref{eq:990}, $\Dpred > I_{\max}$.

If the conditions of Lemma~\ref{lemma:rateless_prior_predictor_lemma} are satisfied, we have for all $Q$ \eqref{eq:688}:
\begin{equation}\begin{split}\label{eq:1011}
R
& \geq
\min \left( \sum_{i=1}^{B+1} \frac{m_i}{n} \cdot I(Q, \overline W_i) - \Dpred, I_{\max} \right)
\\ & \geq
I \left(Q, \sum_{i=1}^{B+1} \frac{m_i}{n} \overline W_i \right) - \Dpred
\\&=
I \left(Q, \overline W \right) - \Dpred ,
\end{split}\end{equation}
where we used the convexity of $I(Q,W)$ with respect to the channel $W$. Maximizing both sides of \eqref{eq:1011} with respect to $Q$ yields the desired result \eqref{eq:917R}.

The conditions of Lemma~\ref{lemma:rateless_prior_predictor_lemma} on $n,K$ remain as conditions of the theorem. The application of Lemma~\ref{lemma:ab_alphabeta} in \eqref{eq:993} yields the following value of $\lambda$:
\begin{equation}\begin{split}\label{eq:988}
\lambda^*
&=
\left( \frac{b_2 \beta}{I_{\max} \alpha} \right)^{\frac{1}{\alpha+\beta}}
=
\left( \frac{\half c_1 \sqrt{\frac{\ln (n)}{n}}}{I_{\max}} \right)^{\frac{2}{3}}
\\ & =
\left( K \cdot |\mathcal{X}| (|\mathcal{X}|-1) \cdot I_{\max}^{-1} \cdot \frac{\ln (n)}{n} \right)^{\frac{1}{3}}
.
\end{split}\end{equation}
This concludes the proof of Lemma~\ref{lemma:C_overlineW_achievability_w_side_info}.
\endofproof

\subsection{Channel knowledge compared to channel estimation}\label{sec:example_prediction_channel2}
In this section we demonstrate the claim made in Section~\ref{sec:arbitrary_var_target_rate}, that even imposing on the synthetic problem only the limitation that the past channels are not given, but need to be estimated, leads to the conclusion $C_2$ is not attainable.

To show this we use an example, based on randomization of the channel sequence. As in Section~\ref{sec:toy_problem}, we assume $I(\hat Q_i, W_i)$ bits are transmitted in time instance $i$ (in other words, this is the gain obtained in retrospect for choosing $\hat Q_i$), however, instead of knowing the full channel sequence, the predictor is only allowed to base its decisions on measurements of the channel input and output, i.e. on the values of $(\vr Y_1^{i-1}, \vr X_1^{i-1})$ where $Y_i$ is the result of $W_i$ operated on $X_i$. It would make sense to also require that $X_i$ be distributed $\hat Q_i(x)$ but this assumption is not required for the counter example.

\begin{figure}
\centering
\ifpdf
  \setlength{\unitlength}{1bp}%
  \begin{picture}(248.46, 134.36)(0,0)
  \put(0,0){\includegraphics{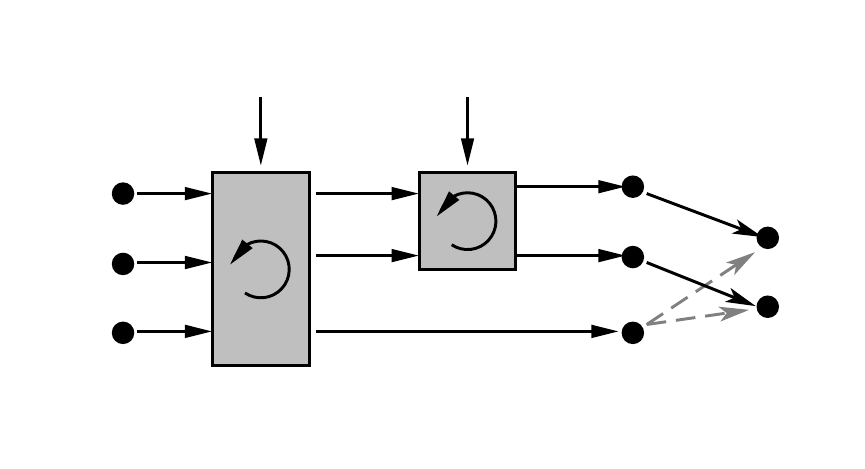}}
  \put(5.67,77.26){\fontsize{9.96}{11.95}\selectfont $0$}
  \put(5.67,55.43){\fontsize{9.96}{11.95}\selectfont $1$}
  \put(5.67,35.59){\fontsize{9.96}{11.95}\selectfont $2$}
  \put(27.50,93.14){\fontsize{9.96}{11.95}\selectfont $X$}
  \put(214.02,79.25){\fontsize{9.96}{11.95}\selectfont $Y$}
  \put(227.91,65.36){\fontsize{9.96}{11.95}\selectfont $0$}
  \put(227.91,43.53){\fontsize{9.96}{11.95}\selectfont $1$}
  \put(89.01,7.81){\fontsize{9.96}{11.95}\selectfont $W_{sr}(Y|X)$}
  \put(75.12,120.91){\fontsize{9.96}{11.95}\selectfont \makebox[0pt]{$r \in \{0,1,2\}$}}
  \put(144.57,120.91){\fontsize{9.96}{11.95}\selectfont \makebox[0pt]{$s \in \{0,1\}$}}
  \put(73.13,110.99){\fontsize{9.96}{11.95}\selectfont \makebox[0pt]{Chosen once}}
  \put(140.60,110.99){\fontsize{9.96}{11.95}\selectfont \makebox[0pt]{Chosen i.i.d}}
  \end{picture}%
\else
  \setlength{\unitlength}{1bp}%
  \begin{picture}(248.46, 134.36)(0,0)
  \put(0,0){\includegraphics{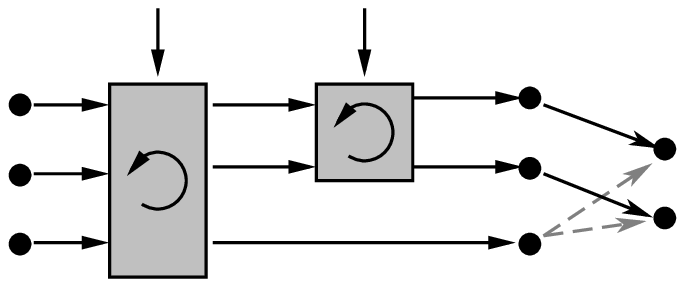}}
  \put(5.67,77.26){\fontsize{9.96}{11.95}\selectfont $0$}
  \put(5.67,55.43){\fontsize{9.96}{11.95}\selectfont $1$}
  \put(5.67,35.59){\fontsize{9.96}{11.95}\selectfont $2$}
  \put(27.50,93.14){\fontsize{9.96}{11.95}\selectfont $X$}
  \put(214.02,79.25){\fontsize{9.96}{11.95}\selectfont $Y$}
  \put(227.91,65.36){\fontsize{9.96}{11.95}\selectfont $0$}
  \put(227.91,43.53){\fontsize{9.96}{11.95}\selectfont $1$}
  \put(89.01,7.81){\fontsize{9.96}{11.95}\selectfont $W_{sr}(Y|X)$}
  \put(75.12,120.91){\fontsize{9.96}{11.95}\selectfont \makebox[0pt]{$r \in \{0,1,2\}$}}
  \put(144.57,120.91){\fontsize{9.96}{11.95}\selectfont \makebox[0pt]{$s \in \{0,1\}$}}
  \put(73.13,110.99){\fontsize{9.96}{11.95}\selectfont \makebox[0pt]{Chosen once}}
  \put(140.60,110.99){\fontsize{9.96}{11.95}\selectfont \makebox[0pt]{Chosen i.i.d}}
  \end{picture}%
\fi
\caption{\label{fig:example_prediction_channel2}%
 An illustration of the generation of the channels $W_{sr}$ in Example~\ref{example:prediction_channel2}.}
\end{figure}

\begin{example}\label{example:prediction_channel2}
Consider a ternary input binary output channel. We will choose the channel randomly, and consider the average gain of the predictor and the reference (since the average regret is a lower bound for the maximum regret). The basic channels are $W_1=\left[ \begin{array}{ccc} \half & 0 & 1 \\ \half & 1 & 0 \end{array} \right]$, $W_2=\left[ \begin{array}{ccc} \half & 1 & 0 \\ \half & 0 & 1 \end{array} \right]$. Note that in the two channels, the first input is useless, and using only the two last inputs yields a rate of $1$ bit/use. We add to this family of channels all 3 possible cyclic rotations of the inputs, and term the channel $W_s^{r}$ ($s=1,2; r=1,2,3$). The resulting channels are depicted in Fig.~\ref{fig:example_prediction_channel2}. Now we generate the sequence of channels as follows: choose $r$ randomly (one for the entire sequence), and choose a random (uniform, i.i.d.) sequence of $s_i$-s. The competitor, knowing $r$, easily selects a prior that optimizes $\sum_i I(Q,W_i)$, since $W_1^r$ and $W_2^r$ have the same optimizer for each $r$, and achieves a rate of 1. Because of the random generation of the sequence $s_i$, for any value of $r$, the channel output $\vr Y_1^{i-1}$ is uniform i.i.d. over $\{0,1\}$ and independent of the input. Therefore the predictor cannot infer any information on $r$ from the input-output distribution. Therefore the best the predictor can do (in terms of optimizing for the worst-case $r$), is place a uniform prior over all 3 inputs, and therefore obtain a rate of $\frac{2}{3}$, i.e. a regret of $\frac{1}{3}$ bit per channel use. By increasing the size of the channel input, this gap can be increased indefinitely.
\end{example}

The conclusion from the example is that $C_2$ cannot be attained universally when actual channel measurements are used.

\subsection{An analysis of the prior quantization approach}\label{sec:prior_quantization_analysis}
In Section~\ref{sec:discussion_prediction_scheme} we mentioned an alternative of using a ``codebook'' of priors, instead of the exponential weighting scheme over the continuum of priors, which was used in this paper. Following is a rough analysis of this approach, for the block-wise variation setting. We first determine the accuracy required of the codebook. Suppose we have two priors $Q_1, Q_2$ with $\| Q_1 - Q_2 \|_{\infty} \leq \Delta$, and for a certain channel $W$ the resulting output distributions are $P_1, P_2$ respectively ($P_m = \sum_x Q_m(x) W(y|x), m=1,2$). We write $I(Q_m, W) = H(P_m) - \sum_x Q_m(x) H(W(\cdot|x))$ (output entropy minus output entropy given the input). Since by definition $\| P_1 - P_2 \|_{\infty} \leq | \mathcal{X} | \cdot  \| Q_1 - Q_2 \|_{\infty}$, by using Lemma~\ref{lemma:false_entropy_Lp_bound} we have $|H(P_1) - H(P_2)| \leq f_{\infty}(| \mathcal{X} | \cdot  \Delta)$, and since the second term in $I(Q_m, W)$ may change by at most $\log |\mathcal{X}| \cdot \Delta$, we have $|I(Q_1,W) - I(Q_2,W)| \leq f_{\infty}(| \mathcal{X} | \cdot  \Delta) + \log |\mathcal{X}| \cdot \Delta \defeq \Delta_I$. Therefore, in order to bound the loss due to the codebook quantization to $\Delta_I = O \left( \frac{\ln n}{n} \right)^{\half}$, we need to have $\Delta = O(n^{-\half})$ (here, $Q_1$ represents the any prior, and $Q_2$ represents the closest point in the codebook). To have a density of $O(n^{-\half})$ per dimension, $N = O \left( n^{\half (|\mathcal{X}|-1)} \right)$ points are required. Now, since $\max_Q \frac{1}{n} \sum_{i=1}^n I(Q,W_i)$ differs from $\max_{m \in {1,\ldots,N}} \frac{1}{n} \sum_{i=1}^n I(Q_m,W_i)$ by at most $\Delta_I$, we can now consider the problem of competing against the $N$ priors (considered as $N$ experts). The best normalized redundancy than can be attained is $O \left( \sqrt{\frac{\ln N}{n}} \right) = O \left( \sqrt{\frac{\ln n}{n}} \right)$ (see the lower bound \cite[Theorem 3.7]{Nicolo} and the upper bound \cite[Corollary 2.2]{Nicolo} in Cesa-Bianchi and Lugosi's book). Note that since the predictor loss and the codebook loss are balanced, we cannot gain by changing the codebook density. However, we have not shown that the bound on $\Delta_I$ is tight.

\subsection{Operation with any positive feedback rate}\label{sec:zero_feedback_rate}
Here we show how the scheme can be modified to operate with any positive feedback rate. Feedback is used in the scheme \S\ref{sec:arbitrary_var_rateless_scheme} for two purposes:
\begin{enumerate}
\item In order to report reception of a rateless block (we use 1 bit per channel use)
\item In order to send the estimated averaged channel $\fpr{W}_i$ after the end of each block (or alternatively, the next prior $\hat Q_{i+1}$).
\end{enumerate}

Suppose feedback is limited to rate $R_\tsubs{FB}$. Instead of reporting successful reception on each symbol, we report it each $N_1 = \lceil \frac{1}{R_\tsubs{FB}} \rceil$ symbols. The price would be wasting up to $N_1$ symbols per block, which essentially form an unused ``gap'' between successful decoding of block $i$ and the start of block $i+1$.

We now give a coarse bound on the number of bits required to represent the estimated averaged channel $\fpr{W_i}$. $\fpr{W_i}$ is completely specified to the transmitter by specifying the empirical distribution $\hat P_{\vr x, \vr y}(x,y)$ which takes at most $(m+1)^{|\mathcal{X}| \cdot |\mathcal{Y}|}$ values for a block of length $m$. Since $m \leq n$, the number of bits is at most $N_2 = \log |\mathcal{X}| \cdot |\mathcal{Y}| \cdot \log(n+1) = O(\ln n)$. These bits can be sent over $\frac{N_2}{R_\tsubs{FB}}$ channel uses at the end of each block, thus forming another unused ``gap'' between the blocks. Overall the gap between blocks is $N_1 + \frac{N_2}{R_\tsubs{FB}} = O \left( \frac{\log n}{R_\tsubs{FB}} \right)$. Since the maximum number of blocks grows sub-linearly in $n$, the overall loss can be made negligible.

Specifically, the effect of the additional gap on the rate can be analyzed using the same technique used to analyze the loss in the last symbol (the transition between \eqref{eq:mi1413rand0} and \eqref{eq:mi1413rand}), and would effectively increase the term $\log \left( \frac{|\mathcal{X}|}{\lambda} \right)$ in $\delta_1$ \eqref{eq:1418} by a factor of the gap $O(\log n)$. Since $K \in \omega(\log n)$ it is easy to see that under the same setting of of the parameters of the scheme, we would still have $\delta_1 \ntoinfty 0$ and $\Delta_C \ntoinfty 0$, and nearly at the same convergence rate.

A delay in the feedback link would simply mean that an additional fixed gap will be added between the blocks, which also does not prevent asymptotical convergence.

\subsection{Generation of the prior using rejection sampling}\label{sec:rejection_sampling_predictor}
As mentioned, implementation of the prediction methods described in this paper, which are based on weighted average over the unit simplex, require the calculation of integrals. In the below, we show an alternative method to generate the same results, using a method based on rejection sampling. Instead of explicitly calculating the predictor $\hat Q$, we describe an algorithm that generates a random variable $X \sim \hat Q$ (which can be used to generate a letter in the random codebook), based on multiple drawings of uniform random variables. The number of random drawings required in this algorithm is polynomial in $n$, but still prohibitively large, so unfortunately it is not practical.

First, any scalar random variable can be derived from a uniform $[0,1]$ random variable by the inverse transform theorem. A generation of the mixture of an exponentially weighted and a uniform distribution such as in \eqref{eq:Qi3382}, only requires to toss a coin with probability $\lambda$, which determines whether $X$ is generated using the exponentially weighted distribution or using a uniform distribution. Therefore the problem of generating the predictors described here \eqref{eq:p404}, \eqref{eq:Qi3382}, boils down to the following problem: we would like to generate a random variable $X$ distributed according to
\begin{equation}\label{eq:2393}
\hat Q = \int w(Q) Q dQ
,
\end{equation}
where
\begin{equation}\label{eq:2398}
w(Q) = \frac{e^{\eta g(Q)}}{\int_{\Delta} e^{\eta g(Q)} d Q}
,
\end{equation}
and where $g(Q)$ is a concave function and is bounded $0 \leq g(Q) \leq n \cdot g_0$. $\Delta$ is the unit simplex (which implicitly refers to the alphabet $\mathcal{X}$). All integrals below are over the unit simplex. Furthermore, we would like to accomplish this without computing any integrals.

The first observation is that instead of generating an $X$ from $\hat Q$ it is enough to generate a the probability vector $Q$ randomly with the probability distribution $w(Q)$ and then generate an $X$ from the (specific) probability distribution $Q$. The last step can be accomplished using the inverse transform theorem. In this case we have:
\begin{equation}\begin{split}\label{eq:2397}
\Pr(X = x)
&=
\underset{Q \sim w(Q)}{\E} \left[ \Pr(X = x | Q) \right]
\\&=
\underset{Q \sim w(Q)}{\E} \left[ Q(x) \right]
=
\int Q(x) w(Q) dQ
.
\end{split}\end{equation}
This leaves us with the problem of generating $Q \sim w(Q)$. This is accomplished by rejection sampling. I.e. we first generate a random variable with a different distribution, and if it does not satisfy a given condition, we ``reject it'' and re-generate it, until the condition is satisfied.

We first generate a probability distribution $P$ uniformly over the unit simplex $\Delta$. There are several algorithms for uniform sampling over the unit simplex \cite{Onn11Simplex}. A simple algorithm, for example, is normalizing a vector of i.i.d. exponential random variables. Define $G(Q) = e^{\eta g(Q)}$, and $a(Q) = \alpha G(Q)$. We will determine $\alpha$ later on such that $\forall Q: \alpha \cdot G(Q) \leq 1$. Having generated $P$, we toss a coin with probability $a(P)$ for ``accept''. If $P$ is accepted, this is the resulting random variable and we set $Q=P$. Otherwise, we draw $P$ again and repeat the process. Let $A$ denote the event of acceptance, and $f_P$ denote the distribution of $P$ which is the uniform distribution over the simplex. The distribution of $Q$ equals the distribution of $P$ given that it was accepted. I.e.:
\begin{equation}\begin{split}\label{eq:2425}
f_Q(q)
&=
f_{P|A}(q)
=
\frac{\Pr \{ A | P=q \} \cdot f_P(q)}{\Pr\{ A \}}
\\&=
\frac{\Pr \{ A | P=q \} \cdot f_P(q)}{\int \Pr \{A | P=q \} \cdot f_P(q) dq}
=
\frac{a(q) \cdot \frac{1}{\vol(\Delta)}}{\int a(q) \cdot \frac{1}{\vol(\Delta)} dq}
\\&=
\frac{G(q)}{\int G(q) dq}
=
\frac{e^{\eta g(q)}}{\int e^{\eta g(q)} dq}
=
w(q)
,
\end{split}\end{equation}
which is the desired distribution.

To determine $\alpha$, suppose we know the maximum of $g(Q)$. This is usually possible since it is a convex optimization problem. Even if this value is not known, a bound on this value will be sufficient. Suppose that $Q^*$ is the maximizer of $g(Q)$ and therefore also of $G(Q)$. Then it is enough to set $\alpha = \frac{1}{G(Q^*)} = e^{-\eta g(Q^*)}$.

An important question from implementation perspective is the average number of iterations required. Since the probability of acceptance $\Pr \{ A \}$ in each iteration is fixed, the number of iterations is a geometrical random variable, with mean $\overline N = \frac{1}{\Pr \{ A \}}$. By Lemma~\ref{lemma:F_exp_weight_UB} we can relate $G(Q^*)$ to $\E G(Q)$ and bound the average number of iterations. Using the lemma we have:
\begin{equation}\begin{split}\label{eq:2394}
g(Q^*)
&\leq
\frac{1}{\eta} \ln \left[ \frac{\displaystyle \int e^{\eta g(Q)} dQ}{\vol(\Delta)} \right] + \frac{d}{\eta} \ln \left( \frac{\eta e n  g_0}{d} \right)
\\& \leq
\frac{1}{\eta} \ln \left(  \E  \left[ G(P) \right] \right) + \frac{d}{\eta} \ln \left( \frac{\eta e n g_0}{d} \right)
,
\end{split}\end{equation}
where $d = |\mathcal{X}|-1$ is the dimension of the unit simplex. We obtain the following bound on $\alpha$:
\begin{equation}\label{eq:2403}
\alpha = e^{-\eta g(Q^*)} \geq \frac{1}{\E  \left[ G(P) \right]} \cdot \left( \frac{\eta e n g_0}{d} \right)^{-d}
,
\end{equation}
and the average number of iterations can be bounded:
\begin{equation}\begin{split}\label{eq:2394x}
\overline N
&=
\frac{1}{\Pr \{ A \}}
=
\frac{1}{\E \left[ \Pr \{ A | P \} \right]}
\\&=
\frac{1}{\E \left[ a(P) \right]}
=
\frac{1}{\alpha \E \left[ G(P) \right]}
\leq
\left( \frac{\eta e n g_0}{d} \right)^{d}
.
\end{split}\end{equation}
Since $\eta$ is polynomial in $n$ and tends to $0$, $\overline N$ grows slower than $n^d$, however this number is still prohibitively large.

The algorithm described is summarized in Table~\ref{tbl:algorithm_for_hatQ}.

\begin{table}
  \centering
\begin{tabular}{|p{8cm}|}
\hline
\textbf{Generation of a random variable $X \sim \hat Q$, \eqref{eq:2393}, \eqref{eq:2398}} \\
\hline
\begin{enumerate}
\item Compute the maximum of $g(Q)$ (a convex optimization problem), or a bound on it.
\item Set $\alpha \leq e^{-\eta \max_Q g(Q)}$.
\item Draw $Q$ uniformly over the unit simplex \cite{Onn11Simplex}. \label{line:2488}
\item Toss a coin and with probability $1 - \alpha e^{\eta g(Q)}$ return to step \ref{line:2488}.
\item Draw $X$ randomly according to the distribution $Q(x)$.
\end{enumerate}
\\
\hline
 \end{tabular}  \caption{An algorithm to generate $X \sim \hat Q$}\label{tbl:algorithm_for_hatQ}
\end{table}

\subsection{Why ``follow the leader'' fails}\label{sec:FL_failure_example}
As noted in Section~\ref{sec:toy_categorization}, the relation of the synthetic prediction problem to prediction under the absolute loss function, implies that the FL predictor cannot be applied to our problem. Here we give a specific example to see why FL fails, based on the channel defined in Section~\ref{sec:toy_categorization}. We construct the following sequence of channels: the channel at $i=1$ is a mixture of $W_0$ with probability $\half$ and a completely noisy channel $Y=\Ber \left( \half \right)$. For this channel $I(Q,W) = \half I(Q,W_0)$. At time $i=2$, the best a-posteriori strategy is $q=0$. The sequence of channels from time $i=2$ onward is the alternating sequence $(W_1,W_0,W_1,W_0,\ldots)$. It is easy to see that the resulting cumulative rates are linear functions of $q$ and thus the optimum is attained at the boundaries of $[0,1]$ and $q_i = (0,1,0,1,\ldots)$. At each time, since the channel that slightly dominates the past is opposite of the channel that is about to appear, the FL predictor chooses the prior that yields the \emph{least} mutual information, and ends up having a zero rate in time instances $i=2,\ldots,n$. On the other hand, by using a uniform fixed prior, a competitor may achieve an average rate of $\half$ over these symbols. Therefore the normalized regret of FL would be at least $\half$, and does not vanish asymptotically.

The problem with the FL predictor is that it takes a decision based on a slight inclination of the cumulative rate toward one of the extremes.

Note that for $|\mathcal{X}|=4, |\mathcal{Y}|=2$, $I(Q,W)$ does not satisfy the Lipschitz condition required in \cite[Theorem 1]{SeqDecision1993} for this strategy to work.


\end{document}